\documentclass[3p,times,11pt]{elsarticle}
\usepackage{algorithm} 
\usepackage{algpseudocode}
\usepackage[titletoc,title]{appendix}
\usepackage{mdframed}
\usepackage{float}
\usepackage{wrapfig,blindtext}
\usepackage{graphicx} 
\usepackage{epstopdf} 
\usepackage{pslatex} 
\usepackage{amsmath,amssymb}
\usepackage{caption}
\usepackage{subcaption}
\usepackage{amsfonts,amsthm} 
\usepackage[english]{babel} 
\usepackage[dvipsnames]{xcolor}
\usepackage{enumitem}
\usepackage{booktabs}
\usepackage{wasysym}
\usepackage{pdfpages}
\usepackage{float}
\usepackage{comment}
\usepackage{soul}

\usepackage{nomencl}
\usepackage{multirow}
\usepackage{soul}
\usepackage{cancel}
\usepackage{ulem}

\usepackage[font=small,labelfont=bf,
   justification=justified,format=plain]{caption}

\usepackage{comment}

\usepackage{nomencl}
  \makenomenclature

\newcommand\norm[1]{\left\lVert#1\right\rVert_2}

\newcommand{\MC}{\mathcal}
\DeclareMathOperator*{\argmin}{arg\,min}

\newcommand{\itermaxLinMS}{\ensuremath{l}}      

\newcommand{\iterIdx}{\ensuremath{n}}           
\newcommand{\normVarIdx}{\ensuremath{v}}        
\newcommand{\varIdx}{\ensuremath{i}}            
\newcommand{\genIdx}{\ensuremath{i}}            

\newcommand{\linearFOM}{\ensuremath{\mathbf{u}}}

\newcommand{\stateVar}{\ensuremath{q}}
\newcommand{\solConsFOM}{\ensuremath{\mathbf{\stateVar}}}
\newcommand{\solPrimFOM}{\ensuremath{\mathbf{\stateVar}_p}}
\newcommand{\solPrimFOMAdaptive}{\ensuremath{\hat{\mathbf{\stateVar}}_p}}
\newcommand{\solConsFOMFunc}[1]{\ensuremath{\solConsFOM \left( #1 \right)}}

\newcommand{\solPrimFOMRef}{\ensuremath{\mathbf{\stateVar}_{p,\text{ref}}}}

\newcommand{\solPrimFOMUnst}{\ensuremath{\mathbf{\stateVar}_p^\prime}}

\newcommand{\solPrimROMFull}{\ensuremath{\widetilde{\mathbf{\stateVar}}_p}}

\newcommand{\QoI}{\ensuremath{\mathbf{\phi}}}
\newcommand{\solPrimROMRed}{\ensuremath{\mathbf{\stateVar}}_{r}}
\newcommand{\solPrimFOMVar}{\ensuremath{\mathbf{\stateVar}_{p,\varIdx}}}

\newcommand{\solPrimROMProj}{\ensuremath{\mathbf{\bar{\stateVar}}_{p}}}
\newcommand{\solPrimROMProjVar}{\ensuremath{\mathbf{\bar{\stateVar}}_{p,\varIdx}}}

\newcommand{\rhsVar}{\ensuremath{f}}
\newcommand{\rhs}{\ensuremath{\mathbf{\rhsVar}}}
\newcommand{\rhsFunc}[1]{\ensuremath{\rhs \left( #1 \right)}}

\newcommand{\gmROM}{\ensuremath{{\widetilde{\boldsymbol{\Gamma}}}}}

\newcommand{\resVar}{\ensuremath{r}}
\newcommand{\res}{\ensuremath{\mathbf{\resVar}}}
\newcommand{\resFunc}[1]{\ensuremath{\res \left( #1 \right)}}

\newcommand{\scaleVarCons}{\ensuremath{P}}
\newcommand{\scaleMatCons}{\ensuremath{\mathbf{\MakeUppercase{\scaleVarCons}}}}
\newcommand{\scaleVarPrim}{\ensuremath{H}}
\newcommand{\scaleMatPrim}{\ensuremath{\mathbf{\MakeUppercase{\scaleVarPrim}}}}

\newcommand{\trialBasisVar}{\ensuremath{V}}

\newcommand{\trialBasisPrim}{{\ensuremath{\mathbf{\MakeUppercase{\trialBasisVar}}_p}}}

\newcommand{\trialBasisPrimN}[1]{{\ensuremath{\mathbf{\MakeUppercase{\trialBasisVar}}^{#1}_p}}}
\newcommand{\testBasisVar}{\ensuremath{W}}

\newcommand{\testBasisPrim}{\ensuremath{{\mathbf{\MakeUppercase{\testBasisVar}}}_{{p}}}}

\newcommand{\testBasisPrimGPOD}{\ensuremath{{\mathbf{\overline{\MakeUppercase{\testBasisVar}}}}_{{p}}}}

\newcommand{\sigmaPOD}{\ensuremath{\widetilde{\sigma}}}

\newcommand{\resApprox}{\ensuremath{\mathbf{\overline{\resVar}}}}
\newcommand{\resBasisVar}{\ensuremath{U}}
\newcommand{\resBasis}{\ensuremath{\mathbf{\MakeUppercase{\resBasisVar}}}}
\newcommand{\sampVar}{\ensuremath{S}}
\newcommand{\sampMat}{\ensuremath{\mathbf{\MakeUppercase{\sampVar}}}}

\newcommand{\trialSpacePrim}{\ensuremath{\MC{\MakeUppercase{\trialBasisVar}}_p}}

\newcommand{\numDOF}{\ensuremath{N}}            
\newcommand{\numElements}{\ensuremath{N_{elem}}}    
\newcommand{\numVars}{\ensuremath{N_{var}}}     
\newcommand{\numSolModes}{\ensuremath{n_p}}     
\newcommand{\numSolModesTotal}{\ensuremath{n_{p,total}}}
\newcommand{\numResModes}{\ensuremath{n_d}}     
\newcommand{\numSamps}{\ensuremath{n_s}}
\newcommand{\numSnaps}{\ensuremath{n_t}}

\newcommand{\timeVar}{\text{t}}                 
\newcommand{\dt}{\ensuremath{\Delta \text{t}}}  
\newcommand{\dTimeVar}{\text{dt}}               

\newcommand{\errROMVsFOM}{\ensuremath{\epsilon}}

\newcommand{\errProjVsFOM}{\ensuremath{\bar{\epsilon}}}

\newcommand{\water}{\ensuremath{\text{H}_2\text{O}}}
\newcommand{\oxygen}{\ensuremath{\text{O}_2}}
\newcommand{\methane}{\ensuremath{\text{CH}_4}}
\newcommand{\carbonDiox}{\ensuremath{\text{CO}_2}}

\newcommand{\lp}{\left(}
\newcommand{\rp}{\right)}

\newcommand{\zeroVec}{\ensuremath{\mathbf{0}}}

\newcommand{\addrOne}[1]{\textcolor{black}{#1}}
\newcommand{\addrTwo}[1]{\textcolor{black}{#1}}

\journal{Journal of Computational Physics, January 2023}

\begin{document}
\topmargin -1.5cm
\textheight 23cm

\begin{frontmatter}

\title{Predictive Reduced Order Modeling of Chaotic Multi-scale Problems Using Adaptively Sampled Projections}

\author{Cheng Huang}
\ead{chenghuang@ku.edu}
\address{University of Kansas, Lawrence, KS}
\author{Karthik Duraisamy}
\ead{kdur@umich.edu}
\address{University of Michigan, Ann Arbor, MI}

\begin{abstract}
An adaptive projection-based reduced-order model (ROM) formulation is presented for model-order reduction of problems featuring chaotic and convection-dominant physics. An efficient method is formulated to adapt the basis at every time-step of the on-line execution  to account for the unresolved dynamics. The adaptive ROM is formulated in a Least-Squares  setting using a variable transformation to promote stability and robustness. An efficient strategy is developed to incorporate non-local information in the basis adaptation,  significantly enhancing the predictive capabilities of the resulting ROMs. A detailed analysis of the computational complexity is presented, and validated. The adaptive ROM formulation is shown to require  negligible offline training and naturally enables both future-state and parametric predictions. The formulation is evaluated on  representative reacting flow benchmark problems, demonstrating that the ROMs are capable of providing efficient and accurate predictions including those involving significant changes in dynamics due to parametric variations, and transient phenomena. A key contribution of this work is the development and demonstration of a comprehensive ROM formulation that targets {\em predictive capability} in chaotic, multi-scale, and transport-dominated problems.

\end{abstract}

\end{frontmatter}

\section{Introduction}
\label{sec:intro}

Even though high-performance computing is enabling high-fidelity simulations of complex multi-scale physics relevant to real engineering systems~\citep{WangFR2017,adityaDNS,OefeleinSupercritical}, many practical computations remain intractable. This is especially true in many-query work flows such as optimization and uncertainty quantification. Projection-based model-order reduction (MOR)~\citep{Lumley1997, Graham1997, Benner_Gugercin_Willcox_PMR2015} has emerged as a promising avenue to construct reduced-order models (ROMs) with reduced computational cost by many orders of magnitude in comparison to the full-order model (FOM)~\citep{McQuarrieOpInf2021,Carlberg2017,HuangMPLSVT2022}. Typically, the construction of projection-based ROMs includes two major steps: 1) An \emph{offline} stage in which a small number of expensive FOM simulations are performed over a range of target parameters to obtain a low-dimensional subspace that approximates the high-dimensional FOM solution space; and 2) The \emph{online} stage which projects the FOM equations onto the low-dimensional subspace, leading to a set of reduced equations for the ROM. Over the past decade,  projected-based ROMs have been successfully demonstrated in many areas such as flow control~\cite{Barbagallo2012, Barbagallo2009, Barbagallo2011}, aeroelasticity~\cite{Lieu2007, Lucia2004}, shape optimization~\citep{Zahr_ParametricROM_Optimization2014,Yano_TopologyOpt_2021,Wen_ShapeOpt_2022,Tezaur_ParametricROM_2022}, aerodynamics~\citep{GrimbergLocalPROM2021_F16,Yano_GoalOrientedAdaptiveROM2020,Du_Yano_DGROM2022}, building digital twins~\citep{McClellan_ROMfor_UAV_DigitalTwin2022,Kapteyn2020}, hypersonics~\citep{BloniganHypersonics}, rocket combustion~\citep{HuangMPLSVT2022,McQuarrieOpInf2021,Swischuk2020AIAAJ_LL,qian2022_3Dcvrc}, and detonation~\citep{Farcas_2022_OpInf_RDE}. 

Despite the many successes, three major issues remain to be addressed to enable the reliable use of projection-based ROMs in parametric {\em predictions} of complex convection-dominated multi-scale problems such as those involving turbulence and combustion:

First, it is well-recognized that projection-based ROMs suffer from stability issues, which may arise from the inherent lack of numerical stability of the projection method itself (e.g. Galerkin projection~\citep{Rempfer2000}), mode truncation (e.g. removal of low-energy spatial modes~\cite{Bergmann2009}), or simplifications of model equations~\citep{Noack2005}. Several remedies for these stability issues have been proposed in the literature, which include but are not limited to: 1) balanced truncation~\citep{Moore1981, Pernebo1982} for linear dynamical systems, 2) eigenvalue reassignment~\cite{Kalashnikova2014}, convex optimization~\citep{Amsallem_ROMstabilization2012}, or a combination of the two~\cite{rezaian2020hybrid} from a control perspective, 3) closure models to account for the effect of truncated ROM dynamics based on the resolved ROM dynamics~\citep{Bergmann2009,Lucia2003,SanVMS2015,IliescuVMS2014,StabileVMS2019,parish2017non,parish2017dynamic,Ahmed_ClosuresForROM2021}, 4) defining appropriate inner products to obtain low-dimensional space~\citep{Rowley2004,Barone2008,Barone2009JCP,Afkham2020}, 5) least-squares residual minimization through Petrov-Galerkin projection~\cite{carlberg2013gnat,Carlberg2017,Carlberg2018Consv,GrimbergPROM2020_stability,HuangMPLSVT2022}, and 6) applying limiters for local stability~\citep{HuangAIAAJ2019,Huang2020_SpeciesLimiters}.

Second, conventional projection-based MOR methods leverage a \emph{linear} low-dimensional subspace, such as that spanned by a proper orthogonal decomposition (POD) basis, to achieve dimensionality reduction. However, linear dimensionality reduction is limited in effectiveness in problems that exhibit a slow decay of Kolmogorov N-width~\citep{Pinkus_KolmogorovNwidth,Greif_KolmogorovNwidth}, common characteristics observed in most convection-dominated multi-scale problems. This specific issue is usually referred to as the \emph{Kolmogorov barrier} and several approaches have been proposed to \textit{break} this barrier:

\begin{itemize}
    \item One group of researchers seek to construct \textit{multiple} local \textit{linear} subspaces, each of which is tailored for a particular region of the state space corresponding to specific dynamics observed in the target problems. In contrast to the conventional approaches which construct one \textit{single} global subspace for the entire state space of interest, the local subspaces provide significant improvement in dimensionality reduction and therefore significant savings in ROM construction, which otherwise would require an impractical number of basis modes using the global subspace. The construction of local subspaces is often achieved by clustering the snapshots into different partitions, each of which represents unique features of the dynamics and can be defined based on parameter space~\citep{Eftang_hpCertified2010,Eftang_ParameterMultiD2012,Peherstorfer_LDEIM2014}, time~\citep{Dihlmanna_ROM_adaptiveTimePart2011,Rapún_localPOD2010,Parish_wLSPG2021,Shimizu_wLSPG2021}, predefined similarity metrics (e.g. k-means~\citep{Peherstorfer_LDEIM2014,AmsallemLocalROM,GrimbergLocalPROM2021_F16,Amsallem_NonlinearMOR_LocalROB2012,Kaiser_2014ClusteredROMMixingLayer}), or dissimilarity criterion (e.g. projection error~\citep{Amsallem_ProjErr_LocalROM2016}). Recently, Geelen and Willcox~\citep{GeelenWillcoxLocalROM} incorporated this idea of localized subspaces in the operator inference framework, which enabled localized data-driven ROMs and achieved significant improvement in efficiency and accuracy. 
    
    \item Another group of researchers pursue to \textit{break} the Kolmogorov barrier by seeking nonlinear manifolds instead of linear subspaces for the low-rank representations of the dynamics. One class of popular approaches leverage transformation of the subspaces/snapshots to construct such manifolds with the goal of recovering low-rank structures with fast Kolmogorov N-width decays, which include the method of freezing~\citep{Ohlberger_2013MethodOfFreezing}, shifted POD~\citep{Reiss_2018sPOD,Schulze_2019sPOD_PulsedDetonation,Reiss_2021_OptMR}, transported subspaces~\citep{Rim_2019TransportedBasis} or snapshots~\citep{Nair_2018_TransportedSnapshots}, and implicit feature tracking~\citep{Alireza_2023ImpTracking}. Moreover, other researchers seek to directly compute the nonlinear manifolds via \textit{either} convolutional autoecoders~\citep{LeeCarlbergNonlinearManifold,KimChoiNonlinearMainfold} \textit{or} explicitly quadratic approximation~\citep{Barnett_QuadraticPROM2022,Geelen_2023_Quadratic,Rutzmoser_2017QuadraticStructural}. 
    
    \item The third group of researchers resorted to adaptive MOR~\citep{Peherstorfer_BreakingKolmogorovBarrier2022,peherstorfer2015online,ZimmermannAdaptiveBasis2018}, which updates the ROM in the online stage to \textit{either} satisfy \textit{a posteriori} error estimation \textit{or} seek an optimal representation of the evaluated dynamics in the target problems. Yano et al.~\citep{Yano_TopologyOpt_2021} devised a trust-region method that informs ROM construction procedure to meet the accuracy conditions for topology optimization problems. Carlberg~\citep{CarlbergAdaptiveBasis2015} enriches the reduced-basis space online by ‘splitting’ a given basis vector into several vectors with disjoint support \textit{a posteriori} without requiring additional FOM solves, analogous to mesh-adaptive h-refinement, which has been further improved by Etter and Carlberg~\citep{EtterAdaptiveBasis2020} via vector-space sieving. Peherstorfer~\citep{PeherstorferADEIM} adapts the affine approximation space by exploiting the \textit{locality} in space and time of propagating coherent structures evaluated from the sampled FOM solutions and recently this method has been extended and demonstrated for a premixed flame problem featuring convection-dominated physic~\citep{Uy_2022_adaptiveROMFlame}. Ramezanian et al.~\citep{Ramezanian_2021_OnTheFlyROM} proposed an \textit{on-the-fly} ROM method for reactive species transport by deriving evolution equations for low-rank components (i.e. a low-dimensional time-dependent manifold) through a variational principle.
\end{itemize}

Third, even with the stability issues and Kolmogorov barrier addressed with the aforementioned approaches, one remaining challenge in projection-based ROMs (or ROM methods in general) lies in their \emph{restricted predictive capabilities} for \emph{convection-dominated problems}. For example, if a ROM is utilized to predict
conditions (e.g. Reynolds number, Mach number, etc.) that are different
from the ones used to train the ROM during the offline stage, there is usually no guarantee regarding the accuracy of the ROM predictions. More importantly, though projection-based ROMs have been successfully demonstrated to reasonably predict the future state of coherent dynamics (e.g. acoustics) in convection-dominated problems such as rocket combustion~\citep{HuangMPLSVT2022,McQuarrieOpInf2021,qian2022_3Dcvrc,Farcas_2022_OpInf_RDE}, these ROMs provide limited predictive capabilities of \addrTwo{non-coherent} features (e.g. turbulence and shocks) in the target problems since essentially every future state is unique. Thus, many of the above techniques can catastrophically fail to be predictive. One possible solution may be to collect a significant amount of FOM training data in the offline stage to construct the ROM, which, however, can eventually make the cost of ROM construction (offline training + online calculations) intractable and counters the purpose of ROM in general. On the other hand, the adaptive MOR discussed above opens a promising avenue to address this limitation by minimizing, or completely eliminating, the offline training stage requirement while building the ROM online, which inherently enhances the predictive capabilities of the ROM and enables \textit{true} predictions of \addrTwo{convection-dominated} features in the problems~\citep{PeherstorferADEIM,Huang_CBROM2022,Arnold-Medabalimi_2022GTMC}. 

In the present work, inspired by the works in adaptive MOR, especially by Peherstorfer~\citep{PeherstorferADEIM}, we develop a comprehensive adaptive projection-based reduced-order model framework for complex multi-scale applications to \textit{break} the Kolmogorov barrier and enable \textit{true} predictive capabilities for projection-based ROM. Detailed evaluations of the technique are presented in challenging reacting flow applications exhibiting convection-dominant features.

The remainder of the paper is organized as follows. Section~\ref{sec:fom} presents the full-order model (FOM) and time discretization. Section~\ref{sec:mplsvt} reviews the formulation for model-form
preserving least-squares with variable transformation (MP-LSVT). Section~\ref{sec:adaptivity} introduces the proposed adaptive ROM formulation. 
 Section~\ref{subsubsec:comp_proc} presents the algorithm and the underlying computational complexity.
 Section~\ref{sec:results} presents numerical results and analysis for ROMs of benchmark reacting flow problems and assesses their robustness, efficiency, and predictive capabilities. In Section~\ref{sec:conclusion}, we provide concluding remarks and perspectives.

\section{Full-Order Model}
\label{sec:fom}

The governing equations of the full-order model (FOM) can be represented as a generic dynamical system
\begin{equation}
    \frac{\text{d} \solConsFOM(\solPrimFOM)}{\dTimeVar} = \rhsFunc{\solPrimFOM, \timeVar} \ \ \ \text{with} \ \ \ \solPrimFOM \left( \timeVar = 0 \right) = \solPrimFOM^0,
     \label{eq:fom_semi_discrete}
\end{equation}
where $\timeVar \in [0,t_f]$ is the solution time, which spans the time interval from $0$ to $t_f$, $\solPrimFOM: [0, t_f] \rightarrow {\mathbb{R}^{\numDOF}}$ is the vector of solution (or state) variables, $\solPrimFOM^0 \in {\mathbb{R}^{\numDOF}}$ is the vector of states to be specified as the initial conditions at $t = 0$, $\solConsFOM: \mathbb{R}^{\numDOF} \rightarrow \mathbb{R}^{\numDOF}$ and $\rhs: \mathbb{R}^{\numDOF} \times [0,t_f] \rightarrow \mathbb{R}^{\numDOF}$ are (typically highly non-linear) functions of  $\solPrimFOM$. $\numDOF$ is the total number of degrees of
freedom in the system (e.g. for finite volume/element method, \addrTwo{$\numDOF = \numElements \times \numVars \times p$}, where $\numElements$ is the total number of elements, $\numVars$ is the number of state variables in each element\addrTwo{, and $p$ represents the order of accuracy of the numerical methods}). For a FOM based on conservation laws, the function, $\solConsFOM(\cdot)$, represents the conservative state. In the context of spatially discretized partial differential equations, the function $\rhs(\cdot)$, represents surface fluxes, source terms, and body forces arising from the spatial discretization of the governing equations.

Different time-discretization methods can be introduced to solve Eq.~\ref{eq:fom_semi_discrete} (e.g. linear multi-step, or Runge--Kutta methods~\citep{ButcherNumMeth}). For all the numerical examples presented in the current paper, we use linear multi-step methods for both FOM and ROM calculations and refer the reader to~\citep{HuangMPLSVT2022} for details. An $\itermaxLinMS$-step version of linear multi-step methods can be expressed as
\begin{equation}
    \resFunc{\solPrimFOM^\iterIdx} \triangleq \solConsFOMFunc{\solPrimFOM^{\iterIdx}} + \sum^{\itermaxLinMS}_{j=1} \alpha_j \solConsFOMFunc{\solPrimFOM^{\iterIdx-j}} - \dt \beta_0 \rhsFunc{\solPrimFOM^\iterIdx, \timeVar^\iterIdx} - \dt \sum^{\itermaxLinMS}_{j=1} \beta_j \rhsFunc{\solPrimFOM^{\iterIdx-j}, \timeVar^{\iterIdx-j}} = 0 \ \ \ \ (n \ge l),
    \label{eq:fom_linear_multi_discrete}
\end{equation}
where $\dt \in \mathbb{R}^{+}$ is the physical time step for the numerical solution, and the coefficients $\alpha_j$, $\beta_j \in \mathbb{R}$ are determined  based on $\itermaxLinMS$. If $\beta_0 = 0$, the method is explicit; otherwise, the method is implicit. $\res: \mathbb{R}^{\numDOF} \rightarrow \mathbb{R}^{\numDOF}$ is defined as the FOM equation residual. The state variables, $\solPrimFOM^\iterIdx$, are solved for at each time step so that $\resFunc{\solPrimFOM^\iterIdx} = \zeroVec$. 

\section{Model-form Preserving Model Reduction for Transformed Solution Variables}
\label{sec:mplsvt}
For problems involving multiscale  phenomena with strong convection and non-linear effects, it is well-recognized that ROM robustness can be a major issue. To address this challenge, we  pursue the model-form preserving least-squares with variable transformation (MP-LSVT) formulation to construct the reduced-order model (ROM).  This methodology is briefly described below -  we refer the reader to~\citep{HuangMPLSVT2022} for further details.

\subsection{Construction of Low-dimensional Subspace for Solution Variables}
\label{subsec:pod}
The state $\solPrimFOM$ in Eq.~\ref{eq:fom_semi_discrete} can be expressed in a trial space $\trialSpacePrim \triangleq \text{Range}(\trialBasisPrim)$ (i.e. a low-dimensional subspace), where $\trialBasisPrim \in \mathbb{R}^{\numDOF \times \numSolModes}$ is the trial basis matrix. Define $\solPrimFOMUnst(\timeVar) \triangleq \solPrimFOM(\timeVar) - \solPrimFOMRef$, where $\solPrimFOMRef$ is a reference state. Possible reference states include the initial FOM solution, $\solPrimFOMRef = \solPrimFOM(\timeVar = \timeVar_0)$, or the time-averaged FOM solution, $\solPrimFOMRef = \frac{1}{\timeVar_1 - \timeVar_0} \int_{\timeVar_0}^{\timeVar_1} \solPrimFOM(\timeVar) \; \dTimeVar$. We then seek a representation $\solPrimROMFull: [0,t_f] \rightarrow \trialSpacePrim$ such that
\begin{equation}
    \scaleMatPrim \lp \solPrimROMFull - \solPrimFOMRef \rp = \trialBasisPrim \solPrimROMRed.
    \label{eq:pod_qp}
\end{equation}
where $\solPrimROMRed: [0,t_f] \rightarrow \mathbb{R}^{\numSolModes}$ is the reduced state with $\numSolModes$ representing the number of trial basis modes. In this work, $\trialBasisPrim$ is computed via the proper orthogonal decomposition (POD)~\citep{lumley1967structure,Berkooz_1993} from the singular value decomposition (SVD), which is a solution to
\begin{equation}
  \min_{\trialBasisPrim \in \mathbb{R}^{\numDOF \times \numSolModes}} || \mathbf{Q}_p - \trialBasisPrim \mathbf{V}^T_p \mathbf{Q}_p ||_F \ \ \ s.t. \ \ \ \mathbf{V}^T_p \trialBasisPrim = \mathbf{I},
    \label{eq:pod_def}
\end{equation}
where $\mathbf{Q}_p$ is a data matrix in which each column is a snapshot of the solution $\solPrimFOMUnst$ at different time instances. A scaling matrix $\scaleMatPrim \in{\mathbb{R}^{\numDOF \times \numDOF}}$ must be applied to $\solPrimFOMUnst$ such that the variables corresponding to different physical quantities in the data matrix $\mathbf{Q}_p$ have similar orders of magnitude. Otherwise, $\mathbf{Q}_p$ may be biased by physical quantities of higher magnitudes (e.g. total energy). In this work, we normalize all quantities by their $L^2$-norm, as proposed by Lumley and Poje~\citep{Lumley1997}
\begin{equation}
    \scaleMatPrim = diag \lp \scaleMatPrim_1, \ldots, \scaleMatPrim_i, \ldots, \scaleMatPrim_{\numElements} \rp,
    \label{eq:pod_norm}
\end{equation} 
where $\scaleMatPrim_i = diag\left( \phi^{-1}_{1,norm}, \ldots , \phi^{-1}_{\numVars,norm} \right)$. Here, $\phi_{\normVarIdx,norm}$ represents the $\normVarIdx^{th}$ state  variable \addrTwo{in $\solPrimFOM$} and
\begin{equation}
    \phi_{\normVarIdx,norm} = {\frac{1}{\timeVar_1 - \timeVar_0}\int^{\timeVar_1}_{\timeVar_0} \frac{1}{\Omega} \int_{\Omega} \phi'^2_\normVarIdx(\mathbf{x}, \timeVar) \; \text{d}\mathbf{x} \; \text{dt}}.
    \label{eq:l2norm}
\end{equation}

\subsection{Least-squares with Variable Transformation}
\label{subsec:mplsvt}
Leveraging the least-squares Petrov-Galerkin (LSPG) projection formulation proposed by Carlberg et al.~\citep{Carlberg2017}, we develop a model-form preserving least-squares formulation for the FOM in Eq.~\ref{eq:fom_semi_discrete}. Our objective is to minimize the fully-discrete FOM equation residual $\res$, defined in Eqs.~\ref{eq:fom_linear_multi_discrete} with respect to the reduced state, $\solPrimROMRed$
\begin{equation}
    \solPrimROMRed^n \triangleq  \argmin_{\solPrimROMRed \in \mathbb{R}^{\numSolModes}} \norm{\scaleMatCons \resFunc{\solPrimROMFull}}^2 \ \ \ \text{with} \ \ \ \solPrimROMFull^0  = \trialBasisPrim \mathbf{V}^T_p \solPrimFOM^0  \ \ \text{\textit{or}}  \ \ \solPrimROMFull^0  = \solPrimFOM^0,
    \label{eq:mplsvt}
\end{equation}
where the approximate solution variables, $\solPrimROMFull = \solPrimFOMRef + \scaleMatPrim^{-1} \trialBasisPrim \solPrimROMRed$. The equation residual, $\res$, is scaled by $\scaleMatCons$ using the $L^2$-norm, similar to the scaling matrix $\scaleMatPrim$ in Eq.~\ref{eq:pod_norm}
\begin{equation}
    \scaleMatCons = diag \lp \scaleMatCons_1, \ldots, \scaleMatCons_i, \ldots, \scaleMatCons_{\numElements} \rp,
    \label{eq:pod_consv_norm}
\end{equation} 
where $\scaleMatCons_i = diag\left( \varphi^{-1}_{1,norm}, \ldots , \varphi^{-1}_{\numVars,norm} \right)$. Here, $\varphi_{\normVarIdx,norm}$ represents the $\normVarIdx^{th}$ evaluated quantity of $\solConsFOMFunc{\solPrimFOM}$ \addrTwo{following the same formulation in Eq.~\ref{eq:l2norm}} such that each equation in $\res$ has similar contributions to the minimization problem in Eq.~\ref{eq:mplsvt}. It is worth pointing out that the treatment of boundary conditions in the MP-LSVT ROM is fully consistent with the FOM in Eq.~\ref{eq:fom_linear_multi_discrete}, which guarantees that the boundary conditions are satisfied in the ROM. 

Following Eq.~\ref{eq:mplsvt}, a reduced non-linear  system of dimension $\numSolModes$ can then be obtained and viewed as the result of a Petrov-Galerkin projection
\begin{equation}
    \lp \testBasisPrim^\iterIdx \rp^T \scaleMatCons \resFunc{\solPrimROMFull^n} = \zeroVec \ \ \ \text{with} \ \ \ \solPrimROMFull^0  = \trialBasisPrim \mathbf{V}^T_p \solPrimFOM^0  \ \ \text{\textit{or}}  \ \ \solPrimROMFull^0  = \solPrimFOM^0,
    \label{eq:mplsvt_proj}
\end{equation}
where $\testBasisPrim$ is the test basis
\begin{equation}
    \testBasisPrim^\iterIdx =  \frac{\partial \scaleMatCons \resFunc{\solPrimROMFull^n}}{\partial \solPrimROMRed^n} = \scaleMatCons \lp \gmROM^\iterIdx - \dt \beta_0 \widetilde{\mathbf{J}}^\iterIdx \gmROM^\iterIdx \rp \scaleMatPrim^{-1} \trialBasisPrim,
    \label{eq:mplsvt_w}
\end{equation}
with $\widetilde{\mathbf{J}}^\iterIdx = \left[\partial \rhs / \partial \solPrimFOM \right]^\iterIdx_{\solPrimFOM = \solPrimROMFull}$ and $\gmROM^\iterIdx = \left[\partial \solConsFOM / \partial \solPrimFOM \right]^\iterIdx_{\solPrimFOM=\solPrimROMFull}$. \addrOne{Similar derivation can be applied to the multi-stage Runge–Kutta method and we refer the readers to Ref.~\cite{HuangMPLSVT2022} for detail.}

\subsection{Hyper-reduction}
\label{subsec:hyper_reduction}

Even though the MP-LSVT method leads to a robust ROM of much lower dimension ($\numSolModes \ll \numDOF$), the evaluations of the non-linear terms remain a bottleneck as they involve $O(\numDOF)$ operations. One remedy to circumvent this issue is via \textit{hyper-reduction} to reduce the computational complexity in evaluating these non-linear terms. Popular hyper-reduction methods include the discrete empirical interpolation method (DEIM)~\citep{chaturantabut2010nonlinear,drmac_qdeim} and gappy POD~\citep{Everson1995_gappyPOD,Willcox2006_gappyPOD,carlberg2013gnat}. The DEIM evaluates the non-linear terms at a small subset of all the components (i.e. sampled) and approximates the terms at the other components (i.e. unsampled) via interpolation in low-dimensional subspaces. On the other hand, gappy POD approximates the non-linear terms at \addrTwo{all} the components via regression rather than via interpolation, which is pursued in the current paper
\begin{equation}
\resApprox \approx \resBasis \lp \sampMat^T \resBasis \rp^{+}  \sampMat^T \res,
\label{eq:hyperreduced_res}
\end{equation}
where $\sampMat \in \mathbb{S}^{\numDOF \times \numSamps}$ is a selection operator that belongs to a class of matrices with $\numSamps$ columns (i.e. sampling points) of the identity matrix, $\mathbf{I} \in \mathbb{I}^{\numDOF \times \numDOF}$, and $\resBasis \in \mathbb{R}^{\numDOF \times \numResModes}$ is a basis set used to approximate the non-linear term, $\res$. Typically, $\resBasis$ is constructed via POD from snapshots of $\res$. However, it is found that setting $\resBasis$ to the trial POD basis $\trialBasisPrim$ in Eq.~\ref{eq:pod_qp} also produces excellent approximations~\citep{HuangMPLSVT2022}; this method is used for all results presented in this paper. By applying sparse sampling and reconstruction to the non-linear equation residual $\res$ in Eq.~\ref{eq:mplsvt} 
\begin{equation}
    \solPrimROMFull^{\iterIdx} \triangleq \argmin_{\solPrimROMFull^{\iterIdx} \in \textrm{Range}(\trialBasisPrim)} \norm{\resBasis \lp \sampMat^T \resBasis \rp^{+} \sampMat^T \scaleMatCons \resFunc{\solPrimROMFull^{\iterIdx}}}^2.
    \label{eq:mplsvt_hyper}
\end{equation}
The resulting test basis $\testBasisPrimGPOD^\iterIdx$  is then given by
\begin{equation}
    \testBasisPrimGPOD^\iterIdx = \frac{ \partial \resBasis \lp \sampMat^T \resBasis \rp^{+} \sampMat^T \scaleMatCons \resFunc{\solPrimROMFull^{\iterIdx}}}{\partial \solPrimROMRed^{\iterIdx}} = \resBasis \lp \sampMat^T \resBasis \rp^{+} \sampMat^T \frac{ \partial \scaleMatCons \resFunc{\solPrimROMFull^{\iterIdx}}}{\partial \solPrimROMRed^{\iterIdx}} = \resBasis \lp \sampMat^T \resBasis \rp^{+} \sampMat^T \testBasisPrim^\iterIdx.
    \label{eq:mplsvt_w_hyper}
\end{equation}
Thus, only $\numSamps$ rows of the test basis $\testBasisPrim^{\iterIdx}$ must be evaluated. This is yet another major step in reducing the number of necessary computations. With the approximated test basis in hand, the hyper-reduced MP-LSVT ROM in physical time becomes
\begin{equation}
    \lp \sampMat^T \testBasisPrim^\iterIdx \rp^T \left[ \lp \sampMat^T \resBasis \rp^{+} \right]^T \lp \sampMat^T \resBasis \rp^{+} \sampMat^T \scaleMatCons \resFunc{\solPrimROMFull^{\iterIdx}} = \zeroVec.
    \label{eq:mplsvt_proj_hyperS}
\end{equation}
Although $[(\sampMat^T \resBasis)^{+}]^T (\sampMat^T \resBasis)^{+} \in \mathbb{R}^{\numSamps \times \numSamps}$ can be precomputed offline, this matrix may become quite large and therefore infeasible to store in memory. In reality \addrTwo{for ROM with static basis}, only $(\sampMat^T \resBasis)^{+} \in \mathbb{R}^{\numResModes \times \numSamps}$ is precomputed offline and loaded into memory at runtime \addrTwo{while if the sampling points or the basis are getting updated, $(\sampMat^T \resBasis)^{+}$ needs to be recomputed online}.

\section{Online Adaptation of Basis and Sampling}
\label{sec:adaptivity}
While the MP-LSVT method improves the robustness and accuracy of the ROM,  predictive capabilities are still restricted by the use of \emph{linear static} basis, which has been shown to be inadequate for predictions in problems with slow \emph{Kolmogorov N-width} decay~\citep{HuangMPLSVT2022,McQuarrieOpInf2021}. Several remedies have been proposed to address this challenge through, for example, nonlinear bases~\citep{LeeCarlbergNonlinearManifold,KimChoiNonlinearMainfold,Geelen_2023_Quadratic,Barnett_QuadraticPROM2022}, and online adaptation~\citep{PeherstorferADEIM,ZimmermannAdaptiveBasis2018} etc. In the current work, the major focus is put on  online adaption methods, which has been demonstrated by the current authors~\citep{Huang_CBROM2022,Arnold-Medabalimi_2022GTMC} necessary to construct \emph{truly} predictive ROMs for chaotic \addrTwo{convection-dominated} fluid flow problems. An ideal  adaption method would  update the trial basis, $\trialBasisPrim$, and the sampling points,  $\sampMat$, during the online \emph{hyper-reduced} ROM calculation (Eq.~\ref{eq:mplsvt_hyper}) simultaneously:
\begin{equation}
    \{ \solPrimROMRed^n, \trialBasisPrimN{n}, \sampMat^{n} \} \triangleq \argmin_{ \solPrimROMRed \in \mathbb{R}^{\numSolModes} , \trialBasisPrimN{\iterIdx} \in \mathbb{R}^{\numDOF \times \numSolModes} , \sampMat^{n} \in \mathbb{S}^{\numDOF \times \numSamps } } \norm{\resBasis \left[ (\sampMat^n)^T \resBasis \right]^{+} (\sampMat^n)^T \scaleMatCons \resFunc{\solPrimROMFull^{\iterIdx}}}^2 \ \ \ \text{with} \ \ \ \resBasis = \trialBasisPrim^\iterIdx
    \label{eq:adaptiverom:def}
\end{equation}
where $\res$ is the fully-discrete FOM equation residual defined in Eq.~\ref{eq:fom_linear_multi_discrete}, $\solPrimROMFull^{\iterIdx} = \solPrimFOMRef + \scaleMatPrim^{-1} \trialBasisPrimN{\iterIdx} \solPrimROMRed^{\iterIdx}$. In the current work, we refer to Eq.~\ref{eq:adaptiverom:def} as \textit{adaptive} ROM formulation (i.e. both basis and sampling are updated online). However, directly solving Eq.~\ref{eq:adaptiverom:def} leads to a prohibitively expensive and potentially intractable problem. Therefore, we pursue a decoupled approach based on a predictor-corrector idea. In the rest of this section, we first use a linear FOM to illustrate the predictor-corrector approach and then provide detailed descriptions of the \textit{adaptive} ROM formulation and algorithm.
\subsection{Illustration of Predictor-corrector Approach}
\label{subsec:adaptivity:predictor-corrector}
For illustrative purposes, consider a one-time-step advancement of a linear FOM
\begin{equation}
\frac{\linearFOM^n - \linearFOM^{n-1}}{ \dt} =  \mathbf{J} \linearFOM^{n},
\label{eq:predictor-corrector:fom}
\end{equation}
where $\linearFOM \in \mathbb{R}^N$ and $\mathbf{J} \in \mathbb{R}^{N \times N}$.
An exact orthogonal linear decomposition can be written as $\linearFOM=\mathbf{V}\linearFOM_r+\mathbf{V}_\bot\linearFOM_{r\bot}$ with $\mathbf{V}^T_\bot \mathbf{V} = 0$, $\mathbf{V}^T_\bot \mathbf{V}_\bot = \mathbf{I}$, and $\mathbf{V}^T \mathbf{V} = \mathbf{I}$ , where \addrTwo{$\linearFOM_r \in \mathbb{R}^{\numSolModes}$} and \addrTwo{$\linearFOM_{r\bot} \in \mathbb{R}^{N-\numSolModes}$} represent the resolved and unresolved reduced states respectively. Assuming that at time step $n-1$, the basis is perfect and fully resolves the state variable $\linearFOM^{n-1}$ (i.e. $\linearFOM_{r\bot}^{n-1}=0$), the FOM equation can be rewritten in terms of the above decomposition as 
 \begin{equation}
 (\mathbf{I} - \dt \mathbf{J}) (\mathbf{V}\linearFOM_r^{n}+\mathbf{V}_\bot\linearFOM_{r\bot}^{n}) =\mathbf{V}\linearFOM_r^{n-1}.
 \label{eq:predictor-corrector:fom_decomp}
 \end{equation}
By projecting Eq.~\ref{eq:predictor-corrector:fom_decomp} onto $\mathbf{V}$ and $\mathbf{V}_\bot$ respectively \addrTwo{with $\mathbf{B}\triangleq \mathbf{I} - \dt \mathbf{J}$}, the exact evolution equations for the resolved and unresolved reduced states then become
\begin{equation}
      \begin{aligned}
      \addrTwo{\mathbf{V}^T \mathbf{B} \mathbf{V} \linearFOM_r^{n} - \dt \mathbf{V}^T \mathbf{J} \mathbf{V}_\bot \linearFOM_{r\bot}^{n}} & \addrTwo{= \linearFOM_r^{n-1}} \\ \addrTwo{\mathbf{V}_\bot^T \mathbf{B} \mathbf{V}_\bot \linearFOM_{r \bot}^{n} - \dt \mathbf{V}_\bot^T \mathbf{J} \mathbf{V} \linearFOM_{r}^{n}} & \addrTwo{= 0.}
     \end{aligned}
\end{equation}
 Thus, the inadequacy of the basis $\mathbf{V}$ to resolve the solution at time step $n$ can result in the generation of unresolved reduced states $\linearFOM_{r \bot}^{n}$. Solving the latter equation  \addrTwo{$\linearFOM_{r \bot}^{n}=(\mathbf{V}_\bot^T \mathbf{B} \mathbf{V}_\bot)^{-1} \dt
\mathbf{V}_\bot^T \mathbf{J} \mathbf{V} \linearFOM_{r}^{n}$} and substituting in the former, we have the following equation for the exact resolved reduced states
\begin{equation}
 \addrTwo{\left[ \mathbf{V}^T \mathbf{B} \mathbf{V} - \dt^2 \mathbf{V}^T \mathbf{J} \mathbf{V}_\bot (\mathbf{V}_\bot^T \mathbf{B} \mathbf{V}_\bot)^{-1} \mathbf{V}_\bot^T \mathbf{J} \mathbf{V} \right] \linearFOM_{r}^{n} = \linearFOM_r^{n-1}.}
 \label{eq:predictor-corrector:exact_reduced_states}
\end{equation}
On the other hand, it can be easily shown that the ROM formulated using the Galerkin method \addrTwo{with $\linearFOM 
\approx \mathbf{V}\bar{\linearFOM}_r^{n}$ substituted into Eq.~\ref{eq:predictor-corrector:fom}} gives the following evolution equation for the (approximate) resolved reduced states $\bar{\linearFOM}_r$
\begin{equation}
    \addrTwo{\mathbf{V}^T \mathbf{B} \mathbf{V} \bar{\linearFOM}_r^{n} = \linearFOM_r^{n-1}.}
    \label{eq:predictor-corrector:g_reduced_states}
\end{equation}
 Similarly, the ROM formulated least squares petrov galerkin (LSPG) gives the following evolution equation for the (approximate) resolved reduced states 
 \begin{equation}
    \addrTwo{\mathbf{V}^T \mathbf{B}^T \mathbf{B} \mathbf{V} \bar{\linearFOM}_r^{n} =  \mathbf{V}^T \mathbf{B} \mathbf{V} \linearFOM_r^{n-1},}
    \label{eq:predictor-corrector:lspg_reduced_states}
\end{equation}
 It shall be pointed out that for a FOM as in Eq.~\ref{eq:predictor-corrector:fom}, the MP-LSVT is equivalent to LSPG as proved by the current author~\citep{HuangMPLSVT2022} and thus only the evolution equation for LSPG is considered here. By comparing Eqs.~\ref{eq:predictor-corrector:exact_reduced_states},~\ref{eq:predictor-corrector:g_reduced_states}, and~\ref{eq:predictor-corrector:lspg_reduced_states}, we then seek to adapt the basis, $\mathbf{V}$, with an augmentation $\boldsymbol{\delta}\mathbf{V}$ such that {\em the subspace is updated to account for the generation of new unresolved scales} at the current time step $n$ (i.e. $\mathbf{V}_\bot\linearFOM_{r\bot}^{n}$)
\begin{equation}
    \mathbf{V}\linearFOM_r^{n}+\mathbf{V}_\bot\linearFOM_{r\bot}^{n} = \linearFOM^n = (\mathbf{V} + \boldsymbol{\delta}\mathbf{V})\bar{\linearFOM}_r^{n},
    \label{eq:predictor-corrector:basis_update}
\end{equation}
which can be \addrTwo{used to} exactly solve \addrTwo{for $\boldsymbol{\delta}\mathbf{V}$} as 
\begin{equation} 
    \boldsymbol{\delta}\mathbf{V} = \frac{( \linearFOM^n - \mathbf{V}\bar{\linearFOM}_r^{n} )(\bar{\linearFOM}_r^n)^T}{||\bar{\linearFOM}_r^{n}||_2^2}. 
    \label{eq:predictor-corrector:delta_basis}
\end{equation}

\subsubsection{State Estimate and Sampling}
\label{subsubsec:adaptivity:StateEstimateAndSampling}
Though Eq.~\ref{eq:predictor-corrector:delta_basis} provides a compact formulation to update the subspace, it still requires access to the exact full-state solution, $\linearFOM^n$, thus impractical for real applications and countering the purpose of ROM in general. Therefore, we formulate a method to \textit{efficiently} estimate the full states leveraging the adaptive sampling algorithm proposed by Peherstorfer~\cite{PeherstorferADEIM}, which only evaluates the full states \addrTwo{at the sampled locations, denoted as $\hat{\linearFOM} \in \mathbb{R}^N$} following Eq.~\ref{eq:predictor-corrector:fom}  defined by the selection operator $\mathbf{S}$ while relying on the reduced states, $\bar{\linearFOM}_r$, and trial basis, $\mathbf{V}$, to approximate the full states (i.e. $\Tilde{\linearFOM} = \mathbf{V} \bar{\linearFOM}_r \approx \linearFOM$) at the unsampled locations, $\mathbf{S}^\ast$ \addrOne{$\in \mathbb{S}^{\numDOF \times (\numDOF-\numSamps)}$} with \addrOne{$\mathbf{S}^\ast = \mathbf{I} \setminus \mathbf{S}$} \addrTwo{(or $\mathbf{S}^\ast \mathbf{S}^{\ast T} + \mathbf{S} \mathbf{S}^{T} = \mathbf{I}$)}
\begin{equation}
    \addrTwo{\mathbf{S}^T \mathbf{B} ( \mathbf{S} \mathbf{S}^T\hat{\linearFOM}^n + \mathbf{S}^\ast \mathbf{S}^{\ast T} \Tilde{\linearFOM}^n ) = \mathbf{S}^T  \linearFOM^{n-1}.}
    \label{eq:StateEstSampling:sampled_full_state_def}
\end{equation}
This leads to the following estimate of the full states at the sampling points~\footnote{Note that the exact solution at the sampling locations is given by
\begin{align*}
\mathbf{S}^T{\linearFOM}^n  = (\mathbf{I} - \dt  \mathbf{J})^{-1} (\mathbf{S}^T+ \dt \mathbf{S}^T  \mathbf{J} \mathbf{S}^\ast \mathbf{S}^{\ast T}  \mathbf{B}^{-1}) \linearFOM^{n-1} 
\end{align*}}
\begin{equation}
    \addrTwo{\mathbf{S}^T\hat{\linearFOM}^n = (\mathbf{S}^T\mathbf{B} \mathbf{S})^{-1} ( \mathbf{S}^T - \mathbf{S}^T \mathbf{B}\mathbf{S}^\ast \mathbf{S}^{\ast T} \mathbf{B}_v^{-1}) \linearFOM^{n-1},}
    \label{eq:StateEstSampling:sampled_full_state}
\end{equation}
where $\mathbf{B}_v^{-1}\triangleq \mathbf{V} (\mathbf{V}^T \mathbf{B} \mathbf{V} )^{-1} \mathbf{V}^T$, and $\mathbf{B}_v^{-1}\triangleq \mathbf{V}(\mathbf{V}^T \mathbf{B}^T \mathbf{B} \mathbf{V} )^{-1} ( \mathbf{V}^T \mathbf{B}^T \mathbf{V} ) \mathbf{V}^T$, for Galerkin and LSPG, respectively. Detailed derivations of Eq.~\ref{eq:StateEstSampling:sampled_full_state} can be found in~\ref{appendix:sampled_full_state}. The trial basis, $\mathbf{V}$, can then be adapted at the sampling locations following Eq.~\ref{eq:predictor-corrector:delta_basis}
\begin{equation} 
    \mathbf{S}^T\boldsymbol{\delta}\mathbf{V} = \frac{\mathbf{S}^T( \hat{\linearFOM}^n - \Tilde{\linearFOM}^n )(\bar{\linearFOM}_r^n)^T}{||\bar{\linearFOM}_r^{n}||_2^2}, 
    \label{eq:StateEstSampling:delta_basis}
\end{equation}
which leverages the estimated full states at a single time step, $\hat{\linearFOM}^n$, to update the basis  while $\mathbf{S}^{\ast T}\boldsymbol{\delta}\mathbf{V} = 0$ at the unsampled locations. This is different from the method formulated by Peherstorfer~\cite{PeherstorferADEIM}, which seeks to incorporate the estimated full states at multiple time steps to update the basis. In addition, we remark that a similar one-step adaptive approach has been pursued by Zimmermann et al.~\citep{ZimmermannAdaptiveBasis2018} based on Grassmann subspace updates. It can then be shown in~\ref{appendix:basis_err} that the difference between the exact basis update (Eq.~\ref{eq:predictor-corrector:delta_basis}) and the above approximate basis update (Eq.~\ref{eq:StateEstSampling:delta_basis}) is determined by
\begin{equation}
||\delta\mathbf{V}_{exact}-\mathbf{S}\mathbf{S}^T{\delta\mathbf{V}}||_2 = \frac{ || \mathbf{e} ||_2 }{ ||\bar{\linearFOM}_r^{n}||_2 },
\label{eq:StateEstSampling:errorBasis}
\end{equation}
where $\mathbf{e}$ is defined as the error in the full-state estimate as derived in~\ref{appendix:StateEstError}
\begin{equation}
\mathbf{e} \triangleq \linearFOM^n - (\mathbf{S} \mathbf{S}^T\hat{\linearFOM}^n + \mathbf{S}^\ast \mathbf{S}^{\ast T} \Tilde{\linearFOM}^n) = ( \mathbf{B}^{-1} - \mathbf{B}_s^{-1} ) ( \mathbf{I} - \mathbf{B} \mathbf{S}^\ast \mathbf{S}^{\ast T} \mathbf{B}_v^{-1} ) \linearFOM^{n-1},
\label{eq:StateEstSampling:error1}
\end{equation}
where $\mathbf{B}_s^{-1}\triangleq \mathbf{S} (\mathbf{S}^T \mathbf{B} \mathbf{S} )^{-1}\mathbf{S}^T$ and more importantly, the accuracy of the basis update is mainly determined by the full-state estimate. By investigating Eq.~\ref{eq:StateEstSampling:error1}, it can be seen that the first term, $( \mathbf{B}^{-1} - \mathbf{B}_s^{-1} )$, on the right-hand side, represents the impact of sampling. This implies that minimizing $(\mathbf{B}^{-1} - \mathbf{B}_s^{-1})$ has the potential to result in more accurate state estimates, and thus accurate basis update. In addition, the second term, $( \mathbf{I} - \mathbf{B} \mathbf{S}^\ast \mathbf{S}^{\ast T} \mathbf{B}_v^{-1} )$, represents the impact of projection at the unsampled points. And it is noted that even if the exact solution $\mathbf{S}^T\linearFOM^n$ is used in the basis update (Eq.~\ref{eq:StateEstSampling:delta_basis}) at the sampled points, instead of the estimate $\mathbf{S}^T \hat{\linearFOM}^n$,
\begin{equation}
\mathbf{e} \triangleq \linearFOM^n - (\mathbf{S} \mathbf{S}^T\linearFOM^n + \mathbf{S}^\ast \mathbf{S}^{\ast T} \Tilde{\linearFOM}^n) = \mathbf{S}^\ast\mathbf{S}^{\ast T}( \mathbf{B}^{-1} - \mathbf{B}_v^{-1} )\linearFOM^{n-1},
\label{eq:StateEstSampling:error2}
\end{equation}
the projection errors, $( \mathbf{B}^{-1} - \mathbf{B}_v^{-1} )$, at the unsampled points, $\mathbf{S}^\ast$, still impose major impacts on the error in the full-state estimate. Therefore, this implies even if the estimate of the full-state solution at the sampling points is exact, the accuracy of the state prediction may be limited by the sampling process. The following section presents a potential remedy. 

\subsubsection{Non-local Information}
\label{subsubsec:adaptivity:NonLocalInfo}
Peherstorfer~\cite{PeherstorferADEIM} has shown that if one assumes `local coherence' of the residual of the approximation in the solution domain, online basis adaptation predicated by a few samples can be effective. Physically, local coherence is relevant when the solution domain contains an isolated sharp feature (such as a pulse or a shock). However, we find that this assumption is invalid in the presence of distributed multi-scale features, even in a linear advection setting. Indeed, in a multi-physics setting such as combustion dynamics which may include thin, disperse reaction regions, long wavelength acoustics and convecting vortices, local coherence cannot be assumed. To address the above issue, we propose a strategy to include non-local effects by periodically estimating the full states at the unsampled locations to reduce the second term in Eq.~\ref{eq:StateEstSampling:error1}, which seeks to evaluate the full states at unsampled locations every $z_s$ time steps ($z_s \geq 1$) with a larger time step, $z_s \dt$
\begin{equation}
 \mathbf{S}^{\ast T}( \mathbf{I} - z_s \dt  \mathbf{J} )( \mathbf{S}^\ast \mathbf{S}^{\ast T}\hat{\linearFOM}^n + \mathbf{S} \mathbf{S}^{T} \hat{\linearFOM}^n ) = \mathbf{S}^{\ast T}  \linearFOM^{n-z_s},
 \label{eq:NonLocInfo:unsampled_full_state_def}
\end{equation}
which leads to the following estimate of the full states at the unsampled points with $\mathbf{B}^{\ast} \triangleq \mathbf{I} - z_s \dt  \mathbf{J}$
\begin{equation}
\mathbf{S}^{\ast T}\hat{\linearFOM}^n = ( \mathbf{S}^{\ast T} \mathbf{B}^{\ast} \mathbf{S}^{\ast} )^{-1} ( \mathbf{S}^{\ast T}  \linearFOM^{n-z_s} - \mathbf{S}^{\ast T} \mathbf{B}^{\ast} \mathbf{S} \mathbf{S}^{T} \hat{\linearFOM}^n ). 
\label{eq:NonLocInfo:unsampled_full_state}
\end{equation}
The basis can then be updated as 
\begin{equation}  
\boldsymbol{\delta}\mathbf{V} = \frac{ (\mathbf{S}\mathbf{S}^T \hat{\linearFOM}^{n} + \mathbf{S^\ast}\mathbf{S}^{\ast T} \hat{\linearFOM}^{n} - \Tilde{\linearFOM}^n ) (\bar{\linearFOM}_r^{n})^T }{||\bar{\linearFOM}_r^{n}||^2}. 
\label{eq:NonLocInfo:delta_basis}
\end{equation}
It should be noted that now if the exact solution $\mathbf{S}^{\ast T}\linearFOM^n$ is used at the unsampled points in the basis update (Eq.~\ref{eq:NonLocInfo:delta_basis}), the error in the full-state estimate becomes
\begin{equation}
\mathbf{e} \triangleq \linearFOM^n - (\mathbf{S} \mathbf{S}^T \linearFOM^n + \mathbf{S}^\ast \mathbf{S}^{\ast T} \linearFOM^n ) = \zeroVec,
\label{eq:NonLocInfo:error2}
\end{equation}
which is not impacted by the projection errors as shown in Eq.~\ref{eq:StateEstSampling:error2}. Then, similar to Eq.~\ref{eq:StateEstSampling:error1}, the error in full-state estimate, if the estimated solution is used at the unsampled points (Eq.~\ref{eq:NonLocInfo:unsampled_full_state}), is given by
\begin{equation}
\mathbf{e} \triangleq [ \mathbf{B}^{-1} - \mathbf{B}_{s^\ast}^{-1}  \mathbf{B}^{z_s - 1} - ( \mathbf{B}_s^{-1} - \mathbf{B}_{s^\ast}^{-1} \mathbf{B}^{\ast} \mathbf{B}_s^{-1} ) ( \mathbf{I} - \mathbf{B} \mathbf{S}^\ast \mathbf{S}^{\ast T} \mathbf{B}_v^{-1}) ] \linearFOM^{n-1},
\label{eq:NonLocInfo:error1}
\end{equation}
where $\mathbf{B}_{s^\ast}^{-1} \triangleq \mathbf{S}^{\ast} \left[ \mathbf{S}^{\ast T} (\mathbf{I} - z_s \dt  \mathbf{J}) \mathbf{S} \right]^{-1} \mathbf{S}^{\ast T}$. A detailed derivation of Eq.~\ref{eq:NonLocInfo:error1} is provided in~\ref{appendix:NonLocInfoStateEstError}\addrOne{, which also shows that Eq.~\ref{eq:NonLocInfo:error1} mitigates the contributions of the projection errors at the unsampled points, $( \mathbf{I} - \mathbf{B} \mathbf{S}^\ast \mathbf{S}^{\ast T} \mathbf{B}_v^{-1} )$, to the total error, $\mathbf{e}$, compared to Eq.~\ref{eq:StateEstSampling:error1}}. 

\subsection{Adaptive ROM Formulation and Practical Considerations}
\label{subsec:adaptivity:algorithm}
With the \textit{predictor-corrector} approach illustrated using a linear FOM above, we now provide details on the formulation to construct adaptive ROMs, which incorporates the one-step basis adaptation (Eq.~\ref{eq:StateEstSampling:delta_basis}) and the non-local full-state evaluation (Eq.~\ref{eq:NonLocInfo:unsampled_full_state}). As highlighted above, due to the intractability of solving Eq.~\ref{eq:adaptiverom:def}, we resort to a decoupled approach that formulates three individual minimization problems to update the reduced states, $\solPrimROMRed^{\iterIdx}$, trial basis, $\trialBasisPrimN{\iterIdx}$, and sampling points, $\sampMat^T_{n}$, with time, respectively. 

First, Eq.~\ref{eq:mplsvt_hyper} is directly leveraged to propagate the reduced states in time from $\solPrimROMRed^{\iterIdx-1}$ to $\solPrimROMRed^{\iterIdx}$. Following the formulation in Eq.~\ref{eq:mplsvt_proj_hyperS}, hyper-reduced MP-LSVT ROM is constructed by setting $\resBasis$ to the trial basis $\trialBasisPrimN{\iterIdx-1}$ for the ease of practical considerations since it is found to provide excellent approximations in the hyper-reduction~\citep{HuangMPLSVT2022}
\begin{equation}
    \lp \sampMat^T_{n-1} \testBasisPrim^\iterIdx \rp^T \left[ \lp \sampMat^T_{n-1} \trialBasisPrimN{\iterIdx-1} \rp^{+} \right]^T \lp \sampMat^T_{n-1} \trialBasisPrimN{\iterIdx-1} \rp^{+} \sampMat^T_{n-1} \scaleMatCons \resFunc{ \solPrimROMFull^{\iterIdx} } = \zeroVec,
    \label{eq:adaptiverom:hROM}
\end{equation}
where $ \solPrimROMFull^{\iterIdx} = \solPrimFOMRef + \scaleMatPrim^{-1} \trialBasisPrimN{\iterIdx-1} \solPrimROMRed^{\iterIdx} $ and it should be noted that the reference state, $\solPrimFOMRef$, and the scaling matrix, $\scaleMatPrim$, are not getting updated in the adaptive ROM formulation.

\subsubsection{Adaptation of the Basis}
\label{subsubsec:BasisAdapt}
Second, following the illustration in section~\ref{subsec:adaptivity:predictor-corrector}, we adopt the \textit{predictor-corrector} approach to adapt the trial basis from $\trialBasisPrimN{n-1}$ to $\trialBasisPrimN{n}$ with an augment, $\delta \trialBasisPrimN{n}  \in \mathbb{R}^{\numDOF \times \numSolModes}$
\begin{equation}
    \trialBasisPrimN{n} = \trialBasisPrimN{n-1} + \delta \trialBasisPrimN{n}.
    \label{eq:adaptiverom:basis_update}
\end{equation}
where $\delta \trialBasisPrimN{n}$ is formulated as the solution to the minimization problem
\begin{equation}
    \delta \trialBasisPrimN{n}  \triangleq  \argmin_{ \delta \trialBasisPrimN{n} \in \mathbb{R}^{\numDOF \times \numSolModes} } \norm{ \solPrimFOMRef + \scaleMatPrim^{-1} ( \trialBasisPrimN{\iterIdx-1} + \delta \trialBasisPrimN{n} ) \solPrimROMRed^{\iterIdx} - \solPrimFOMAdaptive^{\iterIdx} }^2,
    \label{eq:adaptiverom:basis_minimization}
\end{equation}
with $\solPrimROMRed^{\iterIdx}$ obtained by solving Eq.~\ref{eq:adaptiverom:hROM}. It can be easily proved that the one-step adaptive approach illustrated in Eq.~\ref{eq:NonLocInfo:delta_basis} represents a solution satisfying the minimization problem (Eq.~\ref{eq:adaptiverom:basis_minimization}) above
\begin{equation}
    \delta \trialBasisPrimN{n} = \frac{ \lp \solPrimFOMAdaptive^n - \solPrimROMFull^n \rp (\solPrimROMRed^n)^T}{||\solPrimROMRed^n||_2^2},
    \label{eq:adaptiverom:delta}
\end{equation}
where $\solPrimFOMAdaptive^{\iterIdx} \in \mathbb{R}^{\numDOF}$ is the estimated full states based on the FOM residual defined in Eq.~\ref{eq:fom_linear_multi_discrete}. As illustrated in section~\ref{subsubsec:adaptivity:StateEstimateAndSampling}, the evaluations of $\solPrimFOMAdaptive^{\iterIdx}$ are crucial to the accuracy of the basis adaptation, especially on the unsampled locations as shown in section~\ref{subsubsec:adaptivity:NonLocalInfo}. Therefore, to establish an adaptive ROM formulation for multi-scale problems, we seek to incorporate the non-local information in full-state estimates following the procedures in Eqs.~\ref{eq:StateEstSampling:sampled_full_state} and~\ref{eq:NonLocInfo:unsampled_full_state}, which adopts different methods to \textit{efficiently} evaluate the full states at the sampled and unsampled locations. The full states at the \emph{sampled} points, $\sampMat^T_{n-1}\solPrimFOMAdaptive^\iterIdx$, are estimated every time step with a physical time step of $\dt$ (the unknowns in the two equations below are indicated in {\color{blue} blue}) \addrOne{such that the FOM equation residual goes to zero at the \emph{sampled} points, $\sampMat^T_{n-1} \resFunc{ \sampMat_{n-1}{\color{blue}\sampMat^T_{n-1}\solPrimFOMAdaptive^\iterIdx} } = 0$. Following the linear multi-step methods used to define the FOM equation residual in Eq.~\ref{eq:fom_linear_multi_discrete}, we have}
\begin{equation}
    \begin{aligned}
        \sampMat^T_{n-1}\solConsFOMFunc{\sampMat_{n-1} {\color{blue}\sampMat^T_{n-1}\solPrimFOMAdaptive^\iterIdx} + \sampMat_{n-1}^{\ast}\sampMat^{\ast T}_{n-1}\solPrimROMFull^{\iterIdx}, \timeVar^\iterIdx} & + \sum^{\itermaxLinMS}_{j=1} \alpha_j \sampMat^T_{n-1}\solConsFOMFunc{\solPrimROMFull^{\iterIdx-j}} \\
        - \dt \beta_0 \sampMat^T_{n-1}\rhsFunc{\sampMat_{n-1} {\color{blue}\sampMat^T_{n-1}\solPrimFOMAdaptive^\iterIdx} + \sampMat_{n-1}^{\ast}\sampMat^{\ast T}_{n-1}\solPrimROMFull^{\iterIdx}, \timeVar^\iterIdx} &- \dt \sum^{\itermaxLinMS}_{j=1} \beta_j \sampMat^T_{n-1}\rhsFunc{\solPrimROMFull^{\iterIdx-j}, \timeVar^{\iterIdx-j}} = 0. 
    \end{aligned}
    \label{eq:adaptiverom:sampledFullState}
\end{equation}
On the other hand, with $\sampMat^T_{n-1}\solPrimFOMAdaptive^\iterIdx$ determined from Eq.~\ref{eq:adaptiverom:sampledFullState}, the full states at the \emph{unsampled} points, $\sampMat^{\ast T}_{n-1}\solPrimFOMAdaptive^\iterIdx$, are estimated every $z_s$ time steps ($z_s \geq 1$) with a larger physical time step of $z_s \dt$ \addrOne{such that the FOM equation residual goes to zero at the \emph{unsampled} points, $\sampMat^{\ast T}_{n-1} \resFunc{ \sampMat_{n-1}^{\ast}{\color{blue}\sampMat^{\ast T}_{n-1}\solPrimFOMAdaptive^\iterIdx} } = 0$ using the linear multi-step methods in Eq.~\ref{eq:fom_linear_multi_discrete}}
\begin{equation}
    \begin{aligned}
        \sampMat^{\ast T}_{n-1}  \solConsFOMFunc{\sampMat_{n-1}\sampMat^T_{n-1}\solPrimFOMAdaptive^\iterIdx + \sampMat_{n-1}^{\ast} {\color{blue}\sampMat^{\ast T}_{n-1}\solPrimFOMAdaptive^\iterIdx} , \timeVar^\iterIdx} & + \sum^{\itermaxLinMS}_{j=1} \alpha_j  \sampMat^{\ast T}_{n-1} \solConsFOMFunc{\solPrimROMFull^{\iterIdx-j \cdot z_s}} \\ 
        - z_s \dt \beta_0  \sampMat^{\ast T}_{n-1}\rhsFunc{\sampMat_{n-1}\sampMat^T_{n-1}\solPrimFOMAdaptive^\iterIdx+\sampMat_{n-1}^{\ast} {\color{blue}\sampMat^{\ast T}_{n-1}\solPrimFOMAdaptive^\iterIdx} , \timeVar^\iterIdx} &- z_s \dt \sum^{\itermaxLinMS}_{j=1} \beta_j  \sampMat^{\ast T}_{n-1} \rhsFunc{\solPrimROMFull^{\iterIdx-j\cdot z_s}, \timeVar^{\iterIdx-j\cdot z_s}} = 0. 
    \end{aligned}
    \label{eq:adaptiverom:unsampledFullState}
\end{equation}
\addrOne{It is worth highlighting that though the linear multi-step time-discretization methods are used to demonstrate the adaptive ROM formulation, it can be easily extended to Runge-Kutta methods by using the corresponding method to define the FOM equation residual, $\res$, and evaluate the unknowns above}. As illustrated in Eq.~\ref{eq:NonLocInfo:error1}, incorporating the non-local information leads to a lower error in full-state estimates, thus resulting in more accurate basis adaptation. We remark that if $z_s = 1$, the estimated full states, $\solPrimFOMAdaptive^\iterIdx$, recover the (exact) FOM full states but introduces no computational efficiency gain in the adaptive ROM. Therefore, in practice, we seek to evaluate the non-local information with much larger $z_s$ ($\sim O(10)$) in Eq.~\ref{eq:adaptiverom:unsampledFullState} with the goal of reducing the cost while maintaining reasonable accuracy in estimating the non-local full states, $\sampMat^{\ast T}_{n-1}\solPrimFOMAdaptive^\iterIdx$, which makes the selection of $z_s$ crucial for the performance of the adaptive ROM. \addrOne{In fact, it can be seen in Eq.~\ref{eq:adaptiverom:sampledFullState} that the non-local full states, $\sampMat^{\ast T}_{n-1}\solPrimFOMAdaptive^\iterIdx$, remain unchanged when solving for the local full states and vice versa in Eq.~\ref{eq:adaptiverom:unsampledFullState}.} This formulation is motivated \addrOne{and designed} by our observations that in many dynamical systems (especially turbulent reacting flows), the \textit{local} coherence (e.g. turbulence, flame, and shock) often exhibits fast dynamics and requires finer time scales to resolve while the \textit{non-local} coherence (e.g. long-wavelength acoustics and large-scale vortices) features slower dynamics that can be resolved with coarser time scales. \addrOne{Therefore, the \textit{local} coherence (e.g. $\sampMat^T_{n-1}\solPrimFOMAdaptive^\iterIdx$) can be estimated assuming the \textit{non-local} coherence does not evolve much within the finer time scales. On the other hand, this again makes the selection of $z_s$ and the sampling points adaptation strategy as the determinant factors for the adaptive ROM performance. Though formulated based upon implicit assumptions of the coherence features, we still anticipate the adaptive ROM formulation applicable to a wide range of dynamical systems with different time scales for local and non-local coherence, which include turbulent reacting flows, shocks, and etc.}

\subsubsection{Adaptation of the Sampling Points}
\label{subsubsec:SamplingAdapt}
Third, we directly leverage the algorithm in~\citep{PeherstorferADEIM,Uy_2022_adaptiveROMFlame} to update the sampling points from $\sampMat_{n-1}$ to $\sampMat_n$, to minimize the interpolation error, $\mathbf{e}^n_s$, arising from the hyper-reduction
\begin{equation}
   \sampMat_{n} \triangleq \argmin_{ \sampMat^{n} \in \mathbb{S}^{\numDOF \times \numSamps } } \norm{  \mathbf{e}^n_s }^2,
   \label{eq:adaptiverom:sampling_def}
\end{equation}
where $ \mathbf{e}_s \triangleq \solPrimFOMAdaptive^n - \trialBasisPrim^n \lp \sampMat_{n}^{T} \trialBasisPrim^n \rp^{+} \sampMat_{n}^{T} \solPrimFOMAdaptive^n$ with the full state, $\solPrimFOMAdaptive^n$, evaluated based on Eqs.~\ref{eq:adaptiverom:sampledFullState} and~\ref{eq:adaptiverom:unsampledFullState}. For practical implementations, the interpolation error is first evaluated based on the sampling points at time step $n - 1$, $\sampMat_{n-1}$ 
\begin{equation}
   \mathbf{e}^n_s = \solPrimFOMAdaptive^n - \trialBasisPrimN{n} \lp \sampMat_{n-1}^{T} \trialBasisPrim^n \rp^{+} \sampMat_{n-1}^{T} \solPrimFOMAdaptive^n.
   \label{eq:adaptiverom:interpErrNorm}
\end{equation}
The magnitude of each element, $\text{e}^{n}_{s,i}$, in $\mathbf{e}^n_s$ is then examined and arranged in a descending order $ | \text{e}^{n}_{s,\hat{i}_{1}} | \geq \ldots \geq | \text{e}^{n}_{s,\hat{i}_{N_{elem}}} | $, the first $n_s$ indices of which are selected to update $\sampMat_n$. As pointed out by Cortinovis et al.~\citep{Cortinovis_QuasiOptSampling2020}, the updated sampling points are quasi-optimal with respect to an upper bound of the adaptation error. Moreover, we remark that the adaptation of the sampling points requires evaluating the full states, $\solPrimFOMAdaptive^{\iterIdx}$, at all the points, which incurs high computational costs that scale with the total number of degrees of freedom, $\numDOF$, in the FOM. Therefore, similar to the work in~\citep{PeherstorferADEIM,Uy_2022_adaptiveROMFlame}, we choose to adapt the sampling points every $z_s$ time steps, which is consistent with the non-local full-state estimate in Eq.~\ref{eq:adaptiverom:unsampledFullState} and mitigates the penalty of additional expensive evaluations of the full states. Furthermore, it can be readily seen that $z_s$ serves as an important parameter in both basis and sampling point adaptation that determines the accuracy and efficiency of adaptive ROM. In section~\ref{subsubsec:1d:rom_cmp}, we provide detailed investigations on the sensitivity of the ROM performance on $z_s$.

\section{Algorithm and Computational Complexity}
\label{subsubsec:comp_proc}
With practical implementation in mind, we describe the procedure to construct the adaptive ROM,  and also provide a detailed analysis on its computational complexity. 

\subsection{Computational procedure} The adaptive ROM algorithm is summarized in Algorithm~\ref{algorithm:adaptiveROM} with the following inputs:
\begin{enumerate}
    \item $M$: the total number of physical time steps;
    \item $w_\text{init}$: the initial (\textit{offline}) training window size (i.e. the initial number of physical time steps for FOM calculation);
    \item $z_s$: the rate to update sampling points and estimate non-local full states;
    \item $n_s$: the total number of sampling points.
\end{enumerate}

First, the adaptive ROM is initialized as described in lines 2-4. The FOM (Eq.~\ref{eq:fom_linear_multi_discrete}) is solved for a small number of time steps, $w_\text{init} \in \mathbb{N}$ to obtain a collection of full states, $\mathbf{Q}_p^\text{init} = \left[ \solPrimFOM^1 , \ldots , \solPrimFOM^{w_\text{init}} \right]$, which is then used to determine the initial basis, $\trialBasisPrimN{0}$, via POD as described in section~\ref{subsec:pod} and sampling points, $\sampMat_{0}$ via gappy POD. Specifically, in the present paper, we pursue a gappy POD formulation with randomized oversampling~\citep{PeherstorferODEIM} to determine $\sampMat_{0}$. It needs to be remarked that though this step can be considered as an \textit{offline} stage, $w_\text{init}$ is typically negligible compared to the total number of time steps, $M \in \mathbb{N}$ (i.e. $w_\text{init} \ll M$), therefore significantly reducing the \textit{offline} computational cost that involves solving the expensive FOM to collect the full-state information for the \textit{offline/online} ROM methods~\citep{HuangMPLSVT2022,Barnett_QuadraticPROM2022,McQuarrieOpInf2021}. Moreover, the number of time steps for the initial FOM simulation, $w_\text{init}$, is one hyper-parameter in the formulation, the effects of which are investigated further in section~\ref{subsubsec:1d:sensitivity_arom}. Furthermore, once the initial basis and sampling points are established in this initialization step, no additional \textit{offline}-stage FOM simulation is necessary to construct ROMs for another parameter. This arguably ``no \textit{offline} stage" feature inherently enables parametric predictive capabilities in the resulting adaptive ROMs~\citep{PeherstorferADEIM,Ramezanian_2021_OnTheFlyROM}, thus avoiding the common challenges in developing \textit{parametric} ROMs~\citep{Benner_Gugercin_Willcox_PMR2015}, especially for chaotic convenction-dominanted problems exhibiting the Kolmogorov barrier. Such enhanced parametric predictive capabilities are demonstrated in sections~\ref{subsubsec:1d:parametric_arom} and~\ref{subsubsec:2dInjector:transience}. After the initialization, the \textit{online} adaptive ROM calculation starts in line 5, which involves four major steps: 

1. \emph{Propagate reduced states}: solve the ROM for one time step (line 7); 

2. \emph{Estimate the full states}: for every $z_s$ time steps, the full states are estimated at all the points incorporating the non-local information (line 10) \textit{while} otherwise the full states are estimated \textit{only} at the sampled points with the full states approximated using the POD basis at the unsampled points (line 19); 

3. \emph{Update the basis}: for every $z_s$ time steps, the basis is updated at all the points using the one-step adaptive approach (line 12) \textit{while} otherwise the basis is updated only at the sampled points with the zero basis augment (i.e. $\sampMat^{\ast T}_{n-1}\delta \trialBasisPrimN{n} = 0$) at the unsampled points as proved in Eq.~\ref{eq:StateEstSampling:delta_basis} (line 21); and 

4. \emph{Update the sampling points}: for every $z_s$ time steps, the interpolation error, $\mathbf{e}^n_s$, is computed at all the points (line 14) and the sampling points are updated based on the $n_s$ points corresponding to the largest magnitude of the elements in $\mathbf{e}^n_s$ (lines 15 and 16).

\begin{algorithm}
	\caption{Adaptive ROM algorithm} 
	\label{algorithm:adaptiveROM}
	\begin{algorithmic}[1]
	    \State \textbf{Input}: $M, \ w_\text{init}, \ z_s, \ \text{and} \ n_s$
	    \State Solve the FOM (Eq.~\ref{eq:fom_linear_multi_discrete}) for $w_\text{init}$ time steps to collect the full-state data matrix $\mathbf{Q}_p^\text{init}$
	    \State Compute the initial POD basis $\trialBasisPrimN{n}$ from $\mathbf{q}_p$ using Eq.~\ref{eq:pod_qp}
	    \State Compute the initial sampling points $\sampMat_{0}$
		\For {$\iterIdx = w_\text{init}+1,\ldots,M$}
  
            \State \textbf{Propagate the reduced states:} 
            \State \ \ \ \ \ Solve the hyper-reduced MP-LSVT ROM (Eq.~\ref{eq:adaptiverom:hROM}) to obtain $\solPrimROMRed^{\iterIdx}$
            
            \If{ mod($\iterIdx$,$z_s$) == 0 \textit{or} $\iterIdx$ == $w_\text{init} + 1$  }
            
                \State \textbf{Estimate the full states} (at \emph{both} the sampled \emph{and} unsampled points):             
                \State \ \ \ \ \ $\sampMat^{T}_{n-1}\solPrimFOMAdaptive^n$ following Eq.~\ref{eq:adaptiverom:sampledFullState} and $\sampMat^{\ast T}_{n-1}\solPrimFOMAdaptive^n$ following Eq.~\ref{eq:adaptiverom:unsampledFullState}  
                
                \State \textbf{Update the basis} (at \emph{both} the sampled \emph{and} unsampled points): 
                \State \ \ \ \ \ $\trialBasisPrimN{n} = \trialBasisPrimN{n-1} + \delta \trialBasisPrimN{n}$ following Eq.~\ref{eq:adaptiverom:basis_update}

                \State \textbf{Update the sampling points}: 
                \State \ \ \ \ \ Compute the interpolation error, $\mathbf{e}^n_s$, following Eq.~\ref{eq:adaptiverom:interpErrNorm}    
                \State \ \ \ \ \ $[\sim , \hat{\mathbf{i}}]$ = \texttt{sort}($|\mathbf{e}^n_s|$,`descend')    
                \State \ \ \ \ \ $\hat{\mathbf{i}}[1:n_s]$ is selected for $\sampMat^{T}_{n}$
            
            \Else
            
                \State \textbf{Estimate the full states} (\emph{only} at the sampled points): 
                \State \ \ \ \ \ $\sampMat^{T}_{n-1}\solPrimFOMAdaptive^n$ following Eq.~\ref{eq:adaptiverom:sampledFullState} \textit{while} $\sampMat^{\ast T}_{n-1}\solPrimFOMAdaptive^\iterIdx = \sampMat^{\ast T}_{n-1}\solPrimROMFull^{\iterIdx}$
    
                \State \textbf{Update the basis} (\emph{only} at the sampled points): 
                \State \ \ \ \ \ $\sampMat^{T}_{n-1} \trialBasisPrimN{n} = \sampMat^{T}_{n-1} \trialBasisPrimN{n-1} + \sampMat^{T}_{n-1} \delta \trialBasisPrimN{n}$ following Eq.~\ref{eq:adaptiverom:basis_update} \textit{while} $\sampMat^{\ast T}_{n-1}\delta \trialBasisPrimN{n} = 0$ 
            
            \EndIf
      
		\EndFor
	\end{algorithmic} 
\end{algorithm}

 \subsection{Computational complexity} Next, we provide a detailed computational complexity analysis of the adaptive ROM algorithm in terms of floating-point operations (FLOPs). Here, a FLOP refers to any floating-point addition or
multiplication; no distinction is made between the computational cost of either operation. We focus the analysis on the complexity of the FOM and ROM with the implicit time-integration scheme solved using dual time-stepping~\citep{sankaranDualTime1995,Pulliam}, which are used for all the numerical investigations in the current paper. First, we estimate the FLOPs required for FOM calculations, denoted as $\text{FLOP}_\text{FOM}$, for $z_s$ physical time steps, each of which contains $K$ pseudo iterations for dual time-stepping. Details on computing $\text{FLOP}_\text{FOM}$ are provided in~\ref{appendix:flops_fom}. Then, the FLOPs for the conventional static-basis hyper-reduced MP-LSVT ROM (we denote it as \emph{static ROM} for brevity) calculations, denoted as $\text{FLOP}_\text{SROM}$, are estimated for the same number of physical time steps and pseudo iterations as the FOM to establish a baseline for comparison with the adaptive ROM. Details on computing $\text{FLOP}_\text{SROM}$ are provided in~\ref{appendix:flops_srom}. Lastly, the estimated FLOPs for the adaptive ROM, denoted as $\text{FLOP}_\text{AROM}$, to carry out $z_s$ time-step calculations, is provided in Table~\ref{table:flops_adaptiveROM} for each major step in Algorithm~\ref{algorithm:adaptiveROM} with several additional parameters to highlight: $p_1$: the number of pseudo iterations for reduced-state propagation (i.e. solving the hyper-reduced MP-LSVT ROM), and $p_2$: the number of pseudo iterations for full-state estimate. Moreover, it is remarked that there is no additional computational complexity introduced by incorporating the non-local information in the adaptive ROM formulation because the same rate, $z_s$, is adopted to update the sampling points and estimate the non-local full states in the proposed algorithm, the former of which requires full-state estimation at all the points as shown in Eq.~\ref{eq:adaptiverom:interpErrNorm}.

\begin{figure}
	\centering
	\includegraphics[width=0.7\textwidth]{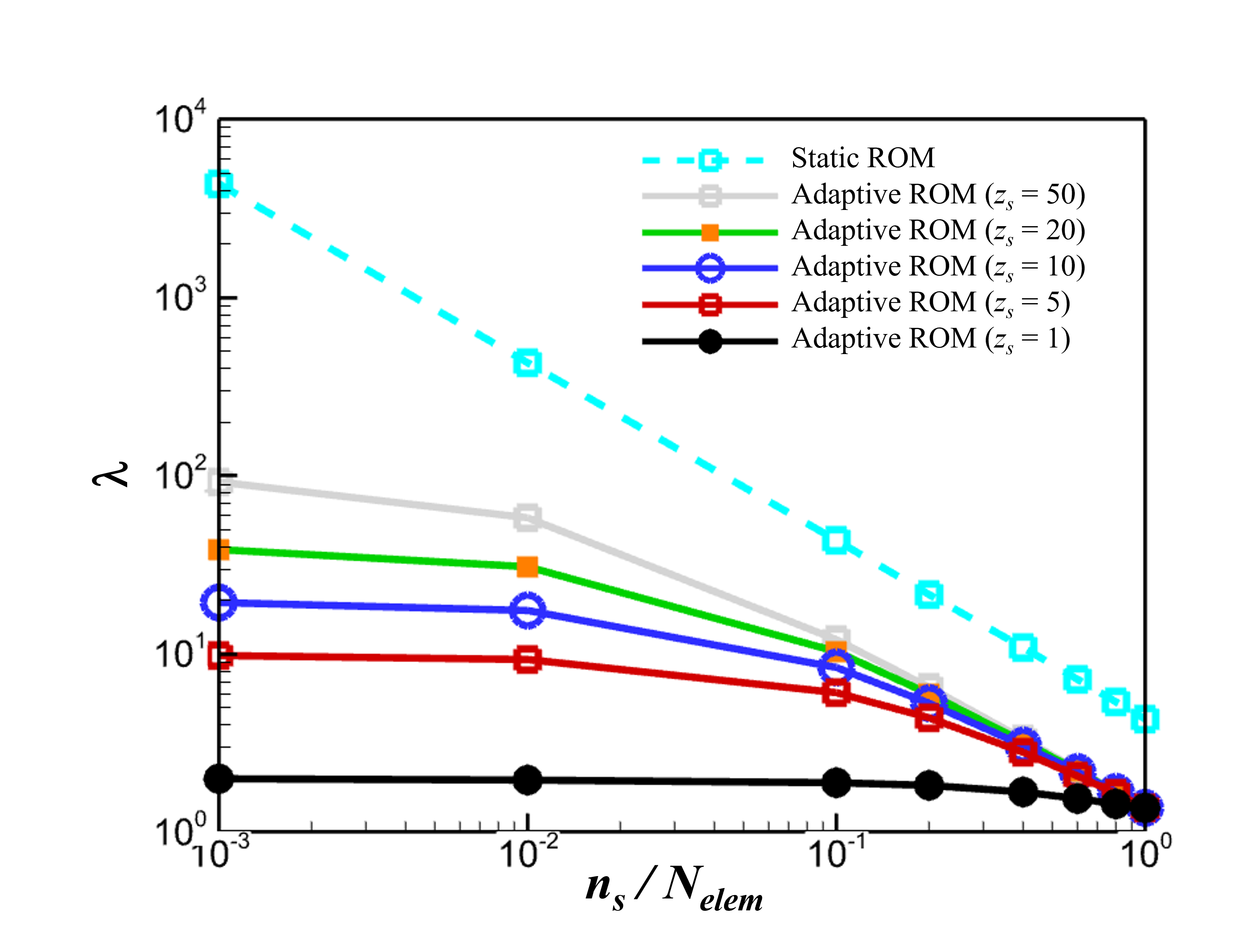}
	\caption{Approximate efficiency gain of static and adaptive ROM (with different $z_s$ adopted) versus sampling points.}
    \label{fig:flops_arom} 
\end{figure}

Furthermore, the ratio of FLOPs between FOM and ROM is defined to quantify the  efficiency gain of the ROM:
\begin{equation}
    \lambda = \frac{\text{FLOP}_\text{FOM}}{\text{FLOP}_\text{ROM}}. 
    \label{eq:ROM_efficiency_gain}
\end{equation}
It can easily shown that for $\numSamps \ll \numElements$ (common for hyper-reduced ROM~\citep{Carlberg2017,HuangMPLSVT2022}) and $\numSolModes \sim O(10)$ (standard for adaptive ROM~\citep{PeherstorferADEIM,Uy_AADEIM2022_flame}), the efficiency gain of adaptive ROM, $\lambda_\text{AROM}$, can be approximated as
\begin{equation}
    \lambda_\text{AROM} \approx \frac{1}{ \frac{p_1}{K \lambda_\text{SROM}}  + \frac{p_2}{z_s K} } \le \frac{z_s K}{p_2},
    \label{eq:AdaptiveROM_efficiency_gain}
\end{equation}
where $\lambda_\text{SROM}$ represents the efficiency gain of static ROM. This can be verified by investigating the efficiency gain for both static and adaptive ROM against the fraction of sampling points, $\numSamps / \numElements$, by selecting $\numSolModes = 5$, $p_1 / K = 0.5$, and $p_2 / K = 0.5$, as shown in Fig.~\ref{fig:flops_arom} with different $z_s$ values used for the adaptive ROM. It can be readily seen that for both static and adaptive ROMs, more efficiency gains can be obtained as the number of sampling points decreases. However, unlike the static ROM, the efficiency gain of which improves \textit{almost linearly} with the number of sampling points, the adaptive ROM's efficiency gain only improves \textit{asymptotically} and saturates towards ${z_s K}/{p_2}$, as proved in Eq.~\ref{eq:AdaptiveROM_efficiency_gain}. On the other hand, it is also expected that efficiency gains can be improved if a lesser number of pseudo iterations are used for the full-state estimate (i.e. smaller $p_2$). Therefore, given the additional operations required to update the basis and sampling points, the adaptive ROM is not expected to achieve the same level of computational efficiency gains as the static ROM. In addition to the number of sampling points, the rate at which sampling points are updated ($z_s$), and the number of pseudo-iterations for full-state estimate, $p_2$ determine the acceleration that the adaptive ROM can achieve.  

\begin{table}
\centering
\begin{tabular}{ll} 
\toprule
\textbf{Operations} & \textbf{Approximate FLOPs} \\
\midrule
Propagate the reduced states & $\frac{p_1}{K} \text{FLOP}_\text{SROM}$ \\
Estimate the full states at sampled points & $[ 1 + ( z_s - 1 ) \frac{\numSamps}{\numElements} ] \frac{p_2}{z_s K} \text{FLOP}_\text{FOM}$ \\
Update the basis & $[N + (z_s - 1) \numVars \numSamps + z_s](4 \numSolModes - 1)$ \\
Update the sampling points & $( \numSolModes + 2 ) \numVars \numElements + ( \numSolModes + 1 ) \numSamps $ \\
\midrule
\textbf{Total} & $ \frac{p_1}{K}\text{FLOP}_\text{SROM} + [ 1 + ( z_s - 1 ) \frac{\numSamps}{\numElements} ] \frac{p_2}{z_s K} \text{FLOP}_\text{FOM} $ \\
& $+ \numVars \numElements (5 \numSolModes + 1)  + ( \numSolModes + 1 ) \numSamps$ \\
& $+ [ (z_s - 1) \numVars \numSamps + z_s ](4 \numSolModes - 1)$ \\
\bottomrule
\end{tabular}
\caption{\label{table:flops_adaptiveROM} Approximated floating-point operations for the adaptive ROM (Algorithm~\ref{algorithm:adaptiveROM}) calculations for $z_s$ time steps, each of which contains $K$ pseudo iterations with $p_1$ used for reduced-state propagation and $p_2$ for full-state estimate.}
\end{table}

\section{Numerical Results and Analysis}
\label{sec:results}

To assess the capabilities of the proposed adaptive ROM formulation in predicting complex multi-scale multi-physics problems, two  reacting-flow problems are considered, both of which exhibit non-stationary dynamics featuring slow decays of Kolmogorov N-width. The first configuration is a one-dimensional, freely propagating, premixed laminar flame. This is a benchmark that was  established by Huang et al.~\citep{HuangMPLSVT2022} to allow extensive investigations of different ROM methods on reacting-flow problems. The 1D problem includes many features that are well known to be challenging for ROM, such as strong convection of sharp-gradient phenomenon (i.e. the flame propagation), nonlinear dynamics from chemical reactions, and multi-physics interactions between flame and acoustics and is available as part of an open source implementation~\cite{wentland2022perform}. The second configuration is a simplified two-dimensional representation of a single injector combustor~\citep{YuJPP}, which exhibits strong chaotic dynamics, which has proven to be difficult or impossible for ROMs to predict. Specifically, this case has been used to assess the accuracy and robustness of different ROM formulations~\citep{HuangAIAAJ2019,McQuarrieOpInf2021}. The computational infrastructure used for the full- and reduced-order models solves conservation equations for mass, momentum, energy and species  transport in a fully coupled way using the in-house CFD code, the General Mesh and Equation Solver (GEMS). GEMS  has been used to model a variety of complex, practical reacting flow problems~\cite{HarvazinskiPoF,HuangLDI}. More details of the FOM equations can be found in~\ref{appendix:fom_eq}. The FOM employs a cell-centered second-order accurate finite volume method for spatial discretization. The Roe scheme~\cite{Roe1981} is used to evaluate the inviscid fluxes and a Green-Gauss gradient reconstruction procedure~\cite{MitchellReconScheme} is used to compute the face gradients and viscous fluxes. A gradient limiter by Barth and Jespersen~\cite{barth1989} is used to preserve monotonicity for flow fields with strong gradients. A ghost cell formulation is used for treatment of boundary conditions. Time integration for all FOM simulations uses the implicit second-order accurate backwards differentiation formula with dual time-stepping. All the FOM and ROM calculations in the rest of this section are performed on one compute node (Dual Intel Xeon 6240R CPU) with 192GB 2933MHz DDR memory.

\subsection{1D Propagating Laminar Flame}
\label{subsec:1DLaminarFlame}
We first consider a one-dimensional, freely propagating, premixed laminar flame. The 1D problem is calculated using the governing equation in Eq.~\ref{eq:fom:governing} with simplified single-step, two-species reaction in which both the reactant and product species are treated as calorically perfect gases with identical molecular weights. Pertinent physical properties are summarized in Table~\ref{table:1DProblem}. The computational domain has a length of 10 mm, discretized with 1000 uniform finite volume cells. This has been confirmed to be sufficient to resolve the flame thickness ($\sim$1 mm). The FOM solution is computed using the second-order accurate backwards differentiation formula with dual time-stepping, and a constant physical time step size of $\dt = 0.01 \; \mu$s. The expression for the chemical reaction source term in Eq.~\ref{eq:fom:source_term} for the reactant follows the Arrhenius form 
\begin{equation}
    \dot{\omega}_\text{Reactant} = -MW_\text{Reactant} \cdot A \text{exp} \left( \frac{-E_A/R_u}{T} \right) \left[ \frac{\rho Y_\text{Reactant}}{MW_\text{Reactant}} \right]^a ,
    \label{1Dsource_term}
\end{equation}
with the pre-exponential factor $A = 2 \times 10^{10}$, the activation energy $E_A/R_u = 24{,}358$ K, and the concentration exponent $a = 1.0$, which is designed to model the conversion of perfectly-mixed reactants to completely-burned products via a one-dimensional flame. Characteristic boundary conditions are specified at the inlet and outlet of the domain while a perturbation of acoustic characteristics $q_{u-c}$ is imposed to introduce pressure oscillations 
\begin{equation}
    q_{u-c} = q_{u-c,ref} [ 1 + A_0 \sin{(2 \pi f t)} ]
    \label{eq:1d_qu-c_forcing}
\end{equation}
where $q_{u-c,ref}$ is the reference acoustic characteristics specified to maintain the nominal pressure at 1MPa in the computational domain, $f$ and $A_0$ represent the frequency and amplitude of the imposed perturbation, respectively. The generated pressure oscillations interact with the flame advection, leading to complex multi-scale interactions. 

\begin{table}
\centering
\begin{tabular}{ lllllll } 
\toprule
Species & MW (g/mol) & $c_p$ (kJ/kg/K) & Pr & Sc & $\mu_{ref}$ (kg/m/s) & $h_{ref}$ (kJ/kg) \\
\midrule
Reactant & 21.32 & 1.538 & 0.713 & 0.62 & 7.35$\times 10^{-4}$ &	-7,432 \\
Product	& 21.32 & 1.538 & 0.713 & 0.62 & 7.35$\times 10^{-4}$ & -10,800 \\
\bottomrule
\end{tabular}
\caption{\label{table:1DProblem}Properties of species reactant and product for 1D Propagating Laminar Flame.}
\end{table}

\begin{figure}
    \centering
    \begin{subfigure}[b]{\textwidth}
         \centering
         \includegraphics[width=0.55\textwidth]{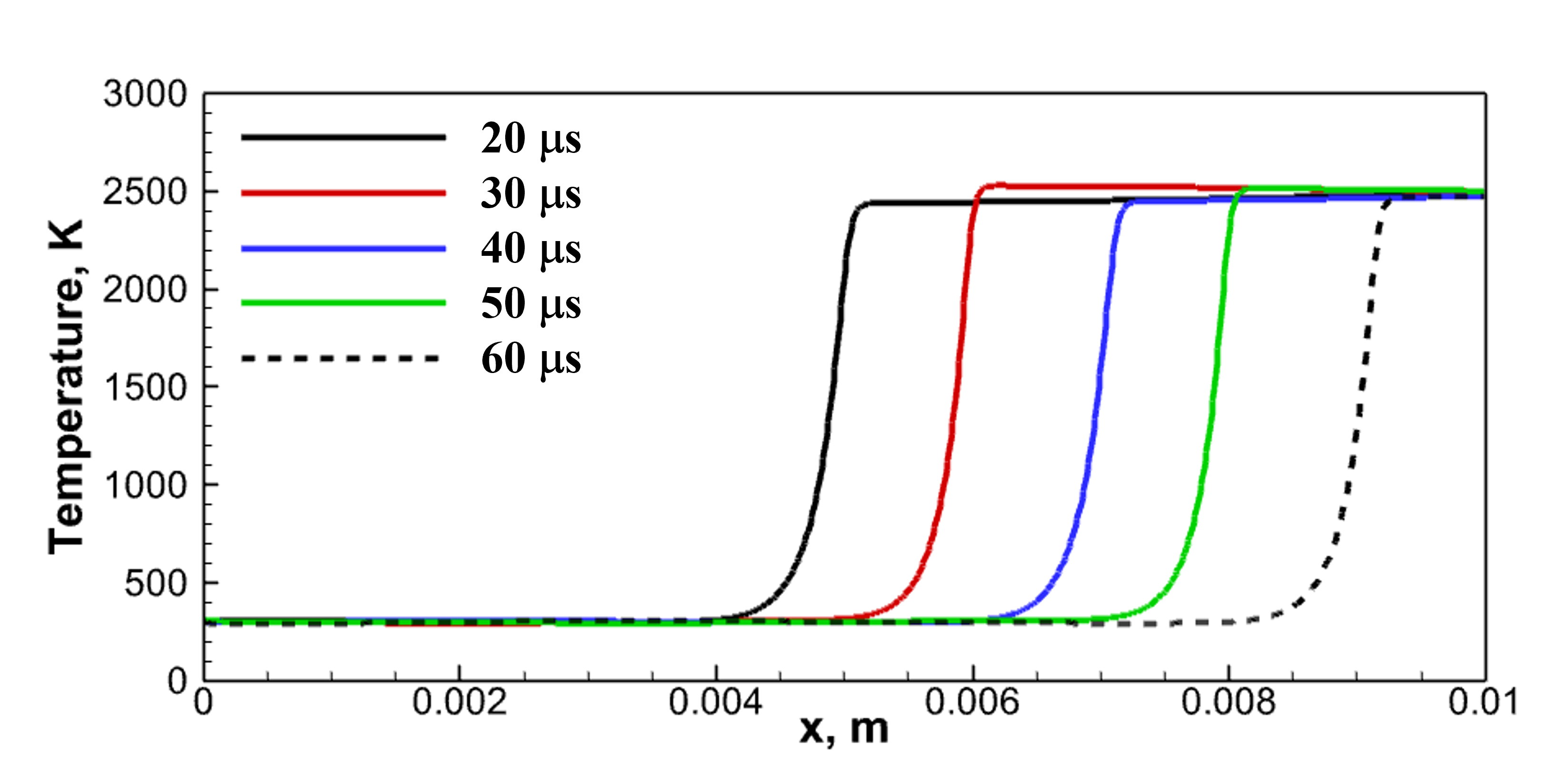}
         \caption{Temperature}
          \label{fig:1d:fom:T}
    \end{subfigure}
    \centering
    \begin{subfigure}[b]{\textwidth}
         \centering
         \includegraphics[width=0.55\textwidth]{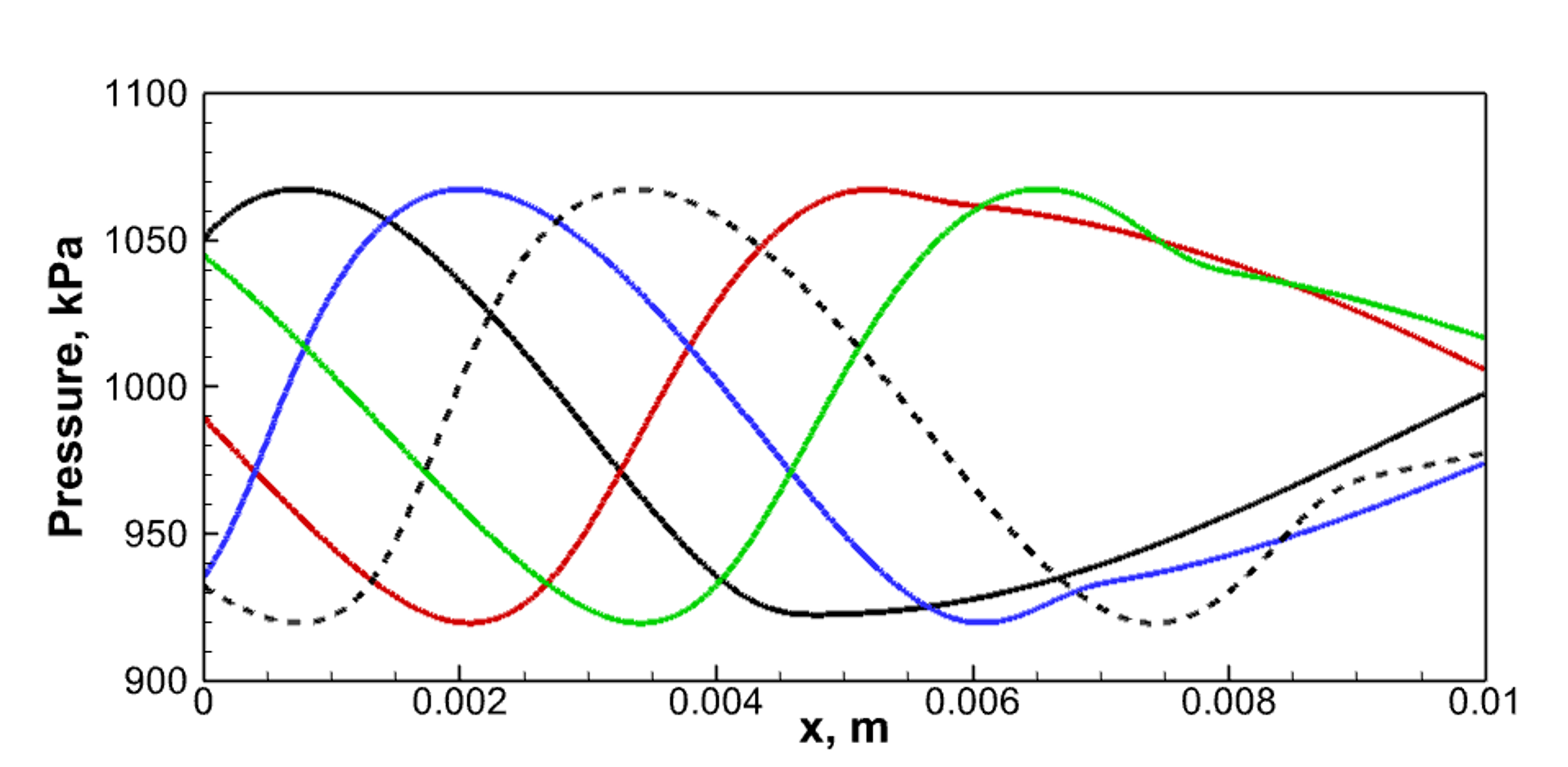}
         \caption{Pressure}
          \label{fig:1d:fom:P}
    \end{subfigure}
	\caption{Representative FOM solutions from 1D propagating laminar flame.}\label{fig:1d:fom} 
\end{figure}

The unsteady solution is advanced from 0 to 65$\mu s$ with solution snapshots collected at every time step. To exclude the initial transients,  solution snapshots from 20 to 65$\mu s$ (a total of 4500 snapshots) are used as the \textit{testing} dataset for the ROM evaluations. Representative temperature and pressure fields from FOM simulation, with an outlet perturbation, generated using $f = 50$kHz and $A_0 = 0.1$ specified in Eq.~\ref{eq:1d_qu-c_forcing}, are shown in Fig.~\ref{fig:1d:fom} at different time instants. The 1D laminar flame problem contains dynamics featuring both local (sharp-gradient flame convection) and global (acoustic oscillations) coherence, which creates an ideal platform to assess the proposed adaptive ROM formulation in section~\ref{subsubsec:comp_proc}. The flame dynamics is dominated by the convection of the sharp flame front accompanied by a temperature rise from 300 to 2500K as seen in Fig.~\ref{fig:1d:fom:T}, analogous to a linear advection problem that is known to exhibit Kolmogorov barrier for ROM development. Moreover, strong acoustic oscillations, indicated by the pressure fields, are introduced by the outlet perturbation, which show significant variations in acoustic wave length as seen from  the temperature variation in Fig.~\ref{fig:1d:fom:P}. Due to the higher sonic velocity in the high-temperature reaction zone, the acoustic wave travels faster, leading to longer wave lengths (right part of the domain) while lower sound speed in low-temperature non-reacting zone results in shorter wave lengths (left part of the domain). In addition, the temperature field is  also influenced by the pressure oscillations, as indicated by the small-amplitude modulations in temperature levels in Fig.~\ref{fig:1d:fom:T}. Though simulated with simplified physical models, this 1D model inherits many features that are challenging for ROM development, which include advection-dominated transport and multi-scale, multi-physics coupling. More importantly, the constructed 1D problem allows evaluations of ROM capabilities in great detail without incurring an exorbitant computational cost.

\subsubsection{POD Characteristics}
\label{subsec:1d:pod}

The POD characteristics of the 1D laminar flame shown in Fig.~\ref{fig:1d:fom} are first investigated to understand how well the POD trial basis represents the FOM dataset, which is evaluated using the POD residual energy 
\begin{equation}
    \text{POD Residual Energy}(\numSolModes), \;\% = \lp 1 - \frac{\sum^{\numSolModes}_{\genIdx=1} \sigmaPOD^2_\genIdx}{\sum^{\numSolModesTotal}_{\genIdx=1} \sigmaPOD^2_\genIdx} \rp \times 100,
    \label{eq:pod:res_energy}
\end{equation}
where $\sigmaPOD_\genIdx$ is the $\genIdx^\text{th}$ singular value of the SVD used to compute the trial basis \trialBasisPrim. The singular values are arranged in descending order. Again, $\numSolModes$ is the number of vectors retained in the POD trial basis, and $\numSolModesTotal$ is the total number of snapshots in the training dataset. In addition, the POD projection error is quantified in
terms of the normalized and variable-averaged $L^2$ error to assess the sufficiency of the POD trial basis in representing, especially in predicting, the target dynamics of interest
\begin{equation}
    \errProjVsFOM^{\iterIdx} = \frac{1}{\numVars} \sum^{\numVars}_{\varIdx=1} \frac{\norm{\solPrimROMProjVar^{\iterIdx} - \solPrimFOMVar^{\iterIdx}}}{\norm{\solPrimFOMVar^{\iterIdx}}} ,
    \label{eq:pod:proj_err}
\end{equation}
where $\solPrimROMProjVar^{\iterIdx}$ represents the $i^\text{th}$ solution variable of the state vector, $\solPrimROMProj^{\iterIdx}$, at time step $\iterIdx$, evaluated as 
\begin{equation}
    \solPrimROMProj^{\iterIdx} = \solPrimFOMRef + \scaleMatPrim^{-1} \trialBasisPrim \trialBasisPrim^T \solPrimFOM^{\iterIdx} ,
    \label{eq:pod:projed_fom}
\end{equation}
following Eq.~\ref{eq:pod_def} with $\solPrimFOM^\iterIdx$ obtained directly from the FOM solutions. We refer to $\solPrimFOM^\iterIdx$ as the projected FOM solution. It noteworthy that the projected FOM solution provides a \textit{best-scenario} evaluation on the accuracy of the resulting ROM (i.e. the ROM solution cannot produce more accurate prediction of the dynamics than the projected FOM solution).

\begin{figure}
     \centering
     \begin{subfigure}[t]{0.45\textwidth}
         \centering
         \includegraphics[width=1.0\textwidth]{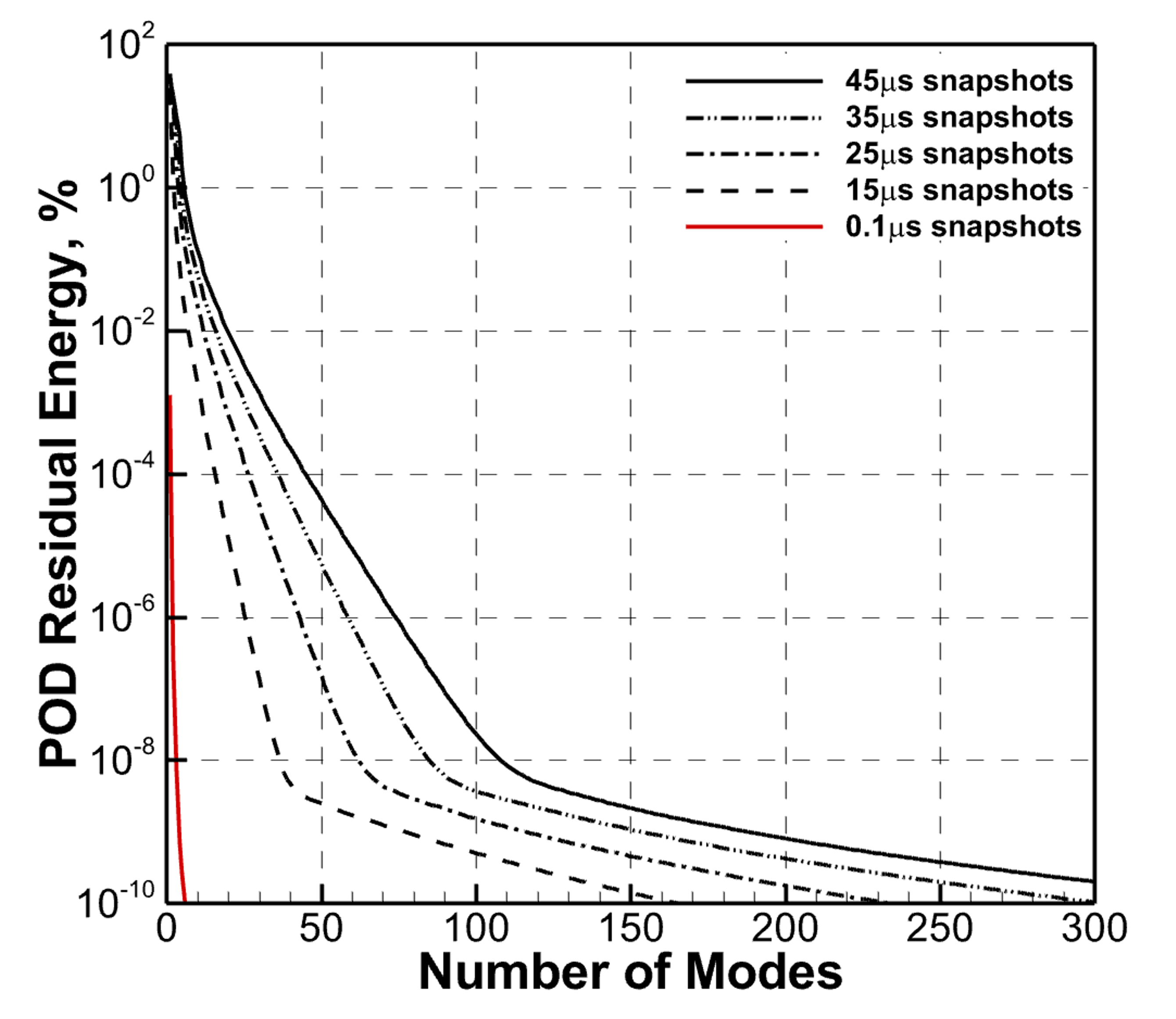}
         \caption{POD residual energy distribution}
         \label{fig:1d:pod_res_energy}
     \end{subfigure}
     \begin{subfigure}[t]{0.45\textwidth}
         \centering
         \includegraphics[width=1.0\textwidth]{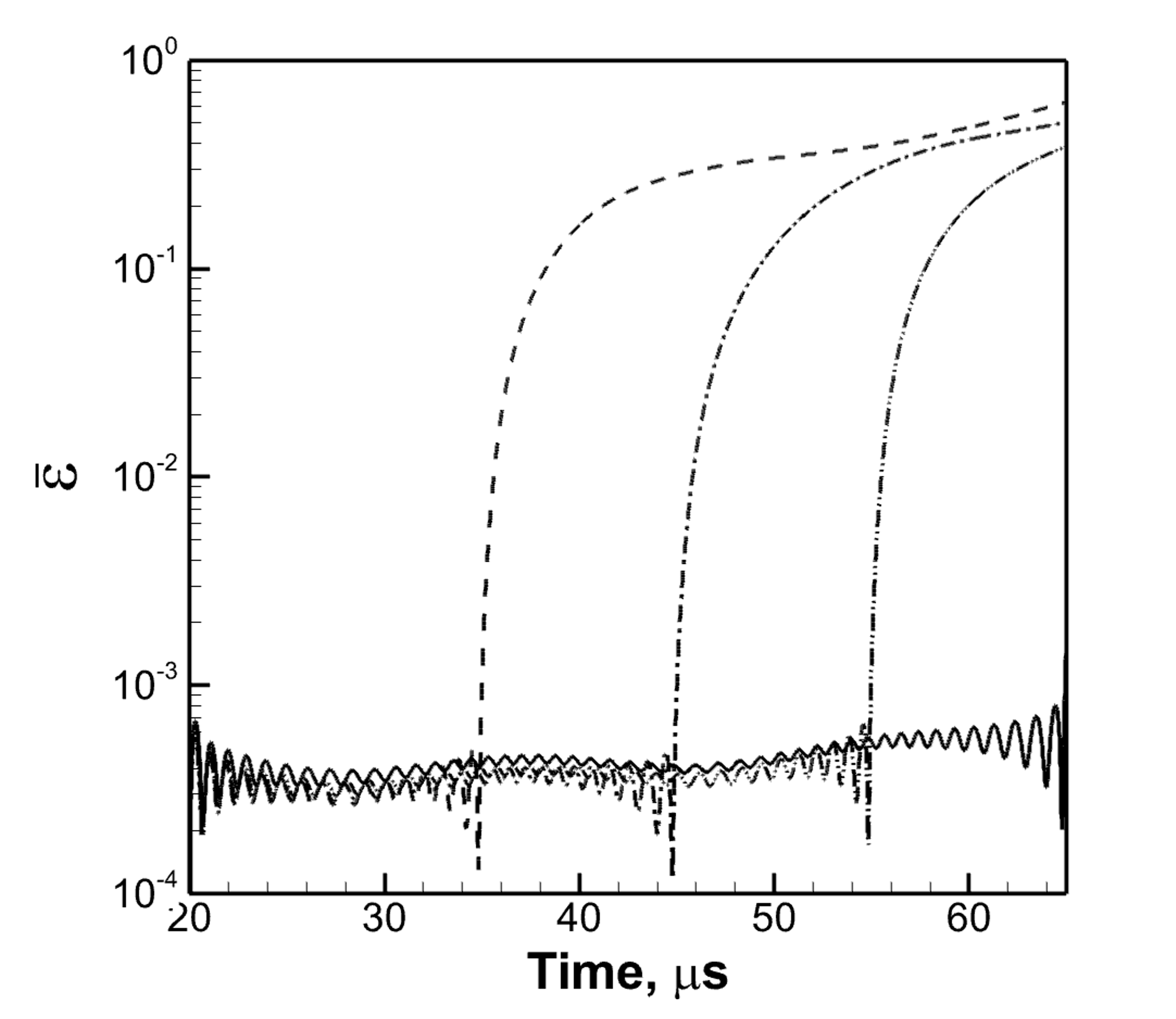}
         \caption{Projection error}
         \label{fig:1d:pod_proj_err}
     \end{subfigure}
     \caption{POD characteristics for the 1D propagating laminar flame.}\label{fig:1d:pod}
\end{figure}

Both the POD residual energy and the projection error are investigated in Fig.~\ref{fig:1d:pod} by including different numbers of FOM snapshots (starting from 20 $\mu s$) in the \textit{training} dataset to construct the POD trial basis, $\trialBasisPrim$ while it shall be reminded that the entire \textit{testing} dataset spans a total of 45$\mu s$. The residual energy is shown as a function of $\numSolModes$ in Fig.~\ref{fig:1d:pod_res_energy}, which reveals the information excluded by the POD representation for a given number of modes. The results show that, by increasing the number of snapshots in the training dataset, the corresponding number of modes to capture the same residual energy level increases proportionally, thus leading to a much slower POD residual energy decay. For example, to retrieve approximately 99.9999\% of the total energy, 16 modes are required for 15$\mu s$ training snapshots, 27 modes for 25$\mu s$, 36 modes for 35$\mu s$, and 45 modes for 45$\mu s$. On the other hand, computing basis using local trajectory in time leads to much faster POD residual energy decay (red line in Fig.~\ref{fig:1d:pod_res_energy}). This reveals that the POD residual energy decays slowly using global trajectories in time (black lines in Fig.~\ref{fig:1d:pod_res_energy}) and does not converge with the amount of training data - common characteristic expected in convection-dominated problems~\citep{McQuarrieOpInf2021}, which indicates the inadequacy of the linear subspace in constructing \textit{predictive} ROMs. Such insufficiencies are  further highlighted by  the  projection errors (Eq.~\ref{eq:pod:proj_err}) for the entire \textit{testing}-dataset period (20 to 65$\mu s$) using the number of modes needed to recover 99.9999\% of the total energy, as shown in Fig.~\ref{fig:1d:pod_proj_err}. Low POD projection errors can be seen within the training period, indicating the dynamics are well-represented by the POD trial basis. However, significant increases (from less than 0.1\% to more than 10\%) in POD projection errors are consistently observed when the trained POD bases are applied to represent dynamics beyond the training period (i.e. \emph{prediction} period), which implies poor predictive capabilities of the resulting ROMs based on these bases. It is noteworthy that the POD projection error remains high in the \emph{prediction} period even though 35$\mu s$ training snapshots (78\% of the \textit{testing} dataset) are included while the entire 45$\mu s$ of snapshots (100\%) are required for accurate representation of the \textit{testing} dataset, which defies the original motivation of developing ROMs.

\subsubsection{Performance of Hyper-reduced ROM: Static vs. Adaptive}
\label{subsubsec:1d:rom_cmp}

Both static and adaptive ROMs are constructed with hyper-reduction via gappy POD with randomized oversampling~\citep{PeherstorferODEIM}, the performance of which are compared in terms of accuracy and efficiency. For consistency, the ROM-performance evaluations are conducted spanning the whole window for the \textit{testing} dataset (20 to 65$\mu s$). The ROM accuracy is measured in terms of the normalized, time-averaged, and variable-averaged $L_2$ error 
\begin{equation}
    \errROMVsFOM =  \frac{1}{\numVars} \sum^{\numVars}_{\varIdx=1} 
 \frac{1}{\numSnaps} \sum^{\numSnaps}_{\iterIdx=1} \frac{\norm{\solPrimROMFull^{\iterIdx} - \solPrimFOMVar^{\iterIdx}}}{\norm{\solPrimFOMVar^{\iterIdx}}} ,
    \label{eq:rom:rom_err}
\end{equation}
where $ \solPrimROMFull^{\iterIdx} = \solPrimFOMRef + \scaleMatPrim^{-1} \trialBasisPrimN{\iterIdx} \solPrimROMRed^{\iterIdx} $ with the reduced state, $\solPrimROMRed^{\iterIdx}$, obtained as the solution to Eq.~\ref{eq:adaptiverom:hROM} for adaptive ROM and Eq.~\ref{eq:mplsvt_proj_hyperS} for static ROM. In addition, the computational efficiency gains, $\lambda$, are reported as the ratio of the FOM computational time to the ROM computational time , similar to Eq.~\ref{eq:ROM_efficiency_gain} \addrOne{, but instead of evaluating the FLOPs, the FOM and ROM computational time is obtained by directly timing the corresponding online calculations}. 

\begin{figure}
     \centering
     \begin{subfigure}[t]{0.7\textwidth}
         \centering
         \includegraphics[width=1.0\textwidth]{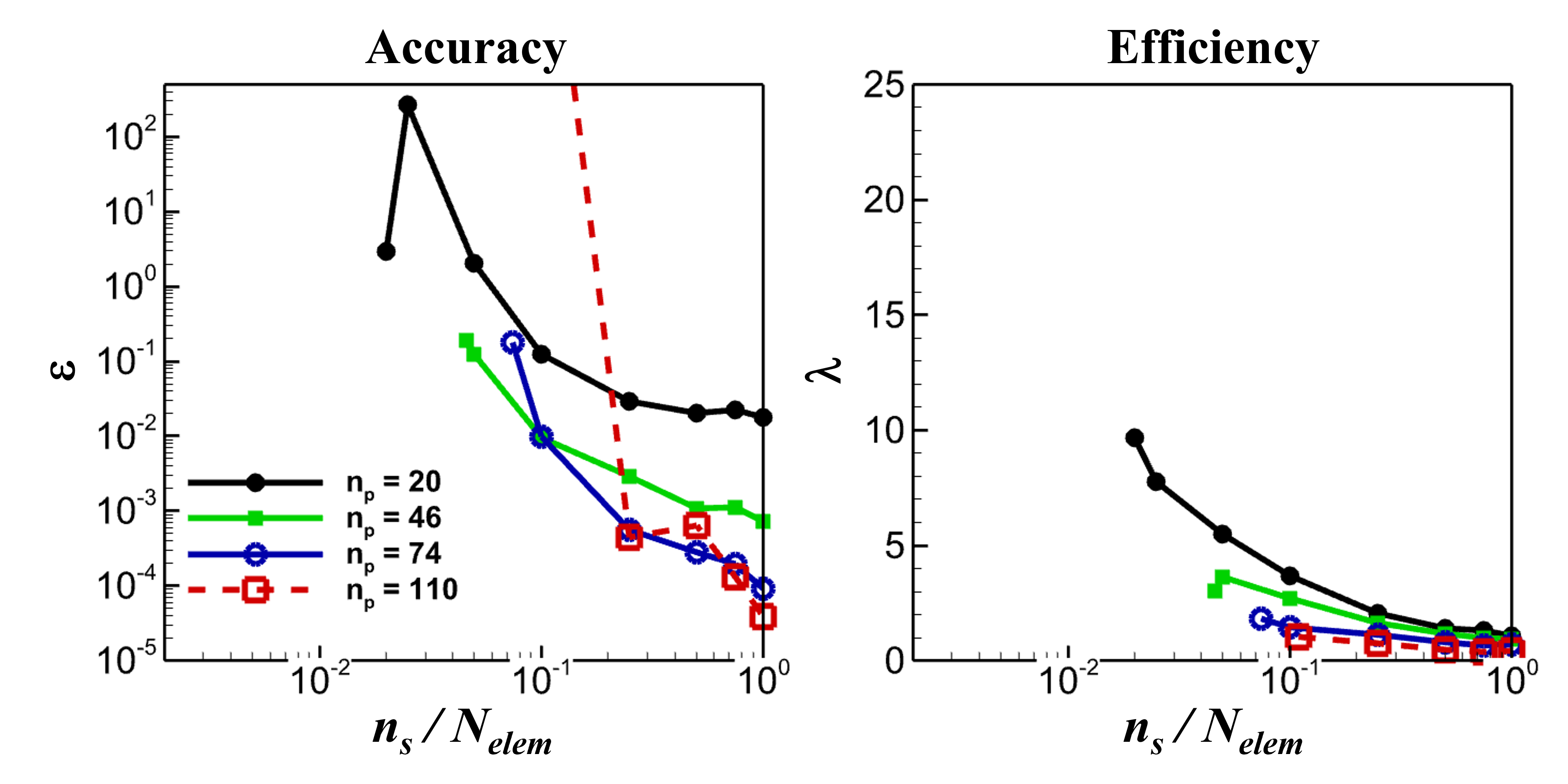}
         \caption{Static ROM}
         \label{fig:1d:srom_err}
     \end{subfigure}     
     \begin{subfigure}[t]{0.7\textwidth}
         \centering
         \includegraphics[width=1.0\textwidth]{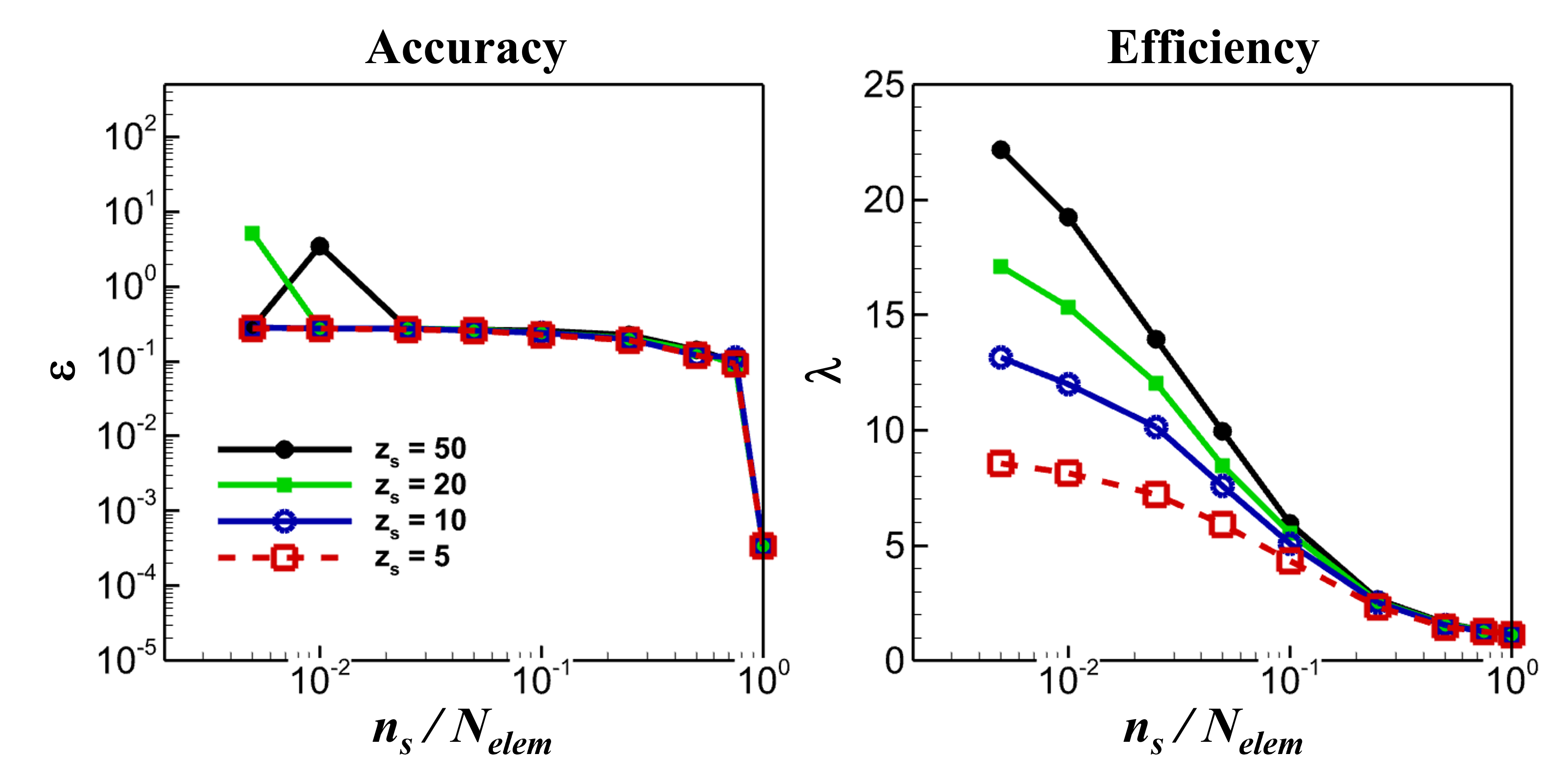}
         \caption{Adaptive ROM \textbf{without} non-local information incorporated}
         \label{fig:1d:aromLocal_err}
     \end{subfigure}
     \begin{subfigure}[t]{0.7\textwidth}
         \centering
         \includegraphics[width=1.0\textwidth]{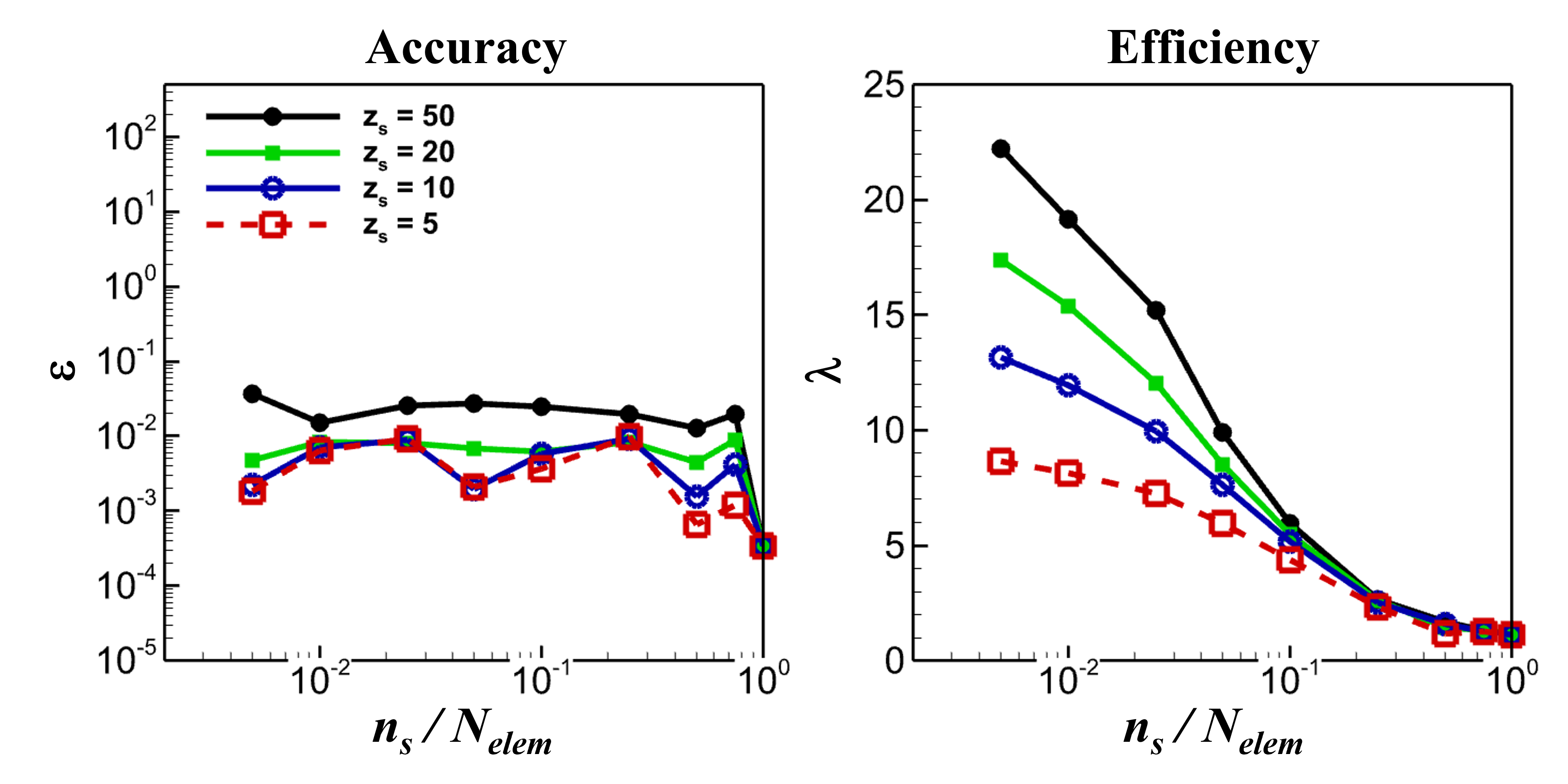}
         \caption{Adaptive ROM \textbf{with} non-local information incorporated}
         \label{fig:1d:arom_err}
     \end{subfigure}
     \caption{Comparisons of accuracy (left) and efficiency (right) between the adaptive ROMs (with and without non-local information incorporated) and static ROMs with respect to the sampling points percentage ($n_s/N$) for the 1D propagating laminar flame (for \textbf{adaptive ROM} - \textit{training} dataset: 20 to 20.1$\mu s$, \textit{testing} dataset: 20 to 65$\mu s$, \textit{and} \textit{prediction} perioid: 20.1 to 65$\mu s$; for \textbf{static ROM} - \textit{training} dataset: 20 to 65$\mu s$, \textit{testing} dataset: 20 to 65$\mu s$, \textit{and} \textit{prediction} perioid: none).}\label{fig:1d:arom_vs_srom}
\end{figure}

Figure~\ref{fig:1d:arom_vs_srom} compares the performance of static and adaptive ROMs with different model parameters and a varying fraction of sampling points, $\numSamps / \numElements$, included for hyper-reduction. Following Algorithm~\ref{algorithm:adaptiveROM}, the adaptive ROMs are initialized with 0.1$\mu s$ snapshots ($w_\text{init} = 10$) in Fig.~\ref{fig:1d:pod_res_energy} and developed with $\numSolModes = 2$ (retrieving 99.9999\% of the total energy), $p_1 / K = 0.5$, and $p_2 / K = 0.5$. As illustrated in section~\ref{subsec:adaptivity:algorithm}, the update rate for sampling points and non-local full-state estimation (i.e. $z_s$) is a determining factor for the performance of the adaptive ROM. Therefore, we choose $z_s$ as the model parameter for adaptive ROM evaluations, and specifically four values ($z_s = 5, \ 10, \ 20, \ \text{and} \ 50$) are selected for investigations. In addition, to obtain a comprehensive assessment of the proposed adaptive ROM formulation, we also develop adaptive ROMs without non-local information incorporated (i.e. no estimation of full states at the unsampled points in Eq.~\ref{eq:adaptiverom:unsampledFullState}). Given the limited prediction that the static ROM can provide beyond the training period as shown in Fig.~\ref{fig:1d:pod_proj_err}, we only derive static ROMs for comparisons with adaptive ROMs using 45$\mu s$ snapshots, the same as the testing dataset and thus resulting in no prediction period. Moreover, we consider four different numbers of trial-basis modes ($\numSolModes = 20, \ 46, \ 74, \ \text{and} \ 110$) as the model parameter for static-ROM construction, which correspond to $10^{-2}$, $10^{-4}$, $10^{-6}$, and $10^{-8} \ \%$ POD residual energy in Fig.~\ref{fig:1d:pod_res_energy}, respectively.

As expected, higher errors and more efficiency gains are observed in all the ROMs as the number of sampling points decreases. As seen in Fig.~\ref{fig:1d:srom_err}, the static ROM performance shows strong dependency on the number of trial-basis modes, $\numSolModes$, which also determines the minimum number of sampling points required to avoid solving an under-determined least-square problem in gappy POD with randomized samples~\citep{PeherstorferODEIM}. In addition, to obtain reasonably accurate reproduction of the testing dataset (e.g. $< 10\%$), at least 10\% of the sampling points are needed, corresponding to less than O(5) efficiency gain. The adaptive ROM results, with and without non-local information incorporated, are compared in Figs.~\ref{fig:1d:aromLocal_err} and~\ref{fig:1d:arom_err}, which readily reveal the benefits of incorporating full-state estimation at the unsampled points in Eq.~\ref{eq:adaptiverom:unsampledFullState} with approximately two orders of magnitude reduction in errors while preserving the efficiency gains.  

Compared to the static ROMs, the adaptive ROMs (with non-local information incorporated) exhibit significantly improved performance. With $\numSolModes = 2$, lower number of sampling points can be used, resulting in higher  computational efficiency gains. On the other hand, the adaptation of the basis and sampling produces highly accurate ROM results with $< 3\%$ errors for the adaptive ROMs in Fig.~\ref{fig:1d:arom_err}. For example, with $z_s = 50$ and 0.5\% sampling points, the adaptive ROM achieves O(22) acceleration in computational time with approximately $3\%$ error compared to the FOM. We remark that even though theoretically higher efficiency gains are expected for static ROMs than the adaptive ROMs based on the analysis in Fig.~\ref{fig:flops_arom}, other metrics, such as accuracy and predictive capability, can largely bias the ROM performance for practical applications. More importantly, it will be emphasized that given the negligible offline training window, the adaptive ROM results are \textit{true} \emph{predictions} of the dynamics in the testing dataset while the static ROM results are merely a \emph{reproduction} of the training data.

Though excellent results are obtained with the proposed adaptive ROM formulation as shown in Fig.~\ref{fig:1d:arom_err}, it is noteworthy that the performance of the resulting adaptive ROM heavily relies on the model parameter $z_s$. With higher $z_s$, the adaptive ROM adopts a larger time step adopted to estimate the full state at the unsampled points in Eq.~\ref{eq:adaptiverom:unsampledFullState} and therefore produces less accurate results. On the other hand, higher $z_s$ leads to less frequent sampling points updates and full-state estimation at all the points (Eq.~\ref{eq:adaptiverom:interpErrNorm}), thus reducing the computational cost and resulting in more efficiency gain. Therefore, as highlighted in sections~\ref{subsubsec:BasisAdapt} and~\ref{subsubsec:SamplingAdapt} above, the selection of $z_s$ reflects the compromise between accuracy and efficiency of the resulting adaptive ROMs.

\subsubsection{Sensitivity of the Adaptive ROM to the Initial Training Window Size}
\label{subsubsec:1d:sensitivity_arom}
\begin{figure}
     \centering
     \includegraphics[width=0.8\textwidth]{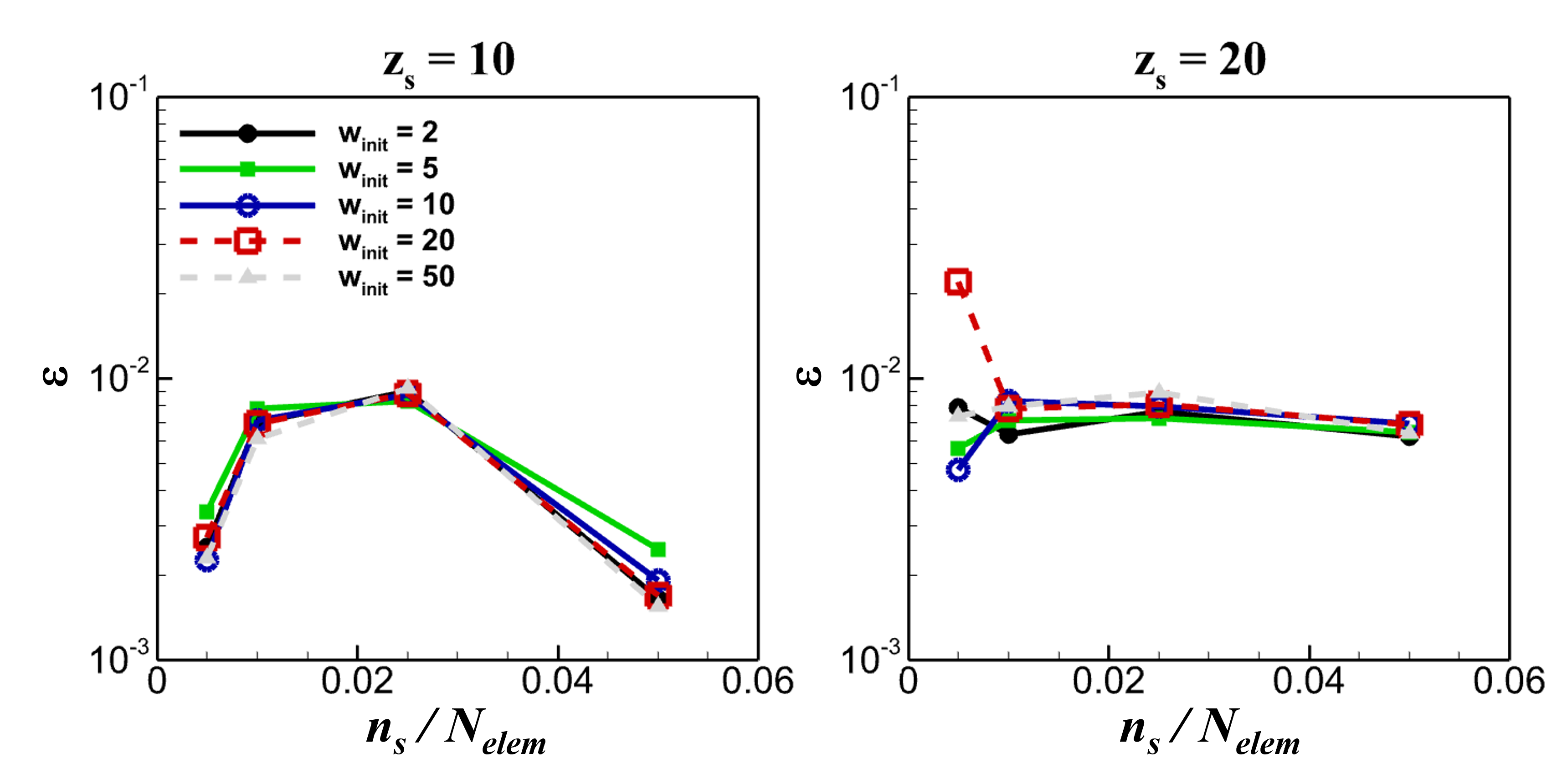}
     \caption{Evaluations of adaptive ROMs' sensitivity to the initial window size ($w_{init}$) for the 1D propagating laminar flame (\textit{training} dataset: 20 to 20+$0.01 w_\text{init} \ \mu s$, \textit{testing} dataset: 20 to 65$\mu s$, \textit{and} \textit{prediction} perioid: 20+$0.01 w_\text{init}$ to 65$\mu s$).}\label{fig:1d:arom_sensitivity}
\end{figure}

In addition to the the update rate for sampling points and non-local full-state estimation, $z_s$, we now investigate the sensitivity of the adaptive ROMs' accuracy to the initial training window size, $w_\text{init}$, based on the error defined in Eq.~\ref{eq:rom:rom_err}. Five values ($w_\text{init} = 2, \ 5, \ 10, \ 20, \ \text{and} \ 50$) are selected for investigations over four sampling points ($\numSamps / \numElements = 0.5, \ 1, \ 2.5, \ \text{and} 5 \% $). The results are shown in Fig.~\ref{fig:1d:arom_sensitivity} with the adaptive ROMs constructed using $\numSolModes = 2$ and two update rates, $z_s = 10$ and $20$, considered. It can be readily seen that overall the accuracy of the resulting adaptive ROMs does not exhibit strong sensitivity to the initial training window size. The prediction errors, $\errROMVsFOM$, remain below $1\%$ for $z_s = 10$ and mostly for $z_s = 20$ (except for $w_\text{init} = 20$ and $\numSamps / \numElements = 0.5\%$, which still remains well below $3\%$). This robustness with respect to the initial training window size is consistent with the analysis of Peherstorfer~\citep{PeherstorferADEIM}, which indicates the flexibility of the adaptive ROM in terms of the \textit{offline} training. Thus, this enables an automated procedure to derive adaptive ROMs in contrast to the conventional offline/online methods that often requires intensive investigations and tuning regarding the \textit{offline} training window length. The computational cost during the offline stage constitutes a major fraction in constructing the offline/online-method-based ROM but is usually not reported in the literature.

\subsubsection{Evaluations of Parametric Predictive Capabilities}
\label{subsubsec:1d:parametric_arom}
\begin{figure}
     \centering
     \includegraphics[width=0.9\textwidth]{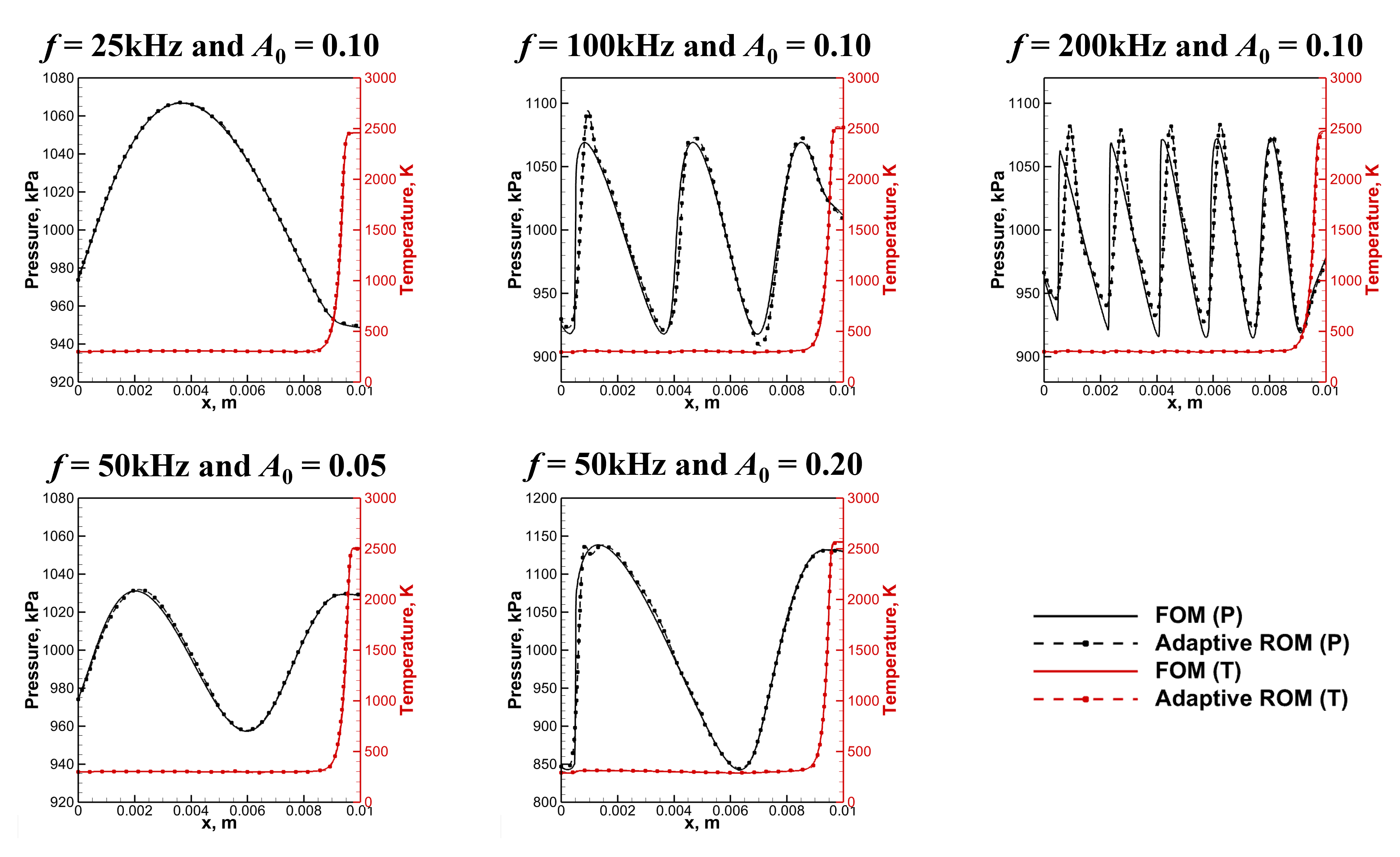}
     \caption{Comparisons of representative FOM and adaptive ROM solutions at 65 $\mu s$ for the 1D propagating laminar flame (\textit{training} dataset: 20 to 20.1$\mu s$ from \emph{FOM performed with $f = 50$kHz and $A_0 = 0.1$}, \textit{testing} dataset: 20 to 65$\mu s$ with different downstream forcing conditions labled on the figures,  \textit{and} \textit{prediction} period: the same as \textit{testing}).}\label{fig:1d:arom_parametric}
\end{figure}

With the adaptive ROM demonstrated to provide accurate \textit{future-state} predictions under the same condition parameter, the final evaluation in this section focuses on the \emph{parametric} predictive capabilities. Specifically, we adopt the adaptive ROM derived in section~\ref{subsubsec:1d:rom_cmp} to predict the dynamics under various acoustic perturbations different values of $f$ and $A_0$ applied in Eq.~\ref{eq:1d_qu-c_forcing}. Specifically, we select the adaptive ROM developed based on 10 snapshots (i.e. $w_\text{init} = 10$) from FOM performed with $f = 50$kHz and $A_0 = 0.1$ with $z_s = 20$ and $\numSamps / \numElements = 0.5\%$ and $1.0\%$. Five FOM simulations are performed with different acoustic perturbations and similar to the evaluations above, the solution snapshots from 20 to 65$\mu s$ of each simulation are used as the \textit{testing dataset} to assess the \emph{parametric} predictive capabilities of the adaptive ROM based on the error in Eq.~\ref{eq:rom:rom_err}, which is summarized in Table~\ref{table:1d:arom_parametric}. It shall be pointed out that the adaptive ROM is initialized \emph{once} (using 10 FOM snapshots obtained with $f = 50$kHz and $A_0 = 0.1$) and then directly applied for parametric predictions with no additional \textit{offline} training as mentioned in section~\ref{subsubsec:comp_proc} above. Therefore, the adaptive ROM results in Table~\ref{table:1d:arom_parametric} are \textit{true} predictions from 20 to 65$\mu s$, which present accurate predictions of flame dynamics under different perturbation frequencies and amplitudes with overall $< 8\%$ errors. In addition, by doubling the sampling points, the adaptive ROM provides more accurate predictions while it also becomes less efficient based on Fig.~\ref{fig:1d:arom_err}. The predictive capabilities of the adaptive ROM are further examined by comparing the pressure and temperature fields between the FOM and adaptive ROM in Fig.~\ref{fig:1d:arom_parametric} at 65 $\mu s$, the last time step of the testing dataset. Excellent agreement in the temperature profiles can be readily seen between FOM and adaptive ROM. Though discrepancies can be observed in pressure profiles, the adaptive ROM predictions match the FOM closely overall and more importantly, the adaptive ROM is capable of capturing the trend of dynamics changes due to parametric variations in acoustic perturbations, a crucial aspect in practical applications.

\begin{table}
\centering
\begin{tabular}{cccc} 
\toprule
$\numSamps / \numElements$ ($\%$) & $f$ (kHz) & $A_0$ & $\epsilon$ \\
\midrule
\multirow{5}{1em}{$0.5$} & 25 & 0.10 & $1.10 \times 10^{-2}$ \\
                         & 50 & 0.05 & $1.17 \times 10^{-2}$ \\
                         & 50 & 0.20 & $7.59 \times 10^{-2}$ \\
                         & 100 & 0.10 & $2.83 \times 10^{-2}$ \\
                         & 200 & 0.10 & $3.19 \times 10^{-2}$ \\
\midrule
\multirow{5}{1em}{$1.0$} & 25 & 0.10 & $9.17 \times 10^{-3}$ \\
                         & 50 & 0.05 & $5.97 \times 10^{-3}$ \\
                         & 50 & 0.20 & $7.63 \times 10^{-2}$ \\
                         & 100 & 0.10 & $1.34 \times 10^{-2}$ \\
                         & 200 & 0.10 & $3.35 \times 10^{-2}$ \\
\bottomrule
\end{tabular}
\caption{\label{table:1d:arom_parametric} Comparisons of the errors in predicting the 1D propagating laminar flame with different acoustic perturbations using adaptive ROM with $w_\text{init} = 10$ (from \emph{FOM performed with $f = 50$kHz and $A_0 = 0.1$}), $z_s = 20$ and $\numSamps / \numElements = 0.5\%$ and $1.0\%$.}
\end{table}

\subsection{2D Reacting Injector}
\label{subsec:2dInjector}
\begin{figure}
	\centering
	\includegraphics[width=1.0\textwidth]{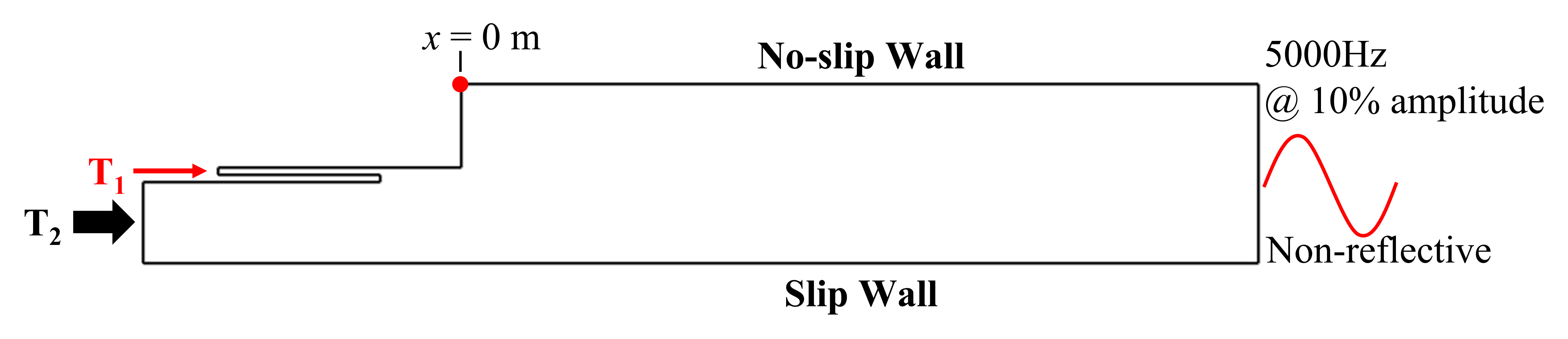}
	\caption{Computational domain of the planar reacting injector simulation (the red dot labels the position to monitor the local pressure time trace).}\label{fig:2d:geometry} 
\end{figure}

 Next, we extend our investigations to a planar representation of the generic laboratory-scale rocket combustor~\cite{YuJPP}, which has been established as a benchmark problem~\citep{Huang_DeepBlueData} for different ROM methods regarding reacting flow applications~\citep{HuangAIAAJ2019,Swischuk2020AIAAJ_LL,McQuarrieOpInf2021,Xu_CPGnet}. The configuration is shown in Fig.~\ref{fig:2d:geometry} and consists of a shear coaxial injector with an outer passage, $T_1$, that introduces fuel near the downstream end of the coaxial inner passage, $T_2$, that feeds oxidizer to the combustion chamber. The $T_1$ stream contains gaseous methane (100\% \methane) at 300K. The $T_2$ stream is 42\% gaseous \oxygen\ by mass and 58\% gaseous \water\ by mass at 700K. Operating conditions in the combustion chamber are maintained similar to conditions in the laboratory combustor~\cite{YuJPP,YuPhD}, with an adiabatic flame temperature of approximately 2700K and a mean chamber pressure of 1.1MPa. Both the $T_1$ and $T_2$ streams are fed with constant mass flow rates, 0.37 and 5.0kg/s, respectively. A non-reflective boundary condition is imposed at the downstream end to allow acoustic waves to properly exit the domain and control acoustic effects on the combustion dynamics. A sinusoidal acoustic perturbation with $f = 5000$Hz and $A_0=0.1$ is imposed at the downstream boundary following Eq.~\ref{eq:1d_qu-c_forcing}. Transport of four chemical species (\methane, \oxygen,  \water, and \carbonDiox) is modeled based on Eq.~\ref{eq:fom:governing}. The chemical reaction is modeled by the global single-step methane-oxygen reaction recommended by Westbrook and Dryer~\cite{WestbrookDryer}: $\methane + 2 \oxygen \rightarrow \carbonDiox + 2 \water$, with all the chemical species treated as thermally perfect gases and a reduced Arrhenius pre-exponential factor ten times smaller~\citep{HuangAIAAJ2019} than the recommended value. Even with this reduced reaction rate, construction of robust and accurate ROMs remains highly challenging, especially in predicting the \addrTwo{non-coherent} dynamics beyond the training window~\citep{HuangAIAAJ2019,McQuarrieOpInf2021}. Similar to the 1D problem in section~\ref{subsec:1DLaminarFlame}, The 2D FOM solution is computed using the second-order accurate  backwards difference formula and dual time-stepping, with a constant physical time step of 0.1$\mu s$. The  mesh consists of a total of 38,523 finite volume cells with 8 solution variables ($p$, $u$, $v$, $T$, $Y_{\methane}$, $Y_{\oxygen}$, $Y_{\water}$, and $Y_{\carbonDiox}$), resulting in a total of 308,184 degrees of freedom.

\begin{figure}
 	\centering
 	\includegraphics[width=1.0\textwidth]{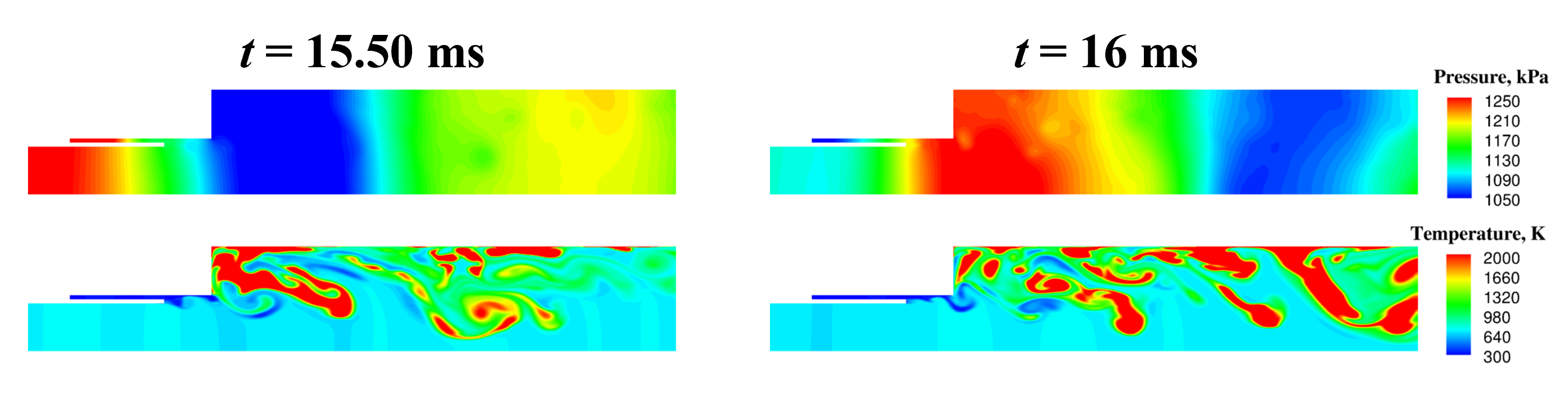}
 	\caption{Representative instantaneous snapshots of pressure (top), temperature (middle) and $CH_4$ mass fraction (bottom) from FOM simulation of the 2D reacting  injector.}\label{fig:2d:fom_snapshots} 
\end{figure}

The unsteady solution of the 2D reacting injector is advanced from 0 to 21ms with solution snapshots from $15$ to $21$ms (a total of 60,000 snapshots) used as the \textit{testing} dataset to evaluate the adaptive ROM's capabilities in predicting long-term dynamics. Two representative instantaneous snapshots of the 2D reacting single injector FOM solutions are shown in Fig.~\ref{fig:2d:fom_snapshots} to demonstrate the overall characteristics of the flow fields and highlight the dominant physics in the problem. The pressure exhibits global dynamics (non-local coherence) over the entire domain with the wave form corresponding to the $5000$Hz downstream perturbation. On the other hand, the combustion dynamics, represented by temperature and \methane mass fraction, are featured with local coherence, which is intermittently distributed in both space and time and spans a wide range of scales, from the small eddies in the shear layers to the large-scale recirculation zone immediately downstream of the dump plane at $x = 0$m. More importantly, strong interactions can be identified between pressure and combustion dynamics. High-temperature pockets are distributed downstream of the combustor when the pressure is high near the dump plane. Alternately, when the pressure is low at the dump plane, high-temperature pockets are concentrated closer to the dump plane. These unique features and interactions introduce varying levels of difficulty in constructing a \textit{truly} predictive ROM, especially in capturing the \addrTwo{non-coherent} dynamics.

\subsubsection{POD Characteristics}
\label{subsubsec:2dInjector:POD}

\begin{figure}
     \centering
     \includegraphics[width=0.6\textwidth]{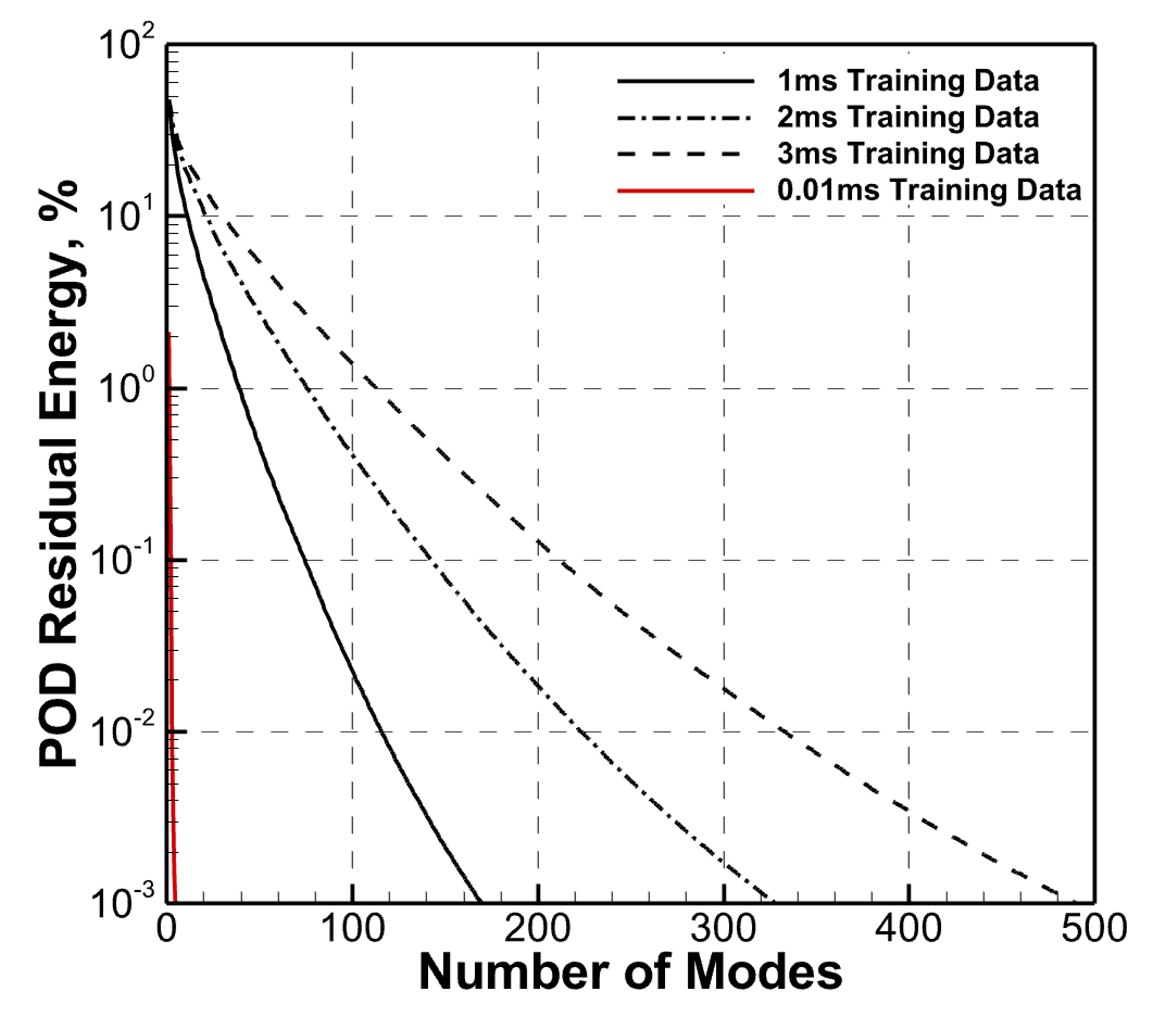}
     \caption{POD characteristics for the 2D reacting injector.}\label{fig:2d:pod}
\end{figure}

The  residual energy (Eq.~\ref{eq:pod:res_energy}) of the POD modes is shown in Fig.~\ref{fig:2d:pod} by including different numbers of FOM snapshots (from $15$ms) in the \textit{training} dataset. Similar to the 1D laminar flame (Fig.~\ref{fig:1d:pod_res_energy}), the POD residual energy does not exhibit convergence with increasing  training data while also showing slower decays due to the increased levels of physical complexity in the 2D reacting injector as observed in Fig.~\ref{fig:2d:fom_snapshots}. Such  characteristics have also been reported by McQuarrie et al.~\citep{McQuarrieOpInf2021} using a different set of solution variables to compute POD modes. The results in Fig.~\ref{fig:2d:pod} show that to recover 99.9\% of the total energy, 75 modes are required for  training snapshots in a 1ms time interval, 143 modes for 2ms, and 212 modes for 3ms while much more modes are needed to recover 99.99\% of the total energy (1ms: 116 modes, 2ms: 224 modes, and 3ms: 334 modes). We remark that many fundamental projection-based ROM methodologies are tested on problems that require  $\sim 10$ trial-basis modes to achieve 99.9\% POD energy~\citep{LeeNonlinearManifold2020,Barone2009JCP,sanANNClosure} while ROMs for more practical engineering systems generally require $\sim 100$ trial-basis modes~\citep{HuangMPLSVT2022,Carlberg2017,GrimbergPROM2020_stability}.

\subsubsection{Performance of the Adaptive ROM}
\label{subsubsec:2dInjector:AROM}

\begin{figure}
     \centering
     \includegraphics[width=0.75\textwidth]{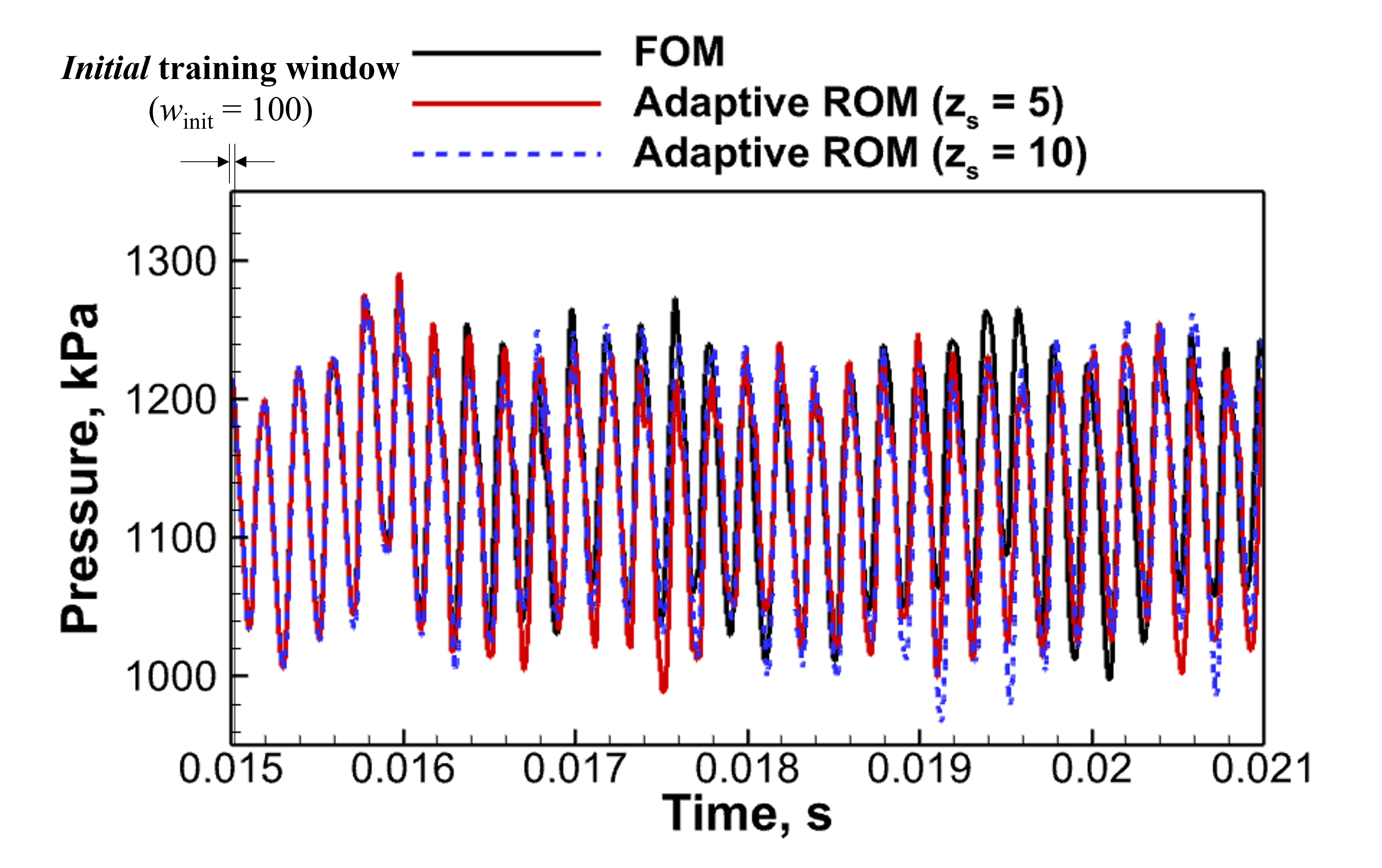}
     \caption{Comparisons of local pressure traces (monitored at the red dot in Fig.~\ref{fig:2d:geometry}) between FOM and adaptive ROM for the 2D reacting injector (\textit{training} dataset: 15 to 15.01ms, \textit{testing} dataset: 15 to 21ms, \textit{and} \textit{prediction} perioid: 15.01 to 21ms).}\label{fig:2d:arom_vs_fom:Plocal}
\end{figure}

The adaptive ROM is initialized using 0.01ms training snapshots ($w_\text{init}=100$) sampled every 10 time steps, which leads to 10  trial basis modes in total. We then  evaluate the adaptive ROM with $\numSolModes = 5$ (recovering $> 99.999\%$ of the total energy as seen in Fig.~\ref{fig:2d:pod}), $\numSamps / \numElements = 1.0\%$, $p_1 / K = 0.5$, $p_2 / K = 0.5$, and two update rates, $z_s = 5 \ \text{and} \ 10$, which produce gains in computational efficiency of $\lambda =$ 9 and 18, respectively. Figure~\ref{fig:2d:arom_vs_fom:Plocal} compares the local pressure traces between the FOM and the adaptive ROMs, which exhibits excellent agreement, and more importantly demonstrates the capabilities of the adaptive ROM in predicting long-term dynamics ($0.01ms$ training versus $\sim 6ms$ prediction). It is noted that even though the pressure oscillations are expected to be periodic following the sinusoidal perturbations in Eq.~\ref{eq:1d_qu-c_forcing}, the pressure trace from FOM still presents non-stationary modulations (e.g. increasing pressure peak values from 15 to 16ms) due to \addrTwo{non-coherent} combustion dynamics (illustrated in Fig.~\ref{fig:2d:fom_snapshots}). This aspect is also well-represented by the adaptive ROMs.

Furthermore, due to the \addrTwo{non-coherent} nature of the dynamics present in the 2D reacting injector (Fig.~\ref{fig:2d:fom_snapshots}), quantitative assessment of the adaptive ROMs following Eq.~\ref{eq:rom:rom_err} might be misleading because a small shift in convection-dominated features, such as the sharp-gradient flames, can often result in large-magnitude errors even though the overall dynamics might still be reasonably represented by the ROM. Therefore, to mitigate the impact of such artifacts on our conclusions, we assess the predictive capabilities of the adaptive ROM based on two important quantities of interest (QoIs), the time-averaged and root-mean-square (RMS) fields of the state variables, which serve as crucial determining factors in many engineering applications
\begin{equation}
    \mathbf{\phi}_{i,\text{average}} = \frac{1}{\numSnaps} \sum^{\numSnaps}_{\iterIdx=1} \mathbf{\phi}_{i}^\iterIdx, \ \ \ \text{and} \ \ \ 
    \mathbf{\phi}_{i,\text{RMS}} = \sqrt{\frac{1}{\numSnaps} \sum^{\numSnaps}_{\iterIdx=1} \left( \mathbf{\phi}_{i}^\iterIdx - \mathbf{\phi}_{i,\text{average}} \right)^2},
    \label{eq:2d:mean_rms}
\end{equation}
where $\numSnaps$ is the total number of snapshots included to calculate the QoI and $\mathbf{\phi}_{i}^\iterIdx$ represents the $i^\text{th}$ solution variable of the state vector from FOM ($\solPrimFOMVar^{\iterIdx}$), projected FOM ($\solPrimROMProjVar^{\iterIdx}$ defined in Eq.~\ref{eq:pod:projed_fom}), or ROM ($\solPrimROMFull^{\iterIdx} = \solPrimFOMRef + \scaleMatPrim^{-1} \trialBasisPrimN{\iterIdx} \solPrimROMRed^{\iterIdx}$), at time step $\iterIdx$. We remark that another remedy to obtain unbiased evaluations of ROM accuracy is through the cross-correlation, which has been devised and demonstrated by Qian et al.~\citep{qian2022_3Dcvrc} for a rocket combustion problem.

\begin{figure}
     \centering
     \begin{subfigure}[t]{0.9\textwidth}
         \centering
         \includegraphics[width=1.0\textwidth]{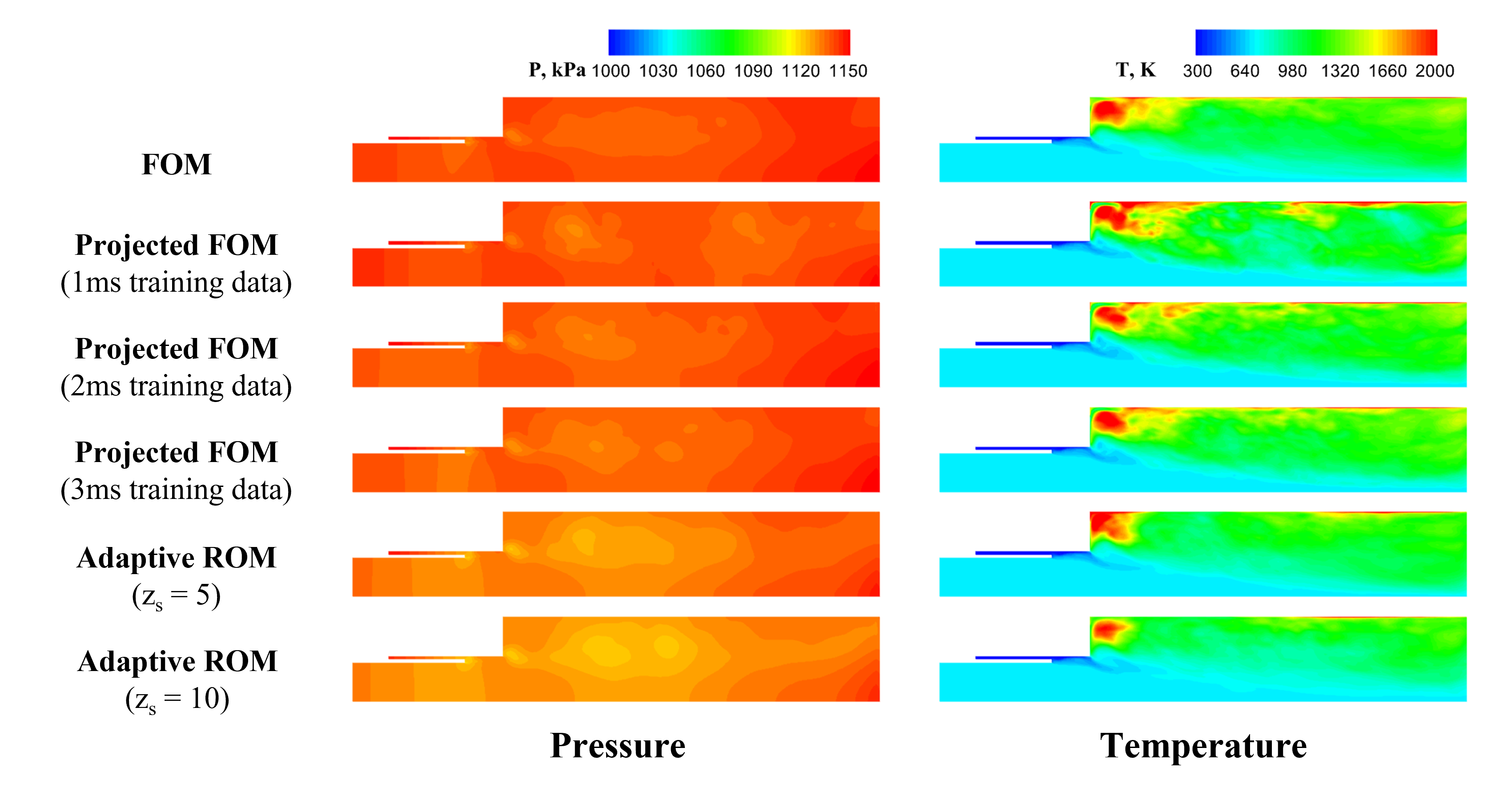}
         \caption{Time-averaged Fields}
         \label{fig:2d:arom_vs_pod_vs_fom:mean}
     \end{subfigure}
     \begin{subfigure}[t]{0.9\textwidth}
         \centering
         \includegraphics[width=1.0\textwidth]{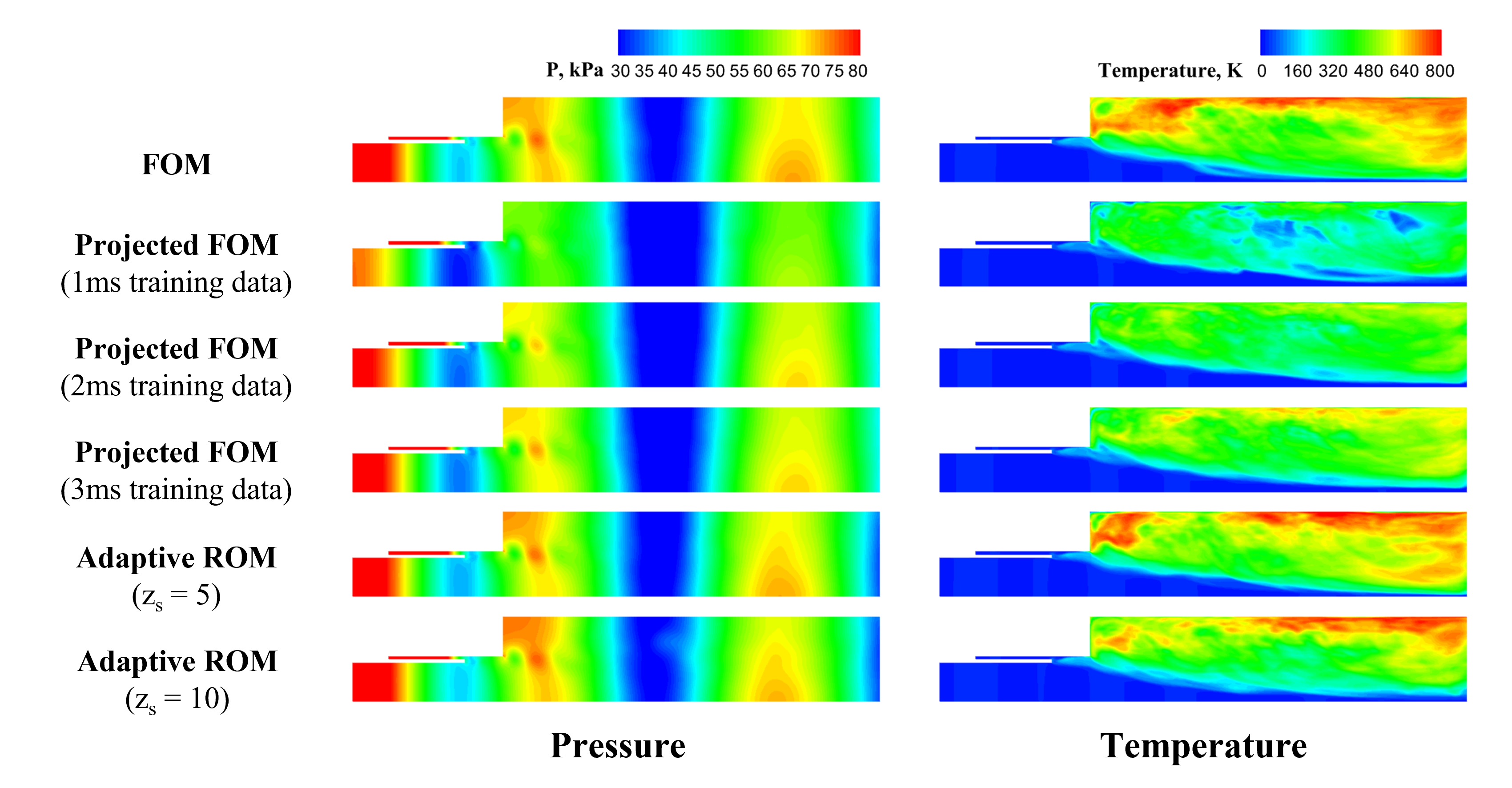}
         \caption{RMS Fields}
         \label{fig:2d:arom_vs_pod_vs_fom:rms}
     \end{subfigure}
     \caption{Comparisons of QoIs (time-averaged and RMS fields) between FOM, projected FOM, and adaptive ROM for the 2D reacting injector.}\label{fig:2d:arom_vs_pod_vs_fom:mean_rms}
\end{figure}

The time-averaged and RMS of pressure and temperature fields are compared in Figs.~\ref{fig:2d:arom_vs_pod_vs_fom:mean_rms}. For comprehensive investigations, we also include the results of projected FOM solutions computed using three sets of training snapshots in Fig.~\ref{fig:2d:pod} with number of modes required to retrieve 99.99\% of the total energy (i.e. 1ms: 116 modes, 2ms: 224 modes, and 3ms: 334 modes). In addition, the  errors of the adaptive ROM in predicting the time-averaged and RMS fields are computed to further quantify the comparisons
\begin{equation}
    \epsilon_{\QoI} = \frac{ | \QoI - \QoI_\text{ref} | }{ | \QoI_\text{ref} | },
    \label{eq:2d:errOnMeanRMS}
\end{equation}
where $\QoI$ represents the QoIs (i.e. either the time-averaged or RMS field) calculated from either the projected FOM or the adaptive ROM, and $\QoI_\text{ref}$ represents the QoIs calculated from FOM. The quantified errors are summarized in Table~\ref{table:2d:arom}. It can be readily seen that both the projected FOM and adaptive ROM solutions are able to represent the time-averaged fields reasonably accurately as shown in Fig.~\ref{fig:2d:arom_vs_pod_vs_fom:mean} with $<$ 1\% for pressure and $<$ 10\% for temperature, which indicates the sufficiency of static ROM in predicting averaged dynamics. However, as seen in Fig.~\ref{fig:2d:arom_vs_pod_vs_fom:rms}, the magnitudes of the RMS fields are largely under-predicted by the projected FOM. Specifically, $> 20\%$ errors are shown in representing temperature RMS fields even though 3ms training data (half of the testing dataset) are included, which indicates the insufficiency of static ROM in predicting unsteady dynamics. On the other hand, the adaptive ROMs are able to provide reasonably accurate predictions ($<$ 8\% for pressure and $<$ 15\% for temperature). It is worth highlighting that higher errors are observed in the pressure RMS predicted by adaptive ROM ($z_s = 5$) than the projected FOM (3ms) while the comparison in Fig.~\ref{fig:2d:arom_vs_pod_vs_fom:rms}(left) seems to indicate the opposite. This can be mostly attributed to the artifacts mentioned above when assessing the errors of convection-dominated features. Therefore, both qualitative comparisons (e.g. in Fig.~\ref{fig:2d:arom_vs_pod_vs_fom:mean_rms}) and quantitative error estimation (e.g. in Table~\ref{table:2d:arom}) are needed for comprehensive assessment of the ROM accuracy in predicting problems featuring convection-dominated physics. Overall, the evaluations in this section demonstrate the capabilities of adaptive ROMs in providing accurate predictions of long-term dynamics.

\begin{table}
\centering
\begin{tabular}{cccc} 
\toprule
QoI & Model & $\epsilon_{P}$ & $\epsilon_{T}$ \\
\midrule
\multirow{5}{10em}{Time-Averaged Fields} & Projected FOM (1ms training)  & $ 2.21 \times 10^{-3}$  & $ 8.66 \times 10^{-2}$ \\
                                         & Projected FOM (2ms training)  & $ 1.59 \times 10^{-3}$  & $ 4.76 \times 10^{-2}$ \\
                                         & Projected FOM (3ms training)  & $ 2.26 \times 10^{-3}$  & $ 3.75 \times 10^{-2}$ \\
                                         & Adaptive ROM ($z_s = 5$)      & $ 6.58 \times 10^{-3}$  & $ 7.39 \times 10^{-2}$ \\
                                         & Adaptive ROM ($z_s = 10$)     & $ 9.78 \times 10^{-3}$  & $ 9.81 \times 10^{-2}$ \\
\midrule
\multirow{5}{10em}{RMS Fields} & Projected FOM (1ms training)  & $ 1.64 \times 10^{-1}$  & $ 4.83 \times 10^{-1}$ \\
                               & Projected FOM (2ms training)  & $ 8.0 \times 10^{-2}$   & $ 3.10 \times 10^{-1}$ \\
                               & Projected FOM (3ms training)  & $ 5.53 \times 10^{-2}$  & $ 2.10 \times 10^{-1}$ \\
                               & Adaptive ROM ($z_s = 5$)      & $ 7.88 \times 10^{-2}$  & $ 1.25 \times 10^{-1}$ \\
                               & Adaptive ROM ($z_s = 10$)     & $ 4.90 \times 10^{-2}$  & $ 1.50 \times 10^{-1}$ \\
\bottomrule
\end{tabular}
\caption{\label{table:2d:arom} Comparisons of the errors in predicting the time-averaged and RMS fields of pressure ($P$) and temperature($T$) using projected FOM and adaptive ROM for the 2D reacting injector.}
\end{table}

\subsubsection{Assessment of Parametric and Transient  Prediction Capabilities}
\label{subsubsec:2dInjector:transience}

\begin{figure}
     \centering
     \begin{subfigure}[t]{0.6\textwidth}
         \centering
         \includegraphics[width=1.0\textwidth]{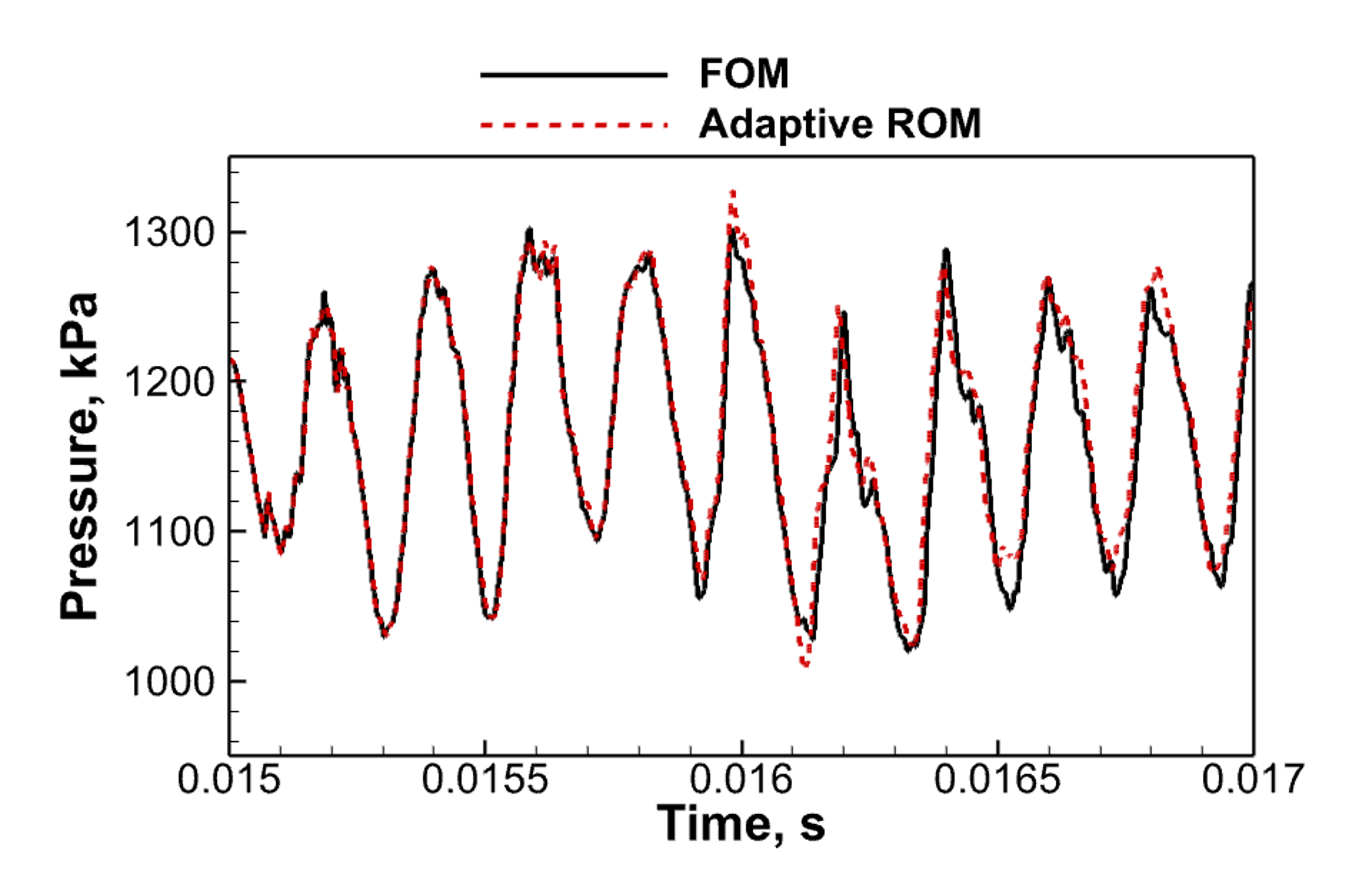}
         \caption{Local Pressure Trace (monitored at the red dot in Fig.~\ref{fig:2d:geometry})}
         \label{fig:2d:arom_mox750_P}
     \end{subfigure}
     \begin{subfigure}[t]{0.8\textwidth}
         \centering
         \includegraphics[width=1.0\textwidth]{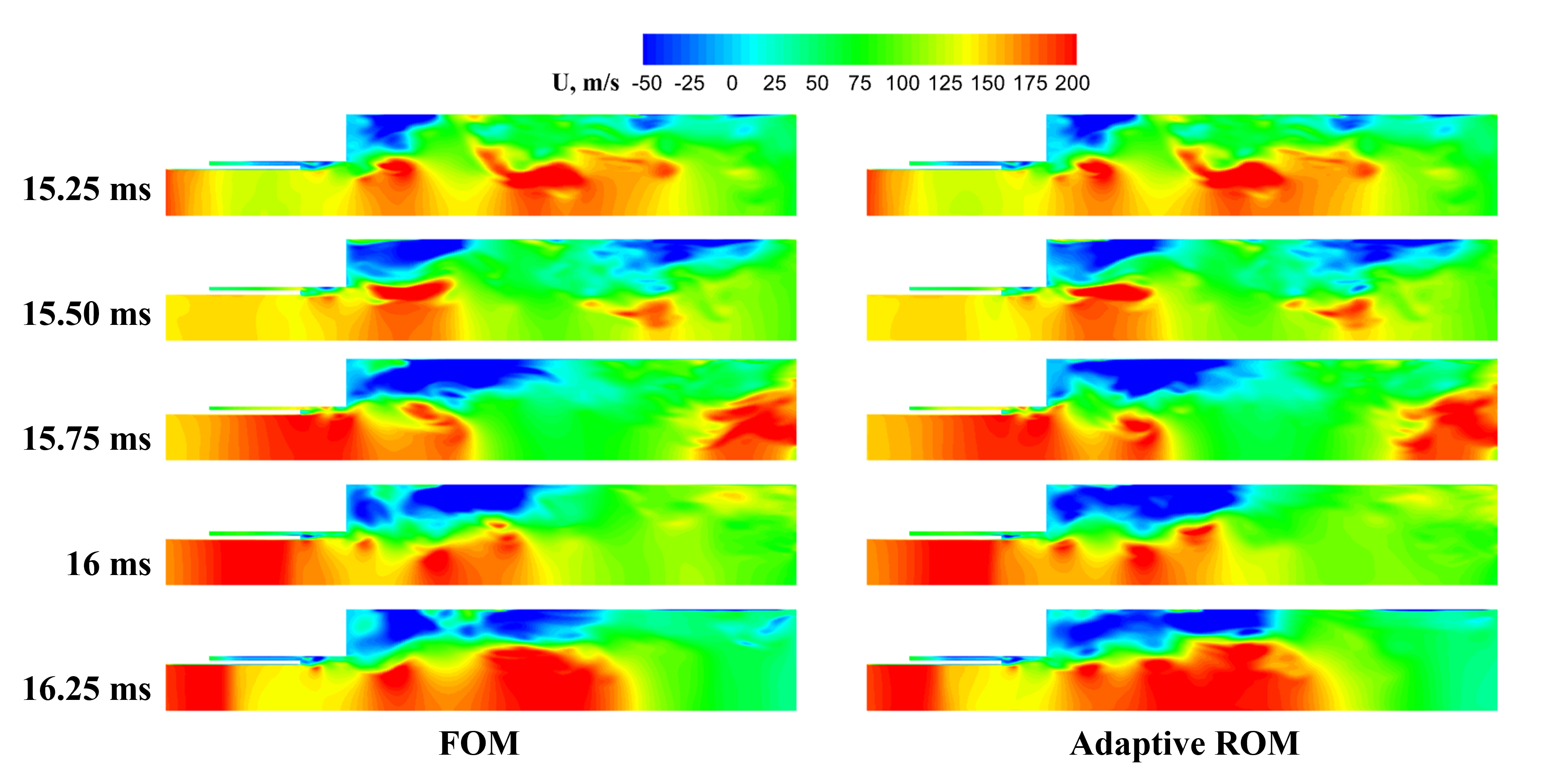}
         \caption{Axial Velocity}
         \label{fig:2d:arom_mox750_U}
     \end{subfigure}
     \begin{subfigure}[t]{0.8\textwidth}
         \centering
         \includegraphics[width=1.0\textwidth]{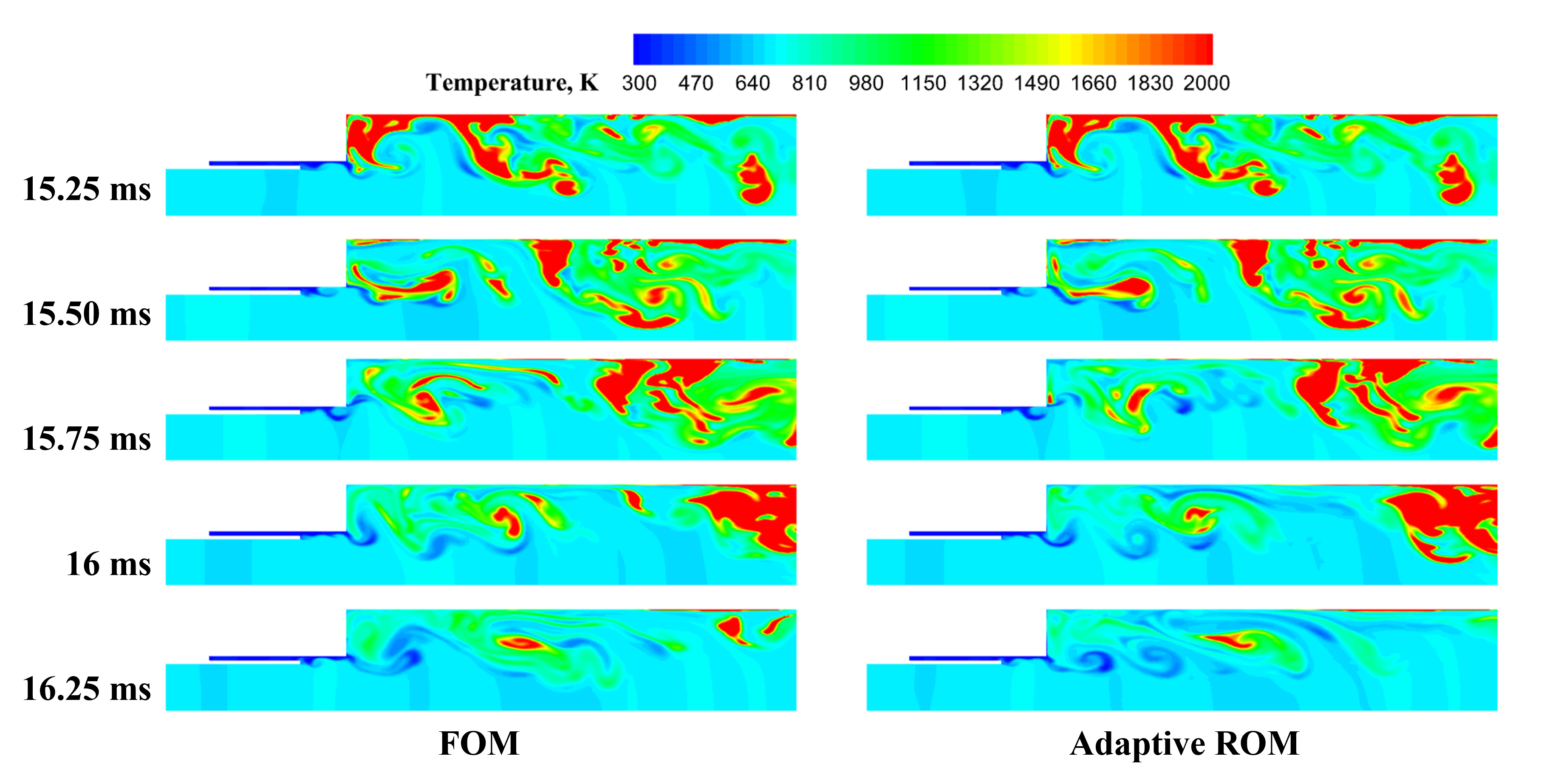}
         \caption{Temperature}
         \label{fig:2d:arom_mox750_T}
     \end{subfigure}
     \caption{Comparisons of local pressure (P) trace, unsteady axial velocity (U) and temperature (T) fields between FOM and adaptive ROM with $\Dot{m}_{ox} = 1.5\Dot{m}_\text{ox,ref}$ for the 2D reacting injector (\textit{training} dataset: 15 to 15.01ms from \emph{FOM performed with $\Dot{m}_{ox} = \Dot{m}_\text{ox,ref}$}, \textit{testing} dataset: 15 to 17ms, \textit{and} \textit{prediction} perioid: 15 to 17ms).}\label{fig:2d:arom_mox750}
\end{figure}

\begin{figure}
     \centering
     \begin{subfigure}[t]{0.6\textwidth}
         \centering
         \includegraphics[width=1.0\textwidth]{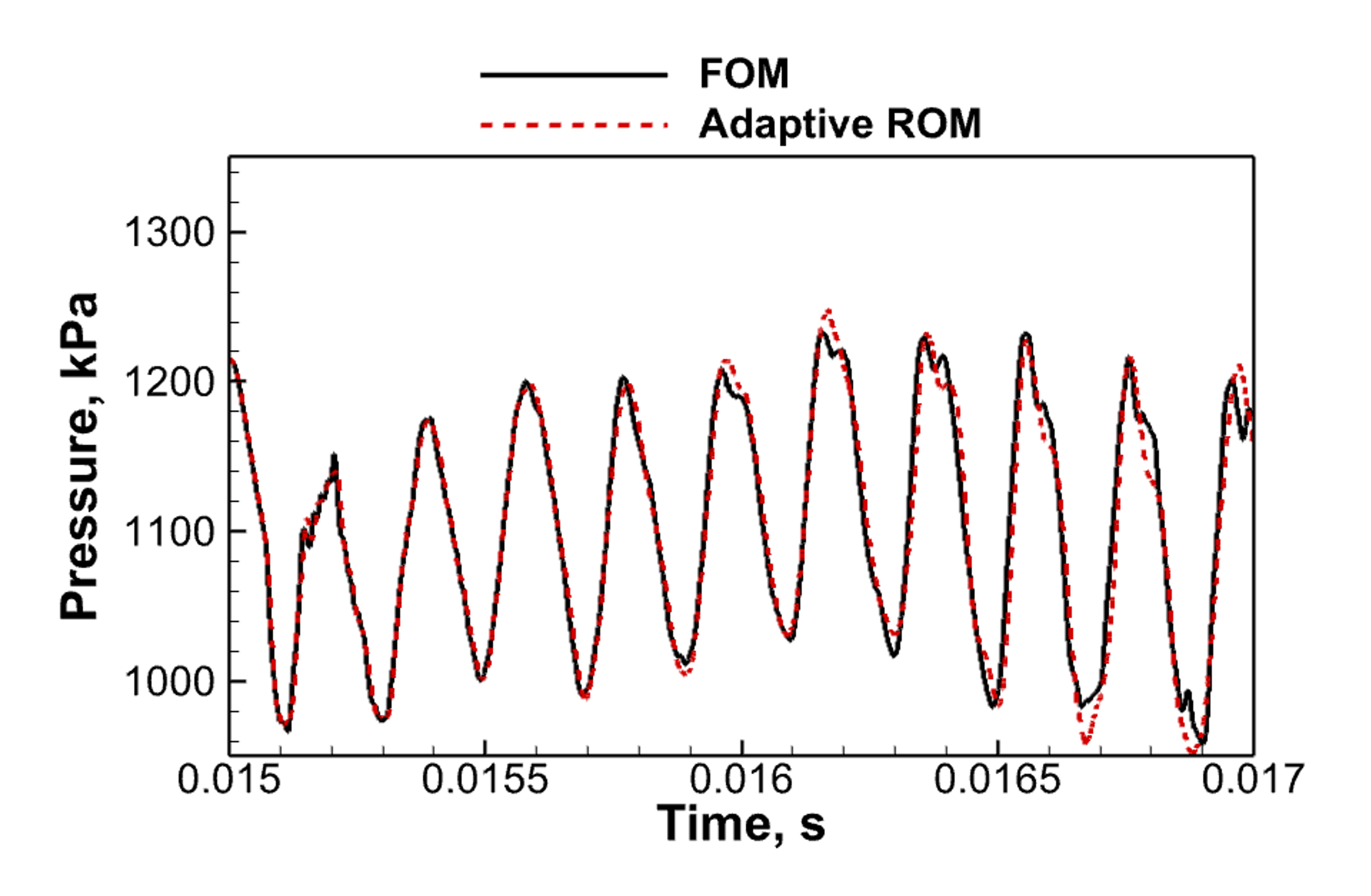}
         \caption{Local Pressure Trace (monitored at the red dot in Fig.~\ref{fig:2d:geometry})}
         \label{fig:2d:arom_mox250_P}
     \end{subfigure}
     \begin{subfigure}[t]{0.8\textwidth}
         \centering
         \includegraphics[width=1.0\textwidth]{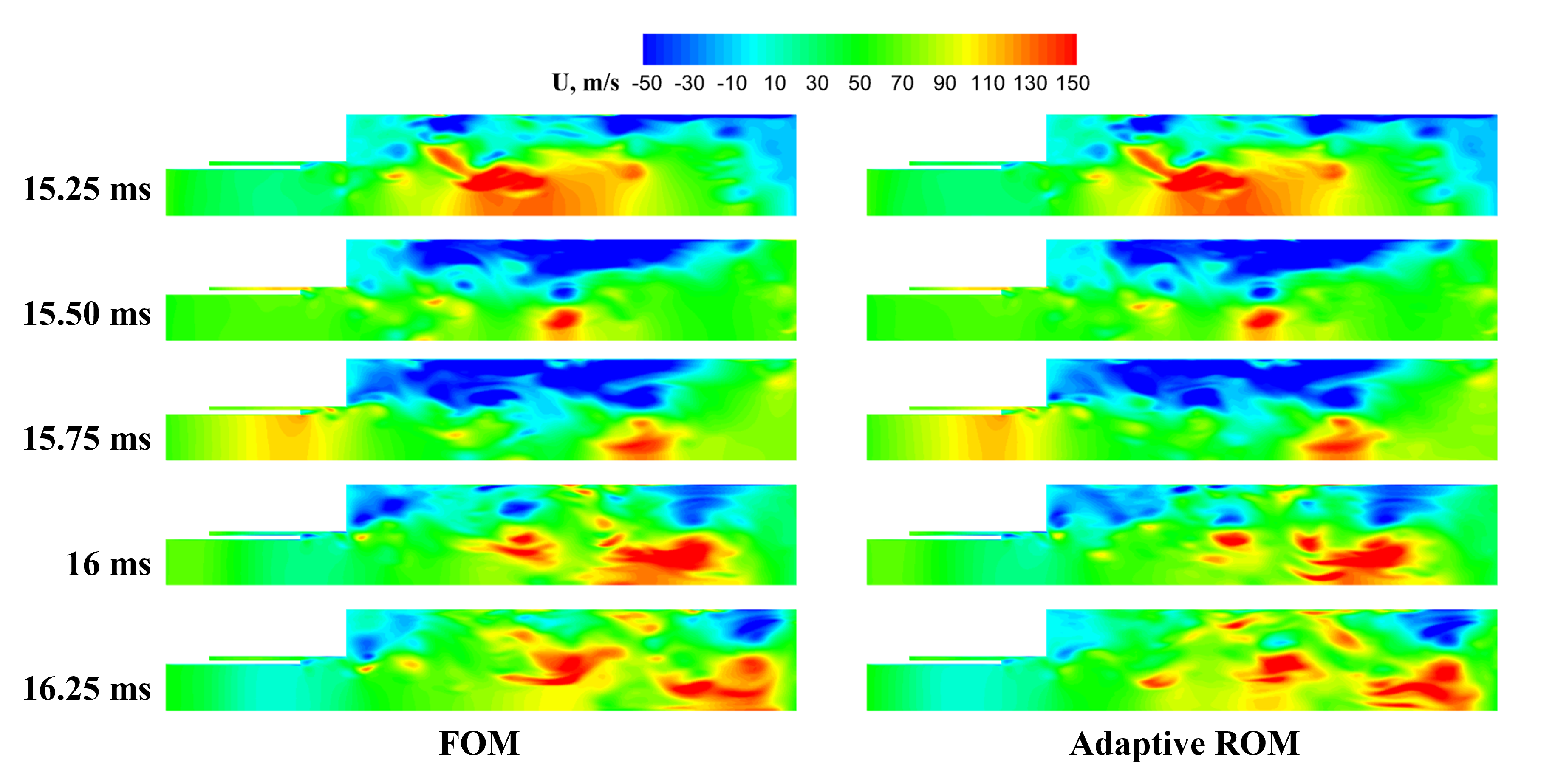}
         \caption{Axial Velocity}
         \label{fig:2d:arom_mox250_U}
     \end{subfigure}
     \begin{subfigure}[t]{0.8\textwidth}
         \centering
         \includegraphics[width=1.0\textwidth]{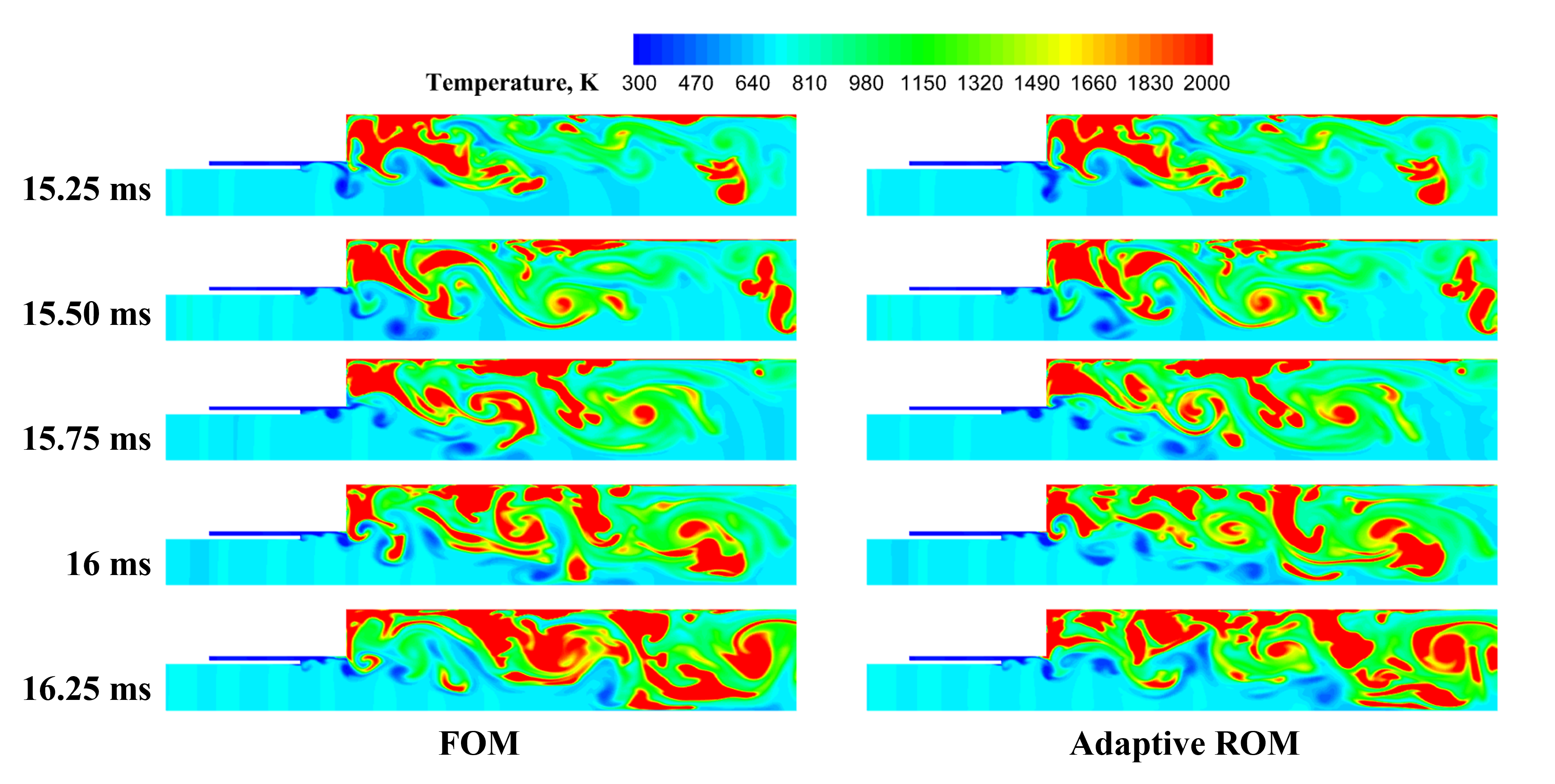}
         \caption{Temperature}
         \label{fig:2d:arom_mox250_T}
     \end{subfigure}
     \caption{Comparisons of local pressure (P) trace, unsteady axial velocity (U) and temperature (T) fields between FOM and adaptive ROM with $\Dot{m}_{ox} = 0.5\Dot{m}_\text{ox,ref}$ for the 2D reacting injector (\textit{training} dataset: 15 to 15.01ms from \emph{FOM performed with $\Dot{m}_{ox} = \Dot{m}_\text{ox,ref}$}, \textit{testing} dataset: 15 to 17ms, \textit{and} \textit{prediction} perioid: 15 to 17ms).}\label{fig:2d:arom_mox250}
\end{figure}

Finally, we assess the capabilities of the adaptive ROM in predicting transient dynamical changes due to parametric variations in operating conditions, a critical aspect in practical engineering applications. The mass flow rates of the oxidizer (i.e. $T_2$ stream in Fig.~\ref{fig:2d:geometry}) impulsively from the reference value ($\Dot{m}_\text{ox,ref}$) to the target of ($\Dot{m}_{ox}$) at 15ms with the perturbations retained as   the  in Fig.~\ref{fig:2d:fom_snapshots}. Such an abrupt change in inlet flow conditions is anticipated to lead to  strong transients and topological changes in the flow. The adaptive ROM, originally trained in the reference mass flow case with $z_s = 10$ is applied to predict the transient dynamics from 15 to 17ms, which is used as \textit{testing} dataset. Specifically, we consider two oxidizer mass flow rates ($\Dot{m}_{ox} = 1.5\Dot{m}_\text{ox,ref} \ \text{and} \ 0.5\Dot{m}_\text{ox,ref}$) for the evaluations in this section. The adaptive ROM results are compared to the FOM in terms of local pressure trace and five representative instantaneous snapshots of axial velocity (U) and temperature (T). The comparisons are shown in Figs.~\ref{fig:2d:arom_mox750} and~\ref{fig:2d:arom_mox250} for $\Dot{m}_{ox} = 1.5\Dot{m}_\text{ox,ref}$ and $\Dot{m}_{ox} = 0.5\Dot{m}_\text{ox,ref}$, respectively. It can readily be seen that by switching to a higher oxidizer mass flow rates, the nominal pressure (the averaged level that the pressure is oscillating about) rises to a higher value (Fig.~\ref{fig:2d:arom_mox750_P}) compared to the baseline (Fig.~\ref{fig:2d:arom_vs_fom:Plocal}) while lowering $\Dot{m}_\text{ox,ref}$ results in a drop in the nominal pressure (Fig.~\ref{fig:2d:arom_mox250_P}), both of which are well matched by the adaptive ROM results. More importantly, it needs to be highlighted that the adaptive ROM successfully predicts the flame extinction (Fig.~\ref{fig:2d:arom_mox750_T}) with the higher oxidizer mass flow rate, which is caused by the significantly increased local strain rates, indicated by the elevated axial velocity in the oxidizer stream (Fig.~\ref{fig:2d:arom_mox750_U}). On the other hand, the adaptive ROM also accurately captures the combustion enhancement (Fig.~\ref{fig:2d:arom_mox250_T}) with the lower oxidizer mass flow rate, which is accompanied with the reduced axial velocity in the oxidizer stream (Fig.~\ref{fig:2d:arom_mox250_U}). Again, it shall be pointed out that the adaptive ROM results in Figs.~\ref{fig:2d:arom_mox750} and~\ref{fig:2d:arom_mox250} are \textit{true} predictions from 15 to 17ms since the adaptive ROM is only initialized \emph{once} (using 10 FOM snapshots obtained with $\Dot{m}_\text{ox} = \Dot{m}_\text{ox,ref}$ in the previous section) and then directly applied for the transient dynamics predictions with no additional \textit{offline} training, similar to the 1D results in section~\ref{subsubsec:1d:parametric_arom}.   

\section{Conclusion}
\label{sec:conclusion}
Projection-based reduced-order models (ROMs) have received increasing attention over the past decade, and have been successfully deployed in some industrial settings. However, when applied to problems involving multi-scale, chaotic and transport-dominated phenomena, predictive capabilities of ROMs have been questionable. This work presents a step towards improving predictive capabilities in such complex applications. Particularly, an adaptive reduced-order model formulation is derived to \textit{break} the Kolmogorov barrier and enable \textit{true} predictions of chaotic and convection-dominant physics. Our formulation builds on recent developments in adaptive~\citep{PeherstorferADEIM,Uy_2022_adaptiveROMFlame} and   stabilized~\citep{HuangMPLSVT2022} model-order reduction and incorporates two main innovations: 1) a compact one-step basis-adaptation method formulated to update the basis to adjust for new unresolved scales in the dynamical systems evaluated on-the-fly; and 2) an efficient full-state estimation strategy developed to incorporate non-local information in ROM adaptation and inherently enables predictions of dynamics exhibiting both local and non-local coherence, which greatly enhances the predictive capabilities of the resulting ROMs. In addition,  the \textit{offline} training  cost is negligible compared to the online calculations. The adaptive procedure  enables accurate and efficient predictions of dynamics well  beyond the training window, and in a parametric setting. A detailed analysis of the computational complexity was also presented.

Comprehensive evaluations of the adaptive ROM formulation were performed  on two representative reacting flow problems, a 1D propagating laminar flame and a 2D reacting injector, characterized by convection-dominated features with multi-scale and multi-physics interactions. For the 1D propagating laminar flame problem, with non-local information incorporated in full-state estimation, the proposed formulation is shown to produce adaptive ROMs with largely improved accuracy while preserving the gains in computational efficiency, compared to the baseline formulation which only considers the local information. The resulting adaptive ROMs significantly outperform the static ROMs trained using the entire testing dataset (i.e. no prediction period) in terms of both accuracy and efficiency. The adaptive ROM is directly applied and successfully demonstrated to provide excellent parametric predictions of the 1D flame under different forcing conditions without any additional \textit{offline} training. Furthermore, the adaptive ROM's capability of predicting long-term \addrTwo{non-coherent} dynamics is demonstrated using a benchmark 2D reacting injector problem. Overall, the adaptive ROMs are able to predict the two QoIs, time-averaged and RMS fields, reasonably accurately over very long time horizons. Moreover, the potential of utilizing the adaptive ROM to predict the transient dynamics due to impulsive changes in operating conditions is successfully illustrated by varying the oxidizer mass flow rates. It is shown that the adaptive ROM is capable of predicting flame extinction and enhancement corresponding to the increase and decrease in oxidizer mass flow rates, respectively. 

With the growing use of large-scale high performance computing in a number of complex problems, the  authors believe  that formulations of this class have an important role to play, not just in many-query applications such as optimization and uncertainty quantification, but also as an in-situ accelerator for high-fidelity simulations. Thus, we envision implementation of this class of techniques as a natural part of future simulation codebases. Towards this end, though these ROMs have been  demonstrated to be truly predictive in problems with challenging features, the  formulation can be improved in several directions: 1) While the authors have realized between 3-4 orders of magnitude cost reduction using static basis ROMs, current adaptive formulations yield between one and two orders of magnitude cost reduction. Thus more detailed investigations are required on the strategies to select sampling points and in hyper-reduction, which has been shown to be important for ROM development~\citep{Wentland_SciTech2021}; 2)  scalable implementation of the adaptive ROM algorithm for large-scale engineering problems requires further development, especially in developing an effective load-balancing strategy to accommodate the adapted sampling points; 3) in the current formulation, the input parameters such as the dimension of the reduced space, the initial training window size, and the number of sampling points are fixed; it can be more beneficial to have them adapted over time.

\section*{Acknowledgments}
{The authors acknowledge many insightful discussions with Profs. Benjamin Peherstorfer (NYU), Karen Willcox (UT Austin), and Charles Merkle (Purdue). This work was funded by the Air Force under the Center of Excellence grant FA9550-17-1-0195, titled Multi-Fidelity Modeling of Rocket Combustor Dynamics (Program Managers: Dr. Mitat Birkan, Dr. Fariba Fahroo, and Dr. Ramakanth Munipalli).}

\appendix
\section{ Governing Equations for Full Order Model}
\label{appendix:fom_eq}

The full order model computations are carried out with an in-house CFD code, the General Equations and Mesh Solver (GEMS), the capabilities of which has been successfully demonstrated in modeling rocket combustion instabilities~\cite{HarvazinskiPoF}. GEMS solves the conservation equations for mass, momentum, energy and species mass fractions in a coupled fashion,
\begin{equation}
    \frac{\partial{Q}}{\partial{t}}+\nabla \cdot\left(\vec{F}-\vec{F_v}\right)=H,
    \label{eq:fom:governing}
\end{equation}
where $Q$ is the vector of conserved variables defined as, $Q=\left(\begin{array}{cccccc}
        \rho & \rho{u} & \rho{v} & \rho{w} & \rho{h^0-p} & \rho{Y_l}\\
    \end{array}\right)^T$
with $\rho$ representing density, $u$, $v$ and $w$ representing velocity field, $Y_l$ representing the $l^{th}$ species mass fraction and the total enthalpy $h^0$ is defined as, $h^0=h+\frac{1}{2}(u^2_i)=\sum_l{h_l{Y_l}}+\frac{1}{2}(u^2_i)$.

The fluxes have been separated into inviscid, $\vec{F}=F_{i}\vec{i}+F_{j}\vec{j}+F_{k}\vec{k}$ and viscous terms, $\vec{F_v}=F_{v,i}\vec{i}+F_{v,j}\vec{j}+F_{v,k}\vec{k}$. And the three inviscid fluxes are,
\begin{equation}
    F_i = \left(\begin{array}{c}
        \rho{u} \\
        \rho{u^2}+p \\
        \rho{uv} \\
        \rho{uw} \\
        \rho{uh^0} \\
        \rho{uY_l} \\
    \end{array} \right), \; F_j = \left(\begin{array}{c}
        \rho{v} \\
        \rho{uv} \\
        \rho{v^2}+p \\
        \rho{vw} \\
        \rho{vh^0} \\
        \rho{vY_l} \\
    \end{array} \right) \; \text{and} \; F_k = \left(\begin{array}{c}
        \rho{w} \\
        \rho{uw} \\
        \rho{vw}+p \\
        \rho{w^2}+p \\
        \rho{wh^0} \\
        \rho{wY_l} \\
    \end{array} \right)
    \label{eq:fom:inviscid_fluxes}
\end{equation}
The viscous fluxes are,
\begin{equation}
    \scalebox{0.8}{$F_{v,i} = \left(\begin{array}{c}
        0 \\
        \tau_{ii} \\
        \tau_{ji} \\
        \tau_{ki} \\
        u\tau_{ii}+v\tau_{ji}+w\tau_{ki}-q_i \\
        \rho{D_{l}}\frac{\partial{Y_l}}{\partial{x}} \\
    \end{array} \right), \; F_{v,j} = \left(\begin{array}{c}
        0 \\
        \tau_{ij} \\
        \tau_{jj} \\
        \tau_{kj} \\
        u\tau_{ij}+v\tau_{jj}+w\tau_{kj}-q_j \\
        \rho{D_{l}}\frac{\partial{Y_l}}{\partial{y}} \\
    \end{array} \right) \; \text{and} \; F_{v,k} = \left(\begin{array}{c}
        0 \\
        \tau_{ik} \\
        \tau_{jk} \\
        \tau_{kk} \\
        u\tau_{ik}+v\tau_{jk}+w\tau_{kk}-q_k \\
        \rho{D_{l}}\frac{\partial{Y_l}}{\partial{z}} \\
    \end{array} \right)$}
    \label{eq:fom:viscous_fluxes}
\end{equation}
where $D_l$ is defined to be the diffusion of the $l^{th}$  species into the mixture. In practice, this is an approximation used to model the multicomponent diffusion as the binary diffusion of each species into a mixture.

The heat flux in the $i^{th}$ direction, $q_i$, is defined as,
\begin{equation}
    q_i = -K\frac{\partial{T}}{\partial{x_i}}+\rho\sum^N_{l=1}D_l\frac{\partial{Y_l}}{\partial{x_i}}h_l+\mathbf{Q}_{\text{source}}
    \label{eq:fom:heat_flux}
\end{equation}
The three terms in the heat flux represent the heat transfer due to the conduction, species diffusion and heat generation from a volumetric source (e.g. heat radiation or external heat source) respectively.

The shear stress, $\tau$ , is also found in the viscous flux and defined in terms of the molecular viscosity and velocity field,
\begin{equation}
    \tau_{ij} = \mu\left(\frac{\partial{u_i}}{\partial{x_j}}+\frac{\partial{u_j}}{\partial{x_i}}-\frac{2}{3}\frac{\partial{u_m}}{\partial{x_m}}\delta_{ij}\right)
    \label{eq:fom:shear_stress}
\end{equation}

The source term, $H$ includes a single entry for each of the species equations signifying the production or destruction of the $l^{th}$ species, $\dot{\omega}_l$, which is determined by the chemical kinetics~\cite{WestbrookDryer},
\begin{equation}
    H=\left(\begin{array}{cccccc}
        0 & 0 & 0 & 0 & 0 & \dot{\omega}_l\\
    \end{array}\right)^T
    \label{eq:fom:source_term}
\end{equation}
\section{Linear Analysis of Adaptivity}

\subsection{Derivation of Eq.~\ref{eq:StateEstSampling:sampled_full_state}}
\label{appendix:sampled_full_state}
Substituting $\Tilde{\linearFOM} = \mathbf{V} \bar{\linearFOM}_r$ to Eq.~\ref{eq:StateEstSampling:sampled_full_state_def}, we have
\begin{equation}
    \mathbf{S}^T \mathbf{B} ( \mathbf{S} \mathbf{S}^T\hat{\linearFOM}^n + \mathbf{S}^\ast \mathbf{S}^{\ast T} \mathbf{V} \bar{\linearFOM}_r^n ) = \mathbf{S}^T  \linearFOM^{n-1},
\end{equation}
with $\mathbf{B}\triangleq \mathbf{I} - \dt \mathbf{J}$. Eq.~\ref{eq:appendix:sampled_full_state} is then rearranged to
\begin{equation}
    ( \mathbf{S}^T \mathbf{B} \mathbf{S} ) \mathbf{S}^T\hat{\linearFOM}^n = \mathbf{S}^T  \linearFOM^{n-1} - \mathbf{S}^T \mathbf{B} \mathbf{S}^\ast \mathbf{S}^{\ast T} \mathbf{V} \bar{\linearFOM}_r^n,
    \label{eq:appendix:sampled_full_state}
\end{equation}
Recalling that the basis is assumed to fully resolve the state variable at time step $n - 1$ (i.e. $\linearFOM^{n-1} = \mathbf{V} \bar{\linearFOM}_r^{n-1}$ \textit{or} $\bar{\linearFOM}_r^{n-1} = \mathbf{V}^T \linearFOM^{n-1}$), based on Eq.~\ref{eq:predictor-corrector:g_reduced_states}, for Galerkin ROM 
\begin{equation}
    \mathbf{V} \bar{\linearFOM}_r^{n} = \mathbf{V} (\mathbf{V}^T \mathbf{B} \mathbf{V} )^{-1} \mathbf{V}^T \linearFOM^{n-1}.
    \label{eq:appendix:g_reduced_states}
\end{equation}
Similarly, based on Eq.~\ref{eq:predictor-corrector:lspg_reduced_states}, for LSPG ROM
\begin{equation}
    \mathbf{V} \bar{\linearFOM}_r^{n} = \mathbf{V} (\mathbf{V}^T \mathbf{B}^T \mathbf{B} \mathbf{V} )^{-1}  (\mathbf{V}^T \mathbf{B}^T \mathbf{V} ) \mathbf{V}^T \linearFOM^{n-1}.
    \label{eq:appendix:lspg_reduced_states}
\end{equation}
For compactness, we define $\mathbf{B}_v^{-1}\triangleq \mathbf{V} (\mathbf{V}^T \mathbf{B} \mathbf{V} )^{-1}\mathbf{V}^T$ and $\mathbf{B}_v^{-1}\triangleq \mathbf{V}(\mathbf{V}^T \mathbf{B}^T \mathbf{B} \mathbf{V} )^{-1} ( \mathbf{V}^T \mathbf{B}^T \mathbf{V} ) \mathbf{V}^T$, for Galerkin and LSPG, respectively, which simplify Eqs.~\ref{eq:appendix:g_reduced_states} and~\ref{eq:appendix:lspg_reduced_states} to
\begin{equation}
    \mathbf{V} \bar{\linearFOM}_r^{n} = \mathbf{B}_v^{-1} \linearFOM^{n-1}.
    \label{eq:appendix:reduced_states}
\end{equation}
Substituting Eq.~\ref{eq:appendix:reduced_states} into Eq.~\ref{eq:appendix:sampled_full_state}, respectively, we can obtain
\begin{equation}
    \mathbf{S}^T\hat{\linearFOM}^n=
    (\mathbf{S}^T\mathbf{B} \mathbf{S})^{-1} ( \mathbf{S}^T - \mathbf{S}^T \mathbf{B} \mathbf{S}^\ast \mathbf{S}^{\ast T} \mathbf{B}_v^{-1}) \linearFOM^{n-1},
\end{equation}

\subsection{Derivation of Eq.~\ref{eq:StateEstSampling:error1}}
\label{appendix:StateEstError}
We start with $\mathbf{e} \triangleq \linearFOM^n - (\mathbf{S} \mathbf{S}^T\hat{\linearFOM}^n + \mathbf{S}^\ast \mathbf{S}^{\ast T} \Tilde{\linearFOM}^n)$ and substitute Eqs.~\ref{eq:appendix:reduced_states} and~\ref{eq:StateEstSampling:sampled_full_state} with $\Tilde{\linearFOM}^n = \mathbf{V} \bar{\linearFOM}^n_r$ and $\linearFOM^{n} = \mathbf{B}^{-1} \linearFOM^{n-1}$ from Eq.~\ref{eq:predictor-corrector:fom}
\begin{align*}
\mathbf{e} & \triangleq \linearFOM^n - (\mathbf{S} \mathbf{S}^T\hat{\linearFOM}^n + \mathbf{S}^\ast \mathbf{S}^{\ast T} \Tilde{\linearFOM}^n) \\
& = \mathbf{B}^{-1} \linearFOM^{n-1} - \mathbf{S} (\mathbf{S}^T\mathbf{B} \mathbf{S})^{-1} ( \mathbf{S}^T - \mathbf{S}^T \mathbf{B} \mathbf{S}^\ast \mathbf{S}^{\ast T} \mathbf{B}_v^{-1}) \linearFOM^{n-1} - \mathbf{S}^\ast \mathbf{S}^{\ast T} \mathbf{B}_v^{-1} \linearFOM^{n-1} \\
& \text{with} \ \mathbf{B}_s^{-1} \triangleq \mathbf{S} (\mathbf{S}^T \mathbf{B} \mathbf{S} )^{-1}\mathbf{S}^T \\
& = ( \mathbf{B}^{-1} - \mathbf{B}_s^{-1} + \mathbf{B}_s^{-1} \mathbf{B} \mathbf{S}^\ast \mathbf{S}^{\ast T} \mathbf{B}_v^{-1} - \mathbf{S}^\ast \mathbf{S}^{\ast T} \mathbf{B}_v^{-1} ) \linearFOM^{n-1} \\
& = ( \mathbf{B}^{-1} - \mathbf{B}_s^{-1} ) ( \mathbf{I} - \mathbf{B} \mathbf{S}^\ast \mathbf{S}^{\ast T} \mathbf{B}_v^{-1} ) \linearFOM^{n-1}.
\end{align*}

\subsection{Derivation of Eq.~\ref{eq:NonLocInfo:error1}}
\label{appendix:NonLocInfoStateEstError}
We start with $\mathbf{e} \triangleq \linearFOM^n - ( \mathbf{S} \mathbf{S}^T \hat{\linearFOM}^n + \mathbf{S}^\ast \mathbf{S}^{\ast T} \hat{\linearFOM}^n )$ and substitute $\mathbf{S}^{\ast T}\hat{\linearFOM}^n = ( \mathbf{S}^{\ast T} \mathbf{B}^{\ast} \mathbf{S}^{\ast} )^{-1} ( \mathbf{S}^{\ast T}  \linearFOM^{n-z_s} - \mathbf{S}^{\ast T} \mathbf{B}^{\ast} \mathbf{S} \mathbf{S}^{T} \hat{\linearFOM}^n )$ from Eq.~\ref{eq:NonLocInfo:unsampled_full_state}
\begin{align*}
\mathbf{e} & \triangleq \linearFOM^n - (\mathbf{S} \mathbf{S}^T\hat{\linearFOM}^n + \mathbf{S}^\ast \mathbf{S}^{\ast T} \hat{\linearFOM}^n) \\
& = \linearFOM^n - \mathbf{S} \mathbf{S}^T\hat{\linearFOM}^n - \mathbf{S}^\ast ( \mathbf{S}^{\ast T} \mathbf{B}^{\ast} \mathbf{S}^{\ast} )^{-1} ( \mathbf{S}^{\ast T}  \linearFOM^{n-z_s} - \mathbf{S}^{\ast T} \mathbf{B}^{\ast} \mathbf{S} \mathbf{S}^{T} \hat{\linearFOM}^n ) \\
& \text{with} \ \mathbf{B}_{s^\ast}^{-1} \triangleq \mathbf{S}^{\ast} (\mathbf{S}^{\ast T} \mathbf{B}^{\ast} \mathbf{S}^{\ast} )^{-1}\mathbf{S}^{\ast T} \\
& = ( \linearFOM^n - \mathbf{B}_{s^\ast}^{-1}  \linearFOM^{n-z_s} ) - ( \mathbf{S} \mathbf{S}^T  - \mathbf{B}_{s^\ast}^{-1} \mathbf{B}^{\ast} \mathbf{S} \mathbf{S}^{T} ) \hat{\linearFOM}^n \\
& \text{with} \ \linearFOM^{n} = \mathbf{B}^{-1} \linearFOM^{n-1} \text{and} \ \linearFOM^{n - z_s} = \mathbf{B}^{z_s - 1} \linearFOM^{n - 1} \text{from Eq.~\ref{eq:predictor-corrector:fom}} \\
& = ( \mathbf{B}^{-1} - \mathbf{B}_{s^\ast}^{-1}  \mathbf{B}^{z_s - 1} ) \linearFOM^{n - 1}  - ( \mathbf{S} \mathbf{S}^T  - \mathbf{B}_{s^\ast}^{-1} \mathbf{B}^{\ast} \mathbf{S} \mathbf{S}^{T} ) \hat{\linearFOM}^n \\
& \text{Substitute $\mathbf{S}^T\hat{\linearFOM}^n = (\mathbf{S}^T\mathbf{B} \mathbf{S})^{-1} ( \mathbf{S}^T - \mathbf{S}^T \mathbf{B}\mathbf{S}^\ast \mathbf{S}^{\ast T} \mathbf{B}_v^{-1}) \linearFOM^{n-1}$ from Eq.~\ref{eq:StateEstSampling:sampled_full_state}} \\
& = ( \mathbf{B}^{-1} - \mathbf{B}_{s^\ast}^{-1}  \mathbf{B}^{z_s - 1} ) \linearFOM^{n - 1}  - ( \mathbf{S} - \mathbf{B}_{s^\ast}^{-1} \mathbf{B}^{\ast} \mathbf{S} ) (\mathbf{S}^T\mathbf{B} \mathbf{S})^{-1} ( \mathbf{S}^T - \mathbf{S}^T \mathbf{B} \mathbf{S}^\ast \mathbf{S}^{\ast T} \mathbf{B}_v^{-1}) \linearFOM^{n-1} \\
& \text{with} \ \mathbf{B}_s^{-1} \triangleq \mathbf{S} (\mathbf{S}^T \mathbf{B} \mathbf{S} )^{-1}\mathbf{S}^T \\
& = ( \mathbf{B}^{-1} - \mathbf{B}_{s^\ast}^{-1}  \mathbf{B}^{z_s - 1} ) \linearFOM^{n - 1}  - ( \mathbf{I} - \mathbf{B}_{s^\ast}^{-1} \mathbf{B}^{\ast} ) ( \mathbf{B}_s^{-1} - \mathbf{B}_s^{-1} \mathbf{B} \mathbf{S}^\ast \mathbf{S}^{\ast T} \mathbf{B}_v^{-1}) \linearFOM^{n-1} \\
& = [ \mathbf{B}^{-1} - \mathbf{B}_{s^\ast}^{-1}  \mathbf{B}^{z_s - 1} - ( \mathbf{B}_s^{-1} - \mathbf{B}_{s^\ast}^{-1} \mathbf{B}^{\ast} \mathbf{B}_s^{-1} ) ( \mathbf{I} - \mathbf{B} \mathbf{S}^\ast \mathbf{S}^{\ast T} \mathbf{B}_v^{-1}) ] \linearFOM^{n-1} \\
& \addrOne{ = [ \mathbf{B}^{-1} - \mathbf{B}_{s^\ast}^{-1}  \mathbf{B}^{z_s - 1} - ( \mathbf{B}^{\ast -1} - \mathbf{B}_{s^\ast}^{-1} ) \mathbf{B}^{\ast} \mathbf{B}_s^{-1} ( \mathbf{I} - \mathbf{B} \mathbf{S}^\ast \mathbf{S}^{\ast T} \mathbf{B}_v^{-1}) ] \linearFOM^{n-1} },
\end{align*}
\addrOne{which shows mitigated contributions from the projection error at the unsampled points, $( \mathbf{I} - \mathbf{B} \mathbf{S}^\ast \mathbf{S}^{\ast T} \mathbf{B}_v^{-1})$, to the total error, $\mathbf{e}$ compared with the one in~\ref{appendix:StateEstError}.}

\subsection{Derivation of Eq.~\ref{eq:StateEstSampling:errorBasis}}
\label{appendix:basis_err}
In this section, we estimate the difference between the exact and the actual basis adaptation with sampling
\begin{align*}
& ||\delta\mathbf{V}_{exact}-\mathbf{S}\mathbf{S}^T{\delta\mathbf{V}}||_2 \\
= & \norm{ \frac{(\linearFOM^n-\Tilde{\linearFOM}^n)(\bar{\linearFOM}_r^n)^T - \mathbf{S}(\mathbf{S}^T \hat{\linearFOM}^{n} -\mathbf{S}^T\Tilde{\linearFOM}^n)(\bar{\linearFOM}_r^n)^T}{||\bar{\linearFOM}_r^{n}||_2^2} } \\
= & \norm{ \frac{(\linearFOM^n - \Tilde{\linearFOM}^n - \mathbf{S}\mathbf{S}^T \hat{\linearFOM}^{n} -\mathbf{S}\mathbf{S}^T \Tilde{\linearFOM}^n)(\bar{\linearFOM}_r^n)^T}{||\bar{\linearFOM}_r^{n}||_2^2} } \\
& \text{with} \ \mathbf{I} - \mathbf{S}\mathbf{S}^T = \mathbf{S}^\ast \mathbf{S}^{\ast T} \\
= & \norm{ \frac{(\linearFOM^n - \mathbf{S}\mathbf{S}^T \hat{\linearFOM}^{n} -\mathbf{S}^\ast \mathbf{S}^{\ast T}\Tilde{\linearFOM}^n)(\bar{\linearFOM}_r^n)^T}{||\bar{\linearFOM}_r^{n}||_2^2} } \\
& \text{with} \ \mathbf{e} \triangleq \linearFOM^n - (\mathbf{S} \mathbf{S}^T\hat{\linearFOM}^n + \mathbf{S}^\ast \mathbf{S}^{\ast T} \Tilde{\linearFOM}^n) \ \text{defined in Eq.~\ref{eq:StateEstSampling:error1}} \\
= & \norm{ \frac{ \mathbf{e} (\bar{\linearFOM}_r^n)^T }{ ||\bar{\linearFOM}_r^{n}||_2^2 } } \\
& \text{using} \norm{\mathbf{x}\mathbf{y}^T} =   \norm{\mathbf{x}} \norm{\mathbf{y}} \ \\
= & \frac{ || \mathbf{e} ||_2 }{ ||\bar{\linearFOM}_r^{n}||_2 }
\end{align*}

\section{Floating-point Operations (FLOPs) Analysis for Computational Models}
Here, we provide details on FLOPs analysis for FOM and static-basis hyper-reduced MP-LSVT ROM, both of which are solved with the implicit time-integration scheme using dual time-stepping method. Both models are propagated for $z_s$ physical time steps, each of which contains $K$ pseudo iterations for dual time stepping. In addition, the notation used for the analysis is presented as follows: 
\begin{enumerate}
    \item $\numDOF$ is the total number of degrees of freedom in the system;
    \item $\numElements$ is the total number of elements/cells;
    \item $\numSamps$ is the total number of sampling elements/cells to construct hyper-reduced ROM;
    \item $\numVars$ is the number of state variables in each element/cell;
    \item $\omega_\mathbf{r}$ is the number of operations to evaluate one nonlinear FOM equation residual;
    \item $\resFunc{\solPrimFOM^\iterIdx}$ in Eq.~\ref{eq:fom_linear_multi_discrete}, which scales with $\numVars$ (i.e. $\omega_\mathbf{r} = a \numVars$ ) and is usually on the order of $O(10^1)$ to $O(10^3)$ for complex systems;
    \item $\omega_{\mathbf{J}_r}$ is the number of operations to evaluate one row in the Jacobian matrix of the nonlinear FOM equation residual, $\mathbf{J}_r = \partial \resFunc{\solPrimFOM} / \partial \solPrimFOM$, which scales with $\numVars$ (i.e. $\omega_{\mathbf{J}_r} = b \numVars$ ) and is usually on the order of $O(10^2)$ to $O(10^4)$ for complex systems, and $\eta$ is the number of operations for sparse-matrix linear solve;
    \item $\numSolModes$ is the number of trial basis modes included to construct ROM.
\end{enumerate}

\subsection{FLOPs analysis for the FOM}
\label{appendix:flops_fom}

The approximated FLOPs for one FOM pseudo iteration is provided in Table~\ref{table:flops_FOM}. Based upon this, the FLOPs, for $z_s$ physical time steps, each of which contains $K$ pseudo iterations for dual time-stepping, are given as
\begin{equation}
    \text{FLOP}_\text{FOM} = z_s K ( a + b + \eta + \frac{4}{\numVars} ) \numVars^2 \numElements
\end{equation}

\begin{table}
\centering
\begin{tabular}{ll} 
\toprule
\textbf{Operations} & \textbf{Approximate FLOPs} \\
\midrule
Evaluate the nonlinear FOM equation residual: $\resFunc{\solPrimFOM^{k-1}}$ & $a \numVars^2 \numElements$ \\
Evaluate the Jacobian matrix: $\mathbf{J}^{k-1}_r = [ \partial \resFunc{\solPrimFOM} / \partial \solPrimFOM ]_{\solPrimFOM = \solPrimFOM^{k-1}}$ & $b \numVars^2 \numElements$ \\
Solve the linearized FOM: $\mathbf{J}^{k-1}_r ( \solPrimFOM^k - \solPrimFOM^{k-1} ) = - \resFunc{\solPrimFOM^{k-1}}$ & $4 \numVars \numElements + \eta \numVars^2 \numElements$ \\
\midrule
\textbf{Total} & $ ( a + b + \eta + \frac{4}{\numVars} ) \numVars^2 \numElements $ \\
\bottomrule
\end{tabular}
\caption{\label{table:flops_FOM} Approximated floating-point operations for FOM calculations in terms of one pseudo iteration from $k-1$ to $k$.}
\end{table}

\subsection{FLOPs analysis for the static-basis hyper-reduced MP-LSVT ROM}
\label{appendix:flops_srom}

The approximated FLOPs for one hyper-reduced MP-LSVT ROM pseudo iteration is provided in Table~\ref{table:flops_ROM}. Based upon this, the FLOPs, for $z_s$ physical time steps, each of which contains $K$ pseudo iterations for dual time-stepping, are given as
\begin{equation}
    \text{FLOP}_\text{SROM} = z_s K \left[ \numVars^2 \numSamps \lp a + b + 2 \numSolModes + \frac{3 \numSolModes + 2\numSolModes^2 + 3}{\numVars} - \frac{\numSolModes^2 + \numSolModes}{\numVars^2} \rp - \numSolModes + \numSolModes^2 + 3 \numSolModes^3 \right]
\end{equation}

\begin{table}
\centering
\begin{tabular}{ll} 
\toprule
\textbf{Operations} & \textbf{Approximate FLOPs} \\
\midrule
Evaluate the residual (at the sampling points): $\sampMat^T \resFunc{\solPrimFOM^{k-1}}$ & $a \numVars^2 \numSamps$ \\
Evaluate the Jacobian matrix (at the sampling points): $\sampMat^T \mathbf{J}_r$ & $b \numVars^2 \numSamps$ \\
Compute the test basis $\testBasisPrimGPOD^{k-1}$ (Eq.~\ref{eq:mplsvt_w_hyper}) & $ \numSamps ( 2 \numVars^2 \numSolModes + \numSolModes \numVars + 2 \numVars \numSolModes^2 - \numSolModes^2 )$ \\
Compute $ \mathbf{r}_s^k = \lp \sampMat^T \testBasisPrim^k \rp^T \left[ \lp \sampMat^T \resBasis \rp^{+} \right]^T \lp \sampMat^T \resBasis \rp^{+} \sampMat^T \scaleMatCons \resFunc{\solPrimROMFull^{k}}$ & $ \numSamps ( 3 \numVars + 2 \numSolModes \numVars - \numSolModes ) + 2 \numSolModes^2 - \numSolModes$ \\
Compute $(\testBasisPrimGPOD^{k-1})^T \testBasisPrimGPOD^\iterIdx$ & $2 \numSolModes^3 - \numSolModes^2$ \\
Solve the linearized ROM: & \\
$ [(\testBasisPrimGPOD^{k-1})^T \testBasisPrimGPOD^{k-1}] ( \solPrimROMRed^{k} - \solPrimROMRed^{k-1} ) = - \mathbf{r}_s^k $ & $ \numSolModes^3 $  \\
\midrule
\textbf{Total} & $  \numVars^2 \numSamps \lp a + b + 2 \numSolModes + \frac{3 \numSolModes + 2\numSolModes^2 + 3}{\numVars} - \frac{\numSolModes^2 + \numSolModes}{\numVars^2} \rp $ \\
& $ - \numSolModes + \numSolModes^2 + 3 \numSolModes^3$ \\
\bottomrule
\end{tabular}
\caption{\label{table:flops_ROM} Approximated floating-point operations for hyper-reduced MP-LSVT ROM calculations in terms of one pseudo iteration from $k-1$ to $k$.}
\end{table}
\bibliographystyle{elsarticle-num}
\bibliography{ref.bib}

\end{document}